\documentclass[final]{IEEEtran}

\usepackage{graphicx}
\usepackage{subfigure}
\usepackage{multirow}
\usepackage{array}
\newcolumntype{P}[1]{>{\centering\arraybackslash}p{#1}}
\newcolumntype{M}[1]{>{\centering\arraybackslash}m{#1}}

\usepackage{amsthm,amssymb,amsmath,graphicx,multirow,color,amsfonts}%
\usepackage[update,prepend]{epstopdf}
\usepackage{multirow}
\usepackage[latin1]{inputenc}
\usepackage{tikz}
\usepackage{bbm} % for \mathbbm{1}
\usepackage{pdfpages}
\usepackage{multirow}
\usepackage{subfig}
\usepackage{comment}

\usepackage{graphicx}
\usepackage{multicol}
\usepackage{cite}

\usepackage{caption}
\usepackage{textcomp}
\usepackage{psfrag}
\usepackage{arydshln}
\usepackage{url}
\usepackage{soul}
\usepackage{graphicx,xcolor}
\usepackage[nolist]{acronym}
\usepackage{algorithm,algorithmic} %algorithm
\usepackage{subfigure}

\usepackage{mathtools,lipsum}
\usepackage{cuted}
\usepackage{amsmath}
\newtheorem{corollary}{Corollary}
\usepackage[capitalise]{cleveref}
\Crefname{equation}{Eq.\!}{Eqs.\!}
\Crefname{figure}{Fig.\!}{Figs.\!}
\Crefname{tabular}{Tab.\!}{Tabs.\!}
\Crefname{section}{Section\!}{Sections.\!}

\def\nb0{{\mathbf{0}}}
\def\nb1{{\mathbf{1}}}

\newtheorem{lemma}{Lemma}

\newtheorem{theorem}{Theorem}

\newtheorem{assumption}{Assumption}
	
\begin{document}

\title{Asymptotic SINR Analysis of Linear Detectors in Uplink Massive NTN MIMO Transmissions
}
\author{
Xue Zhang, {\em Member, IEEE}, Abla Kammoun, {\em Member, IEEE}, \\ and Mohamed-Slim Alouini, {\em Fellow, IEEE}
\thanks{The authors are with Computer, Electrical and Mathematical Sciences
and Engineering (CEMSE) Division, Department of Electrical and Computer
Engineering, King Abdullah University of Science and Technology (KAUST), Thuwal 23955-6900, Saudi Arabia. (e-mail: xue.zhang@kaust.edu.sa; abla.kammoun@kaust.edu.sa; slim.alouini@kaust.edu.sa). 
}
\vspace{-8mm}
}
\maketitle

\vspace{-0.8cm}

\begin{abstract}
This paper examines the performance degradation of receivers caused by imperfect channel state information (CSI), arising from training-based and semi-blind channel estimation techniques. We focus on the uplink of massive multiple-input multiple-output (MIMO) systems operating in time-division duplexing (TDD) mode, considering the asymptotic regime where the number of base station (BS) antennas grows large while the number of users remains fixed. Leveraging random matrix theory (RMT), we analyze the asymptotic behavior of the signal-to-interference-plus-noise ratio (SINR) under linear MRC and MMSE detection schemes. We derive closed-form approximations of the achievable rates, prove their asymptotic tightness, and validate their accuracy through simulations for systems of finite dimensions. Numerical results under realistic non-terrestrial network (NTN) conditions further demonstrate that semi-blind channel estimation enables MRC and MMSE receivers to outperform their training-based counterparts, while also quantifying the number of pilot symbols required to achieve a desired SINR level.
\end{abstract}


\section{Introduction}
\label{sec_introdution}
Massive multiple-input multiple-output (MIMO) systems, where hundreds of base station (BS) antennas simultaneously serve tens of single-antenna users, have demonstrated significant potential in improving spectral efficiency. However, the use of large antenna arrays makes conventional techniques like multi-user maximum likelihood (ML) detection in the uplink and sophisticated precoding or beamforming strategies in the downlink computationally prohibitive~\cite{ngo2013energy}. Fortunately, when the number of BS antennas far exceeds the number of users, simple linear detectors for the uplink and linear precoders for the downlink can achieve near-optimal performance with significantly lower complexity~\cite{marzetta2010noncooperative}. This is because, in such regimes, the random channel vectors between users and the BS become nearly orthogonal, and the effects of fast fading, intra-cell interference, and uncorrelated noise are substantially mitigated~\cite{marzetta2016fundamentals}.

In uplink massive MIMO systems, commonly used low-complexity linear detection algorithms include maximum ratio combining (MRC), zero-forcing (ZF), and minimum mean square error (MMSE) detection~\cite{rusek2012scaling,ngo2013energy,hoydis2013massive}. The MRC detector maximizes the desired signal power but does not suppress multi-user interference, making it more suitable for low-power transmission scenarios. In contrast, the ZF detector effectively eliminates multi-user interference by inverting the channel matrix. However, it neglects noise amplification and may experience serious performance degradation when the channel matrix is ill-conditioned. The MMSE detector strikes a balance between noise suppression and interference mitigation by maximizing the signal-to-interference-plus-noise ratio (SINR). As a result, it can achieve near-optimal performance across a wide range of transmit power levels.

However, the practical implementation of massive MIMO systems remains challenging, since users do not communicate directly with each other, joint data encoding at the user side is not possible~\cite{yang2015fifty}. Consequently, the suppression of multi-user interference must be handled entirely at the BS, requiring highly accurate channel state information (CSI) to reliably separate and detect each user's signal in the uplink transmission, or to precisely precode the transmitted symbols in the downlink. The availability of accurate CSI depends critically on the efficiency of the channel estimation process. Pilot-based channel estimation methods~\cite{hassibi2003much} are widely adopted due to their robustness and simplicity, but they impose substantial training overhead. Blind channel estimation techniques~\cite{noh2014new} can reduce or eliminate this overhead by exploiting the statistical properties of the transmitted data, but they typically suffer from parameter ambiguities and higher computational complexity. Semi-blind channel estimation~\cite{srinivas2019iterative}, which combines limited pilot information with data-aided techniques, has emerged as a promising alternative that offers improved estimation accuracy with moderate overhead. This is particularly attractive in high-dimensional massive MIMO systems. 

This paper aims to explicitly characterize the performance degradation resulting from imperfect CSI, which stems from pilot-based/semi-blind channel estimation techniques. Focusing on the regime where the number of antennas becomes large while the number of users remains fixed, we utilize random matrix theory (RMT) to analyze the asymptotic behavior of the signal-to-interference-plus-noise ratio (SINR) under simple linear detection schemes in uplink massive MIMO systems operating in time-division duplexing (TDD) mode. The analysis also applies to FDD uplink transmission, as it does not rely on channel reciprocity. Here, RMT serves as a powerful mathematical tool for studying the limiting behavior of large-dimensional random matrices, enabling accurate performance analysis in such high-dimensional systems. The following subsection provides a brief overview of related work, followed by a detailed presentation of our main contributions.

\subsection{Related Work}

\subsubsection{Semi-blind channel estimation} 
A substantial body of literature has explored semi-blind channel estimation schemes in multi-user MIMO systems. Among these approaches, ML estimation techniques, particularly those based on expectation-maximization (EM) algorithms, have attracted significant attention. The study in~\cite{nayebi2017semi} examined two EM-based schemes for semi-blind channel estimation with and without channel priors. To further enhance computational efficiency,~\cite{al2021semi} employed eigenvalue decomposition to accelerate the EM algorithm with a Gaussian prior and leveraged the discrete prior of data symbols to derive a tractable version of the EM procedure. In addition to EM-based methods, iterative techniques have also been investigated to compute ML channel estimators. In this context,~\cite{abuthinien2008semi} proposed an iterative two-level optimization loop for jointly estimating channel coefficients and training sequences in MIMO systems. Beyond ML estimators, another class of methods extends blind estimation techniques by incorporating training sequence information to resolve the inherent matrix ambiguities of blind approaches. For instance,~\cite{lawal2023semi} introduced a semi-blind structured signal subspace algorithm to reduce computational complexity and minimize the required training sequence size. Similarly, \cite{rekik2024fast} developed a blind and semi-blind subspace-based algorithm designed to reduce the computational cost, accelerate convergence, and achieve accurate channel estimates with a smaller sample size. When the number of BS antennas significantly exceeds the number of users,~\cite{zhang2026dynamic} proposes a semi-blind method based on random matrix theory (RMT), which provides a simple, closed-form, and easily implementable channel estimator by minimizing a regularized least-squares cost function that combines training-based and blind information, with the regularization parameter optimized using RMT.

\par

\subsubsection{RMT for SINR}
RMT has been widely used to analyze SINR in massive MIMO systems. In \cite{hoydis2013massive}, the authors present a unified performance analysis of the SINR for both uplink and downlink in massive MIMO TDD systems, accounting for imperfect channel estimation, pilot contamination, antenna correlation, and path loss, in the regime where both the number of antennas and users are large. In~\cite{zhang2014power}, tractable expressions for the achievable uplink rates with ZF and MRC are derived in the large-antenna limit, along with approximations that remain accurate for finite antenna systems (with a large number of antennas and a fixed number of users). In~\cite{huang2015uplink}, full-duplex multi-cell massive MIMO systems are analyzed, and a deterministic approximation of the uplink achievable rate with MRC is obtained using RMT, showing that both BS-to-BS and self-interference asymptotically vanish. The work in~\cite{sifaou2016max} studies uplink and downlink transmission in large-scale single-cell multi-user MIMO systems, focusing on low-complexity linear precoders and receivers that maximize the minimum SINR with RMT-based closed-form approximations. In~\cite{sanguinetti2018theoretical}, achievable rates are approximated for single-cell linear processing with uncorrelated Rician fading channels, including line-of-sight (LoS) and non-LoS (NLoS) components, assuming the number of antennas and users grow with a fixed ratio. The authors of \cite{aghashahi2023mu} investigate SINR maximization in multi-user massive MIMO systems with multiple reconfigurable intelligent surfaces (RISs), deriving optimal RIS phase shifts that achieve near-optimal performance even with few base station antennas. In contrast to the aforementioned studies, our work considers the channel as a deterministic quantity rather than assuming that it follows a prescribed statistical model. Randomness is accounted for only through the noise. This feature makes our analysis applicable to realistic channel realizations, as will be illustrated later in the simulation results.

\par

\subsection{Main Contributions}
In this work, we consider a realistic system model that accounts for imperfect channel estimation. To the best of our knowledge, this is the first study to provide a unified, channel-distribution-agnostic SINR performance analysis of uplink massive MIMO systems in the asymptotic regime where the number of BS antennas grows large while the number of users remains fixed. The main contributions are summarized as follows:

\begin{itemize}
\item We derive asymptotically tight approximations of the SINR for linear MRC and MMSE detectors based on both training-based and semi-blind channel estimators. These approximations are computationally efficient and remain accurate for practical system dimensions.
\item We provide a theoretical comparison of the SINR achieved by MRC and MMSE receivers under training-based and semi-blind channel estimation. The asymptotic expressions quantify the contributions of the desired signal, multi-user interference, and effective noise under both estimation approaches.
\item We conduct numerical simulations in realistic non-terrestrial network (NTN) environments to validate the proposed asymptotic expressions through simulations. In particular, the results demonstrate that semi-blind channel estimation enables MRC/MMSE receivers to outperform their training-based counterparts, and that MMSE detection achieves higher SINR than MRC. Furthermore, we quantify the number of pilot symbols required to achieve a given target SINR.
\end{itemize}

\subsection{Organization and Notation}
The remainder of this paper is organized as follows. Section~\ref{sec_model} introduces the system model, including the received signal model and the SINR formulation under linear MRC and MMSE detection. Section~\ref{sec_anaysis} presents the asymptotic analysis of the SINR for MRC and MMSE detectors in the large-scale system regime. Section~\ref{sec_sim} reports simulation results that validate the proposed approach. Finally, Section~\ref{sec_conclusion} concludes the paper.

\textit{Notations}: The vectors (matrices) are denoted by lower-case (upper-case) boldface characters. $\mathbf{I}_{M}$ denotes the $M\times M$ identity matrix, and $\mathbf{0}$ stands for a zero matrix with appropriate dimension. $\mathbb{E}[\cdot]$ represents the expectation operator. The trace of $\mathbf{A}$ are denoted by $\mathrm{tr}(\mathbf{A})$. Superscripts $T$, $H$, and $\ast$ denote the transpose, conjugate transpose, and complex conjugate, respectively. $[\mathbf{A}]_{i,j}$ stands for the $(i,j)$ entry of $\mathbf{A}$. $\mathrm{diag}(\mathbf{a})$ represents a diagonal matrix that employs the elements of $\mathbf{a}$ as its diagonal entries. $|\cdot|$ denotes the absolute value. $\|\cdot\|_F$ and $\|\cdot\|_2$ stand for the Frobenius and $l_2$ norms, respectively. $\overset{a.s.}{\longrightarrow}$ represents almost sure (a.s.) convergence. $a_n\asymp b_n$ denotes $a_n-b_n\overset{a.s.}{\longrightarrow}0$ for two random sequences $a_n$, $b_n$. $\mathrm{Re}\{\cdot\}$ represent real part.

\section{System Model}
\label{sec_model}
%We consider a single-cell uplink system in which a BS equipped with $M$ antennas serves $K$ single-antenna users, where $K < M$. In this section, we present the received signal model and define the SINR for linear detection based on estimated channels.

\subsection{Received Model}
We consider uplink transmissions in an orthogonal frequency-division multiplexing (OFDM) system. Each subcarrier is assumed to experience a block flat-fading channel that remains constant over a coherence block of $N$ symbols and changes independently from one block to another. Let ${\bf h}_1,\ldots,{\bf h}_K \in \mathbb{C}^{M}$ denote the channel vectors between the $K$ single-antenna users and the $M$-antenna BS. The received signal at the BS on a given subcarrier and at time instant $n$ can be modeled as~\cite{nayebi2017semi,zhang2025fundamental}
\begin{align}
    \mathbf{y}(n) = \mathbf{H}\mathbf{s}(n) + \mathbf{v}(n), \label{received signal}
\end{align}
where $\mathbf{H}=[{\bf h}_1,\ldots,{\bf h}_K] \in \mathbb{C}^{M\times K}$ is the channel matrix, $\mathbf{s}(n)=[s_1(n),\ldots,s_K(n)]^{T}$ denotes the transmitted symbol vector. For $n=0,\ldots,L-1$, $\mathbf{s}(n)$ corresponds to known pilot (training) symbols, while for $n=L,\ldots,N-1$, it contains unknown data symbols. The noise vector $\mathbf{v}(n)\sim\mathcal{CN}(\mathbf{0},\sigma_v^{2}\mathbf{I}_M)$ represents additive white Gaussian noise.

Let $\mathbf{S}_p = [\mathbf{s}(0), \dots, \mathbf{s}(L-1)] \in \mathbb{C}^{K \times L}$ denote the matrix of known training symbols, and $\mathbf{S}_d = [\mathbf{s}(L), \dots, \mathbf{s}(N-1)] \in \mathbb{C}^{K \times (N - L)}$ denote the matrix of unknown data symbols. Similarly, define the received signal during the training and data phases as ${\bf Y}_p=[{\bf y}(0),\cdots,{\bf y}(L-1)]$ and 
$\mathbf{Y}_d = [\mathbf{y}(L), \dots, \mathbf{y}(N-1)].$
Using this notation, the received signal model over the entire coherence block can be written compactly    
\begin{align}\mathbf{Y}_p &= \mathbf{H} \mathbf{S}_p + \mathbf{V}_p, \ \ \text{training period}, \\
    \mathbf{Y}_d &= \mathbf{H} \mathbf{S}_d + \mathbf{V}_d, \ \ \text{data transmission},
\end{align}
where $\mathbf{V}_p$ and $\mathbf{V}_d$ are the corresponding noise matrices, with i.i.d. columns distributed as $\mathcal{CN}(\mathbf{0}, \sigma_v^2 \mathbf{I}_M)$. 
We assume that the training sequences are orthogonal and satisfy the normalization condition
$
\mathbf{S}_p \mathbf{S}_p^H = aL \mathbf{I}_K,$
where $a > 0$ denotes the average training power per user. The unknown data symbol vectors $\mathbf{s}(n)$ for $n = L, \ldots, N-1$ are assumed independent across time and users, with covariance $P_s {\bf I}_K$.

%Let $\mathbf{S}_p = [\mathbf{s}(0), \ldots, \mathbf{s}(L-1)]$ and $\mathbf{S}_d = [\mathbf{s}(L), \ldots, \mathbf{s}(N-1)]$ denote the matrices of training and data symbols within a coherence block, respectively. Accordingly, the overall transmitted symbol matrix can be written as $\mathbf{S}=[\mathbf{S}_p,\mathbf{S}_d]$ with corresponding received matrix $\mathbf{Y}=[\mathbf{Y}_p,\mathbf{Y}_d]$ and noise matrix $\mathbf{V}=[\mathbf{V}_p,\mathbf{V}_d]$, respectively. Furthermore, we assume that the training sequences are orthogonal and satisfy the normalization condition $\mathbf{S}_p\mathbf{S}_p^H=aL\mathbf{I}_K$, where $a > 0$ denotes the average training power per user. 

\subsection{Channel Estimation}
In this work, channel estimation is carried out at the base station. We consider two estimation approaches, namely the conventional training-based least-squares (LS) estimator and the semi-blind estimator proposed in~\cite{zhang2026dynamic}. In the sequel, we recall the main principles of these two techniques.
\subsubsection{Training-based estimator }
After correlating the received signal with  the training sequence of user $k$, the BS builds the training-based estimator given by:
%Using the training-based method, the estimated channel vector for the $k$th user is expressed as   
\begin{align}\label{h_k^tr}
    \hat{\mathbf{h}}_k^{\mathrm{tr}} = \mathbf{h}_k + \frac{1}{aL}\mathbf{V}_p\mathbf{z}_k, \ \ k=1,\cdots,K,
\end{align}
where $\mathbf{z}_k=[s_k(1),\ldots,s_k(L)]^H$.
\subsubsection{Semi-blind based estimator}
We consider the semi-blind subspace-based estimator proposed in~\cite{zhang2026dynamic}. 
This approach combines pilot-based information with second-order statistical information extracted from the received data samples. The latter is extracted from the sample covariance matrix of the received signal vectors $\mathbf{y}(n)$ for $n=L,\ldots,N-1$, which is given by
\begin{equation}
\hat{\mathbf{R}}=\frac{1}{N-L}\sum_{n=L}^{N-1}\mathbf{y}(n)\mathbf{y}(n)^{H}.\label{eq:empirical_covariance_matrix}
\end{equation}
Letting $\hat{\bf U}_s=[\hat{\bf u}_{s_1},\cdots,\hat{\bf u}_{s_K}]$ be the $K$ leading eigenvectors of $\hat{\bf R}$, and ${\eta}_1\geq \cdots\geq {\eta}_K$ their corresponding eigenvalues, the semi-blind estimator writes as:
\begin{align}
    \hat{\mathbf{h}}_k^{\mathrm{sb}} =& \left(\hat{\lambda}\mathbf{I}_M+(1-\hat{\lambda})\hat{\mathbf{U}}_s\hat{\mathbf{U}}_s^H\right)\Big(\mathbf{h}_k+\frac{1}{aL}\mathbf{V}_p\mathbf{z}_k\Big) \notag\\
    =& \hat{\lambda}\hat{\mathbf{h}}_k^{\mathrm{tr}}+(1-\hat{\lambda})\hat{\mathbf{U}}_s\hat{\mathbf{U}}_s^H\hat{\mathbf{h}}_k^{\mathrm{tr}},
\end{align}
where $\hat{\lambda}$ is the regularization parameter computed as  
\begin{align}
    \hat{\lambda} = 1-\frac{\sigma_v^2\alpha}{a\beta\frac{1}{K}\sum_{i=1}^K(1-\sigma_v^2\hat{z}_i\hat{t}_i)+\sigma_v^2\alpha},  
\end{align}
with $\hat{z}_i=\frac{1-c/\hat{t}_i^2}{1+c/\hat{t}_i}$, and 
\begin{align}
    \hat{t}_i = \frac{\eta_i/\sigma_v^2-1-c+\sqrt{(\eta_i/\sigma_v^2+1-c)^2-4\eta_i/\sigma_v^2}}{2}. 
\end{align}
Here, $c=M/(N-L)$.

\subsection{SINR with linear detection}
To detect the uplink symbols, we assume that the BS employs a linear receiver. Let $\mathbf{G}=[\mathbf{g}_1,\ldots,\mathbf{g}_K]\in\mathbb{C}^{M\times K}$ denote the linear detection matrix. The received signal is then linearly processed using ${\bf G}$ in order to produce estimates of the $K$ transmitted data streams, i.e.,
\begin{align}\label{r(n)}
    \mathbf{r}(n) &= \mathbf{G}^H\mathbf{y}(n) \notag\\
    &= \mathbf{G}^H \mathbf{H} \mathbf{s}(n) + \mathbf{G}^H \mathbf{v}(n), 
    \quad n=L,\ldots,N-1.
\end{align}
Focusing on the output corresponding to the $k$th user, we obtain:
\begin{align}
    r_k(n) = \mathbf{g}_k^H \mathbf{h}_k s_k(n)
    + \sum_{\substack{i=1,i \neq k}}^{K} \mathbf{g}_k^H \mathbf{h}_i s_i(n)
    + \mathbf{g}_k^H \mathbf{v}(n), 
\end{align}
where the first term represents the desired signal component, the second term captures the multi-user interference, and the third term corresponds to the additive noise. For the training-based estimator, the detection vector $\mathbf g_k$ is independent of the current data symbols and noise, causing the cross terms to vanish in expectation. The resulting SINR is given by
\begin{equation}
\mathrm{SINR}_k(n)
=
\frac{\mathbb E[|\mathbf g_k^H\mathbf h_k s_k(n)|^2]}
{\sum\limits_{\substack{i=1\\i\neq k}}^K
\mathbb E[|\mathbf g_k^H\mathbf h_i s_i(n)|^2]
+\mathbb E[|\mathbf g_k^H\mathbf v(n)|^2]}.
\label{eq:SINR_expression}
\end{equation}
For the semi-blind receiver, however, $\mathbf g_k$ depends on the sample signal subspace estimated from the same data block and is therefore statistically coupled with the current data symbols and noise. Consequently, the cross terms cannot be neglected a priori. To account for this dependence, we decompose the detector output $r_k(n)=\mathbf g_k^H\mathbf y(n)$ into a component correlated with the desired symbol $s_k(n)$ and an uncorrelated residual. The resulting effective SINR is defined as
\begin{equation}
\mathrm{SINR}_k(n)
=
\frac{|\mathbb E[\mathbf g_k^H\mathbf y(n)s_k^\ast(n)]|^2}
{P_s\mathbb E[|\mathbf g_k^H\mathbf y(n)|^2]
-|\mathbb E[\mathbf g_k^H\mathbf y(n)s_k^\ast(n)]|^2}.
\label{eq:SINR_effective_y}
\end{equation}
This definition captures the statistical dependence introduced by semi-blind channel estimation. Moreover, the expectation is taken over the distribution of the transmitted symbol and the additive Gaussian noise. Several linear detectors can be employed at the BS. In this work, we focus on the MRC and MMSE detectors. Under perfect CSI, the MMSE detector is known to be optimal in the sense that it maximizes the output SINR. In practice, however, the channel is not perfectly known and must be estimated. In such situations, the MRC and MMSE detectors are implemented by replacing the true channel with its estimate. In this work, we consider both training-based and semi-blind channel estimation methods. For each estimation approach, we construct the corresponding MRC and MMSE detectors and derive the associated SINR expressions.
%The objective of this paper is therefore to investigate the impact of channel estimation accuracy on the performance of the MRC and MMSE detectors.
%Note that the linear detection matrix $\mathbf{G}$ depends on the channel $\mathbf{H}$. In practice, the channel must be estimated at the base station, which is denoted by $\hat{\mathbf{H}} = [\hat{\mathbf{h}}_1, \ldots, \hat{\mathbf{h}}_K]$. In the following, we review widely used linear multi-user receivers. Since the ZF structure is similar to that of the MMSE, for simplicity, we focus on the two conventional linear detectors: MRC and MMSE.

\textit{1) For MRC}: The MRC receiver aims to maximize the signal-to-noise ratio (SNR) of each user's data stream, without explicitly mitigating multi-user interference. The detection vector for the $k$th user is defined as $\hat{\mathbf{g}}_k^{\mathrm{MRC}} = \hat{\mathbf{h}}_k$, where $\hat{\bf h}_k$ is either $\hat{\bf h}_k^{\text{tr}}$ or $\hat{\bf h}_k^{\text{sb}}$ which results in $\mathrm{SINR}_k^{\mathrm{MRC}}$. 
%\begin{align}\label{true_SINR_MRC}
%    &\mathrm{SINR}_k^{\mathrm{MRC}}(n)= \notag\\
%    &\frac{\mathbb{E}\Big[\Big|(\hat{\mathbf{g}}_k^{\mathrm{MRC}})^H\mathbf{h}_ks_k(n)\Big|^2\Big]}{\sum\limits_{i=1\atop i\neq k}^K\mathbb{E}\left[\left|(\hat{\mathbf{g}}_k^{\mathrm{MRC}})^H\mathbf{h}_is_i(n)\right|^2\right] + \mathbb{E}\left[\left|(\hat{\mathbf{g}}_k^{\mathrm{MRC}})^H\mathbf{v}(n)\right|^2\right]}. 
%\end{align}

\textit{2) For MMSE}: The MMSE receiver is designed to minimize the mean-squared error between the transmitted uplink signals and the detected signals. The MMSE detection vector for the $k$th user is given by:
$\hat{\mathbf{g}}_k^{\mathrm{MMSE}} = \Big(\hat{\mathbf{H}}\hat{\mathbf{H}}^H + \frac{\sigma_v^2}{P_s} \mathbf{I}_M\Big)^{-1}\hat{\mathbf{h}}_k$, where $\hat{\bf H}=[\hat{\bf h}_1,\cdots,\hat{\bf h}_K]$ denotes the channel matrix estimate with $\hat{\bf h}_k$ being either $\hat{\bf h}_k^{\text{tr}}$ or $\hat{\bf h}_k^{\text{sb}}$. The SINR at the output of the MMSE detector thus becomes $\mathrm{SINR}_k^{\mathrm{MMSE}}$.
%\begin{align}
%    &\mathrm{SINR}_k^{\mathrm{MMSE}}(n)= \notag\\
%    &\frac{\mathbb{E}\left[\left|(\hat{\mathbf{g}}_k^{\mathrm{MMSE}})^H\mathbf{h}_ks_k(n)\right|^2\right]}{\sum\limits_{i=1\atop i\neq k}^K\mathbb{E}\left[\left|(\hat{\mathbf{g}}_k^{\mathrm{MMSE}})^H\mathbf{h}_is_i(n)\right|^2\right] + \mathbb{E}\left[\left|(\hat{\mathbf{g}}_k^{\mathrm{MMSE}})^H\mathbf{v}(n)\right|^2\right]}.  
%\end{align}

\par

\section{Asymptotic Analysis}
\label{sec_anaysis}
Since the SINR expressions for MRC and MMSE detectors are difficult to evaluate for finite-dimensional systems, we analyze their behavior in the large-system regime. Specifically, we adopt the following assumption:
\begin{assumption}[Growth rate regime]\label{ass2_parameter}
    We assume that the number $K$ of users is fixed, while the number $M$ of antennas, the number $L$ of pilot symbols, and the length $N$ of the whole transmission block go to infinity at the same rate, that $M, N, L, N-L\rightarrow+\infty$ with $M/N\rightarrow \alpha$, $L/N\rightarrow \beta$, and $M/(N-L)\rightarrow c$ for certain $\alpha,\beta,c>0$. The number of $K$ is independently of $M$ and $N-L$. Under this asymptotic regime, we assume that $\lim\sup \|{\bf H}{\bf H}^{H}\|<\infty$. %the channel vectors are normalized such that $\|\mathbf h_k\|_2=\mathcal{O}(1)$, $k=1,\ldots,K$. Since $K$ remains fixed, $\lim\sup \|{\bf H}{\bf H}^{H}\|<\infty$.}, %we assume that $\lim\sup \|{\bf H}{\bf H}^{H}\|<\infty$.
    
    %while $t_1>\ldots>t_K$ with $t_K>\sqrt{c}$ for all large $M$. 
\end{assumption}

\begin{assumption}[Separability condition] \label{ass:separability} Let ${\bf R}$ be the covariance matrix of ${\bf y}(n), n=L,\cdots,N-1$, that is ${\bf R}=P_s{\bf H}{\bf H}^{H}+\sigma_v^2{\bf I}_M$. Use the following eigenvalue decomposition of ${\bf R}$:
\begin{equation}
{\bf R}=\sigma_v^2\Big(\sum_{k=1}^{K}t_k{\bf u}_{sk}{\bf u}_{sk}^{H}+{\bf I}_M\Big), \label{eq:covariance_matrix}
\end{equation}
with $t_1,\cdots, t_K$ positive.  We assume that $t_1>t_2\cdots>t_K>\sqrt{c}$ for all large $M,N$ and $L$.
\end{assumption}
\begin{assumption}{Symbol data distribution}
We assume that the transmitted data-symbol matrix ${\bf S}_d$ consists of bounded i.i.d. entries with equal second moment. More precisely, $\forall i=1,\cdots,K, j=1,\cdots,N-L$ $\mathbb{E}[|[{\bf S}_d]_{i,j}|^2]=P_s, i=1,\cdots,K, j=1,\cdots,N-L$. and there exists a constant $C>0$ such that, for all $n=L,\ldots,N$ and $k=1,\ldots,K$,
$
|[{\bf S}_d]_{i,j}|\leq C .
$
\label{ass:boundedness}
\end{assumption}
This assumption is naturally satisfied in practical digital communication systems, where the transmitted symbols are drawn from a finite modulation constellation, such as quadruple phase shift keying (QPSK) or quadrature amplitude modulation (QAM).
In the following, we adopt the semi-blind channel estimation method proposed in~\cite{zhang2026dynamic} to estimate the channel matrix. Based on this estimate, we will derive deterministic approximations $\overline{\mathrm{SINR}}_{\mathrm{sb},k}^{\mathrm{MRC}}$ and $\overline{\mathrm{SINR}}_{\mathrm{sb},k}^{\mathrm{MMSE}}$ for the corresponding SINRs $\mathrm{SINR}_{\mathrm{sb},k}^{\mathrm{MRC}}$ and $\mathrm{SINR}_{\mathrm{sb},k}^{\mathrm{MMSE}}$, respectively. For comparison, we will also derive the corresponding deterministic equivalents $\overline{\mathrm{SINR}}_{\mathrm{tr},k}^{\mathrm{MRC}}$ and $\overline{\mathrm{SINR}}_{\mathrm{tr},k}^{\mathrm{MMSE}}$ based on the training-based channel estimation method. The key idea behind deriving these asymptotic expressions is to decompose the SINR into three components: desired signal power, interference power, and noise power. For each component, we obtain its deterministic equivalent, which collectively yields the final asymptotic expression for the SINR.  

In the context of massive MIMO systems, the impact of channel estimation accuracy on SINR performance has been extensively studied for conventional training-based schemes~\cite{hoydis2013massive}. However, the results obtained in this paper for the training-based case differ from most existing works. In particular, prior analyses typically rely on stochastic channel models and express the resulting SINR in terms of second-order channel statistics, such as the spatial correlation matrix. In contrast, we treat the channel as a deterministic quantity, which leads to SINR expressions that depend explicitly on the channel realization itself.

This modeling choice provides an alternative performance characterization from a practical standpoint. Indeed, realistic propagation channels generally deviate from ideal Gaussian assumptions, and the spatial correlation matrix can be difficult to estimate accurately in dynamic environments. By avoiding this intermediate statistical description and expressing the SINR directly as a function of the channel, the proposed analysis provides a complementary characterization of receiver performance under the considered channel conditions.

\subsection{Asymptotic SINR for training-based method}
The training-based MRC detection vector for the $k$th user is given by 
\begin{align}
    \hat{\mathbf{g}}_{\mathrm{tr},k}^{\mathrm{MRC}}=\hat{\mathbf{h}}_k^{\mathrm{tr}}.     
\end{align}
The following theorem provides the deterministic equivalent of the corresponding SINR in the large-system regime.

\begin{theorem}\label{theorem_tra_MRC}
Under Assumption~\ref{ass2_parameter} and \ref{ass:boundedness}, and employing an MRC receiver with a training-based channel estimation method, we have
\begin{align}
    \mathrm{SINR}_{\mathrm{tr},k}^{\mathrm{MRC}} - \overline{\mathrm{SINR}}_{\mathrm{tr},k}^{\mathrm{MRC}} \overset{a.s.}{\longrightarrow} 0, 
\end{align}
where
\begin{align}\label{training_MRC_asy}
    \overline{\mathrm{SINR}}_{\mathrm{tr},k}^{\mathrm{MRC}} = \frac{P_s\|\mathbf{h}_k\|_2^4}{P_s\|\mathbf{h}_k^H\mathbf{H}_k\|_2^2 + \sigma_v^2\|\mathbf{h}_k\|_2^2 + \frac{\sigma_v^4\alpha}{a\beta}}, 
\end{align}
with $\mathbf{H}_k=[\mathbf{h}_1,\ldots,\mathbf{h}_{k-1},\mathbf{h}_{k+1},\ldots,\mathbf{h}_{K}]$.
\begin{proof}
    See the Appendix~\ref{app_tra_MRC}.
\end{proof}
\end{theorem}

For the MMSE receiver with training-based channel estimation, it is convenient to define the auxiliary matrix $\hat{\mathbf{C}}^{\mathrm{tr}}$ as 
\begin{align}
    \hat{\mathbf{C}}^{\mathrm{tr}} =& \Big(\hat{\mathbf{H}}^{\mathrm{tr}}(\hat{\mathbf{H}}^{\mathrm{tr}})^H+\frac{\sigma_v^2}{P_s}\mathbf{I}_M\Big)^{-1},
\end{align}
then the training-based MMSE detection vector for the $k$th user can be expressed as
\begin{align}
    \hat{\mathbf{g}}_{\mathrm{tr},k}^{\mathrm{MMSE}}=\hat{\mathbf{C}}^{\mathrm{tr}}\hat{\mathbf{h}}_k^{\mathrm{tr}}.
\end{align}
We then present the deterministic equivalent of the SINR for the training-based MMSE receiver.

\begin{theorem}\label{theorem_tra_MMSE}
Under Assumption~\ref{ass2_parameter}- \ref{ass:boundedness}, and employing an MMSE receiver with a training-based channel estimation method, we have
\begin{align}
    \mathrm{SINR}_{\mathrm{tr},k}^{\mathrm{MMSE}} - \overline{\mathrm{SINR}}_{\mathrm{tr},k}^{\mathrm{MMSE}} \overset{a.s.}{\longrightarrow} 0, 
\end{align}
where $\overline{\mathrm{SINR}}_{\mathrm{tr},k}^{\mathrm{MMSE}}$ is expressed as 
\begin{align}\label{training_MMSE_asy}
    &\overline{\mathrm{SINR}}_{\mathrm{tr},k}^{\mathrm{MMSE}}= \notag\\
    &\frac{P_s|(\mathbf{t}_k^{\mathrm{tr}})^H\mathbf{h}_k|^2}{P_s\|(\mathbf{t}_k^{\mathrm{tr}})^H\mathbf{H}_k\|_2^2 + \sigma_v^2\|\mathbf{t}_k^{\mathrm{tr}}\|_2^2 + \frac{\sigma_v^4\alpha}{a\beta}\|\mathbf{e}_k^T-\mathbf{f}_k^{\mathrm{tr}}\bar{\mathbf{D}}^{\mathrm{tr}}\|_2^2 },
\end{align}
with $\mathbf{t}_k^{\mathrm{tr}}=\mathbf{h}_k-\mathbf{H}\bar{\mathbf{D}}^{\mathrm{tr}}(\mathbf{f}_k^{\mathrm{tr}})^H$, $\mathbf{f}_k^{\mathrm{tr}}=\mathbf{h}_k^H\mathbf{H}+\frac{\sigma_v^2\alpha}{a\beta}\mathbf{e}_k^T$, and $\bar{\mathbf{D}}^{\mathrm{tr}} = \Big(\mathbf{H}^H\mathbf{H}+\Big(\frac{\sigma_v^2\alpha}{a\beta}+\frac{\sigma_v^2}{P_s}\Big)\mathbf{I}_K\Big)^{-1}$. Here, $\mathbf{e}_k$ denotes a column vector with one at the $k$-th position and zeros elsewhere.
\begin{proof}
    See the Appendix~\ref{app_tra_MMSE}.
\end{proof}
\end{theorem}

The deterministic equivalents in Theorems~\ref{theorem_tra_MRC}
and~\ref{theorem_tra_MMSE} reveal explicitly the impact of training-based
channel estimation on the uplink SINR. In both cases, the SINR denominator
contains the usual multiuser interference and filtered noise terms, together
with an additional penalty proportional to
$\sigma_v^4\alpha/(a\beta)$. This term captures the residual uncertainty due
to channel estimation errors and vanishes in the perfect-CSI regime. For MRC,
this penalty appears as an additive scalar term, reflecting the fact that MRC
does not actively suppress interference nor compensate for channel uncertainty.
In contrast, for MMSE detection, the penalty is weighted by
$\|\mathbf e_k^T-\mathbf f_k^{\rm tr}\bar{\mathbf D}^{\rm tr}\|_2^2$, showing
that the effect of channel estimation errors is shaped by the MMSE filtering
structure. 
These expressions also highlight the fundamental difference between MRC and
MMSE detection. The MRC receiver maximizes the desired signal power but treats
multiuser interference passively, whereas the MMSE receiver balances signal
enhancement, interference suppression, thermal noise, and channel uncertainty.

\subsection{Asymptotic SINR for semi-blind method}
The semi-blind-based MRC detection vector for the $k$th user is 
\begin{align}
    \hat{\mathbf{g}}_{\mathrm{sb},k}^{\mathrm{MRC}}=\hat{\mathbf{h}}_k^{\mathrm{sb}} = \hat{\lambda}\hat{\mathbf{h}}_k^{\mathrm{tr}}+(1-\hat{\lambda})\hat{\mathbf{U}}_s\hat{\mathbf{U}}_s^H\hat{\mathbf{h}}_k^{\mathrm{tr}}.
\end{align}
Unlike the training-based channel estimator, the semi-blind estimator depends  on both the transmitted symbol vectors ${\bf s}(n)$, for $n = L, \ldots, N-1$, and the additive Gaussian noise ${\bf v}(n)$, for $n = L, \ldots, N-1$, through the estimated signal subspace $\hat{\bf U}_s$. This intrinsic dependence significantly complicates the derivation of asymptotic equivalents for the SINR expression in \eqref{eq:SINR_expression} for both MRC and MMSE, as it requires carefully handling the correlation between the channel estimate, the transmitted data, and the noise. To address this challenge, new technical results are established in Lemma \ref{first-order-signal-noise_bis} and \ref{second-order-signal-noise_bis} and Corollary \ref{corollary1} (see Appendix~\ref{app_technical_results_basis}), which explicitly account for this dependency and form the basis of the asymptotic analysis.
%The following theorem establishes the deterministic equivalent of the corresponding SINR in the large-system regime.
In the sequel, we provide the following theorem establishing the deterministic equivalent of the corresponding SINR for the MRC receiver. 
\begin{theorem}\label{theorem_sb_MRC}
Under Assumption~\ref{ass2_parameter}-\ref{ass:boundedness} and employing an MRC receiver with a semi-blind channel estimation method, we have 
\begin{align}
    \mathrm{SINR}_{\mathrm{sb},k}^{\mathrm{MRC}} - \overline{\mathrm{SINR}}_{\mathrm{sb},k}^{\mathrm{MRC}}\xrightarrow[]{a.s.}0, 
\end{align}
where $\overline{\mathrm{SINR}}_{\mathrm{sb},k}^{\mathrm{MRC}}$ is expressed as (\ref{semiblind_MRC_asy}) 
\begin{figure*}[t]
\begin{align}\label{semiblind_MRC_asy}
    \overline{\mathrm{SINR}}_{\mathrm{sb},k}^{\mathrm{MRC}} = \frac{P_s|\mathbf{h}_k^H\mathbf{F}\mathbf{h}_k|^2}{P_s\|\mathbf{h}_k^H\mathbf{F}_s\mathbf{H}\|_2^2 + \sigma_v^2\mathbf{h}_k^H(\mathbf{F}_n+2\mathbf{F}_{sv})\mathbf{h}_k + \bar{\lambda}^2\frac{\sigma_v^4\alpha}{a\beta}-P_s|\mathbf{h}_k^H\mathbf{F}\mathbf{h}_k|^2}, 
\end{align} 
\hrulefill
\end{figure*}
with $\mathbf{F} = \bar{\lambda}\mathbf{I}_M+(1-\bar{\lambda})\mathbf{U}_s(\pmb{\Lambda}_s+\pmb{\Lambda}_{vs})\mathbf{U}_s^H$, $\mathbf{F}_s = \bar{\lambda}\mathbf{I}_M+(1-\bar{\lambda})\mathbf{U}_s\pmb{\Lambda}_s\mathbf{U}_s^H$, $\mathbf{F}_n = \bar{\lambda}^2\mathbf{I}_M+2\bar{\lambda}(1-\bar{\lambda})\mathbf{U}_s\pmb{\Lambda}_n\mathbf{U}_s^H+(1-\bar{\lambda})^2\mathbf{U}_s\bar{\pmb{\Lambda}}_n\mathbf{U}_{s}^H$, $\mathbf{F}_{sv}=c\bar{\lambda}(1-\bar{\lambda})\mathbf{U}_s\pmb{\Lambda}_s\mathbf{U}_s^H+c(1-\bar{\lambda})^2\mathbf{U}_s\pmb{\Lambda}_s^2\mathbf{U}_s^H$ and $\overline{\lambda} = 1- \frac{\alpha\sigma_v^2}{\frac{a\beta\sigma_v^2}{P_sK}\sum_{k=1}^K(1-z_{sk})t_k+\alpha\sigma_v^2}$. Here, $\pmb{\Lambda}_s=\mathrm{diag}([z_{s1},\ldots,z_{sK}])$ with $z_{sk}=\frac{1-c/t_k^2}{1+c/t_k}$, $\pmb{\Lambda}_{vs}=\mathrm{diag}([z_{vs1},\ldots,z_{vsK}])$ with $z_{vsk}=c/t_kz_{sk}$, $\pmb{\Lambda}_n=\mathrm{diag}([z_{n1},\ldots,z_{nK}])$ with $z_{nk}=1-c/t_k^2$, and $\bar{\pmb{\Lambda}}_n =\mathrm{diag}([\bar{z}_{n1},\ldots,\bar{z}_{nK}])$ with $\bar{z}_{nk}= \frac{(t_k^2-c)(t_k(t_k+c)^2+c^2(t_k^2-c))}{t_k^3(t_k+c)^2}$.
\begin{proof}
    See the Appendix~\ref{app_sb_MRC_res}. 
\end{proof}
\end{theorem}

\par 

For the MMSE receiver with semi-blind channel estimation, we define the auxiliary matrix $\hat{\mathbf{C}}^{\mathrm{sb}}$ as
\begin{align}
    \hat{\mathbf{C}}^{\mathrm{sb}} =& \Big(\hat{\mathbf{H}}^{\mathrm{sb}}(\hat{\mathbf{H}}^{\mathrm{sb}})^H+\frac{\sigma_v^2}{P_s}\mathbf{I}_M\Big)^{-1},
\end{align}
then the semi-blind MMSE detection vector for the $k$th user is given by
\begin{align}
    \hat{\mathbf{g}}_{\mathrm{sb},k}^{\mathrm{MMSE}}=\hat{\mathbf{C}}^{\mathrm{sb}}\hat{\mathbf{h}}_k^{\mathrm{sb}}, 
\end{align}
The following theorem provides the deterministic equivalent of the SINR for the semi-blind MMSE receiver in the large-system regime.

\begin{theorem}\label{theorem_sb_MMSE}
Under Assumption~\ref{ass2_parameter}-\ref{ass:boundedness} and employing an MMSE receiver with a semi-blind channel estimation method, we have 
\begin{align}
    \mathrm{SINR}_{\mathrm{sb},k}^{\mathrm{MMSE}} - \overline{\mathrm{SINR}}_{\mathrm{sb},k}^{\mathrm{MMSE}}\xrightarrow[]{a.s.}0, 
\end{align}
where $\overline{\mathrm{SINR}}_{\mathrm{sb},k}^{\mathrm{MMSE}}$ is expressed as (\ref{semiblind_MMSE_asy}) 
\begin{figure*}[t]
\begin{align}\label{semiblind_MMSE_asy}
    \overline{\mathrm{SINR}}_{\mathrm{sb},k}^{\mathrm{MMSE}} = \frac{P_s\left|(\mathbf{t}_k^{\mathrm{sb}})^H\mathbf{F}\mathbf{h}_k\right|^2}{P_s\left\|(\mathbf{t}_k^{\mathrm{sb}})^H\mathbf{F}_s\mathbf{H}\right\|_2^2 + \sigma_v^2(\mathbf{t}_k^{\mathrm{sb}})^H(\mathbf{F}_n+2\mathbf{F}_{sv})\mathbf{t}_k^{\mathrm{sb}}+ \bar{\lambda}^2\frac{\sigma_v^2\alpha}{a\beta}\|\mathbf{e}_k^T-\mathbf{f}_k^{\mathrm{sb}}\bar{\mathbf{D}}^{\mathrm{sb}}\|_2^2-P_s\left|(\mathbf{t}_k^{\mathrm{sb}})^H\mathbf{F}\mathbf{h}_k\right|^2}. 
\end{align}
\hrulefill
\end{figure*}
with $\mathbf{t}_k^{\mathrm{sb}} = \mathbf{h}_k - \mathbf{H}\bar{\mathbf{D}}^{\mathrm{sb}}(\mathbf{f}_k^{\mathrm{sb}})^H$, $\mathbf{f}_k^{\mathrm{sb}} = \mathbf{h}_k^H\mathbf{M}_s\mathbf{H}+\bar{\lambda}^2\frac{\sigma_v^2\alpha}{a\beta}\mathbf{e}_k^T$, $\bar{\mathbf{D}}^{\mathrm{sb}} = \Big(\mathbf{H}^H\mathbf{M}_s\mathbf{H} + \frac{\alpha\sigma_v^2}{a\beta}\bar{\lambda}^2\mathbf{I}_K + \frac{\sigma_v^2}{P_s}\mathbf{I}_K\Big)^{-1}$. Here, $\mathbf{M}_s=\bar{\lambda}^2\mathbf{I}_M+(1-\bar{\lambda}^2)\mathbf{U}_s\pmb{\Lambda}_s\mathbf{U}_s^H$, and $\mathbf{e}_k$ is the vector with a $1$ in the $k$-th coordinate and zeros elsewhere.
\end{theorem}

\begin{proof}
    See the Appendix~\ref{app_sb_MMSE}. 
\end{proof}
Theorem~\ref{theorem_sb_MRC} characterizes the impact of semi-blind channel estimation on the asymptotic SINR of the MRC receiver. In contrast to training-based estimation, the desired signal term is characterized by $\mathbf F$, which accounts for the additional contribution arising from the statistical dependence between the channel estimate and the received data. The multiuser interference term is governed by $\mathbf F_s$, reflecting the effect of the estimated signal subspace on the effective channel correlations. Moreover, the noise-related contribution involves both $\mathbf F_n$ and $\mathbf F_{sv}$, where the latter captures additional effects associated with the dependence introduced by semi-blind estimation. The residual training-error term is scaled by $\bar{\lambda}^2$, illustrating the role of the semi-blind weighting parameter in the resulting SINR.

For the MMSE receiver, Theorem~\ref{theorem_sb_MMSE} shows that the desired signal and multiuser interference terms are characterized by $\mathbf F_s$, while the noise-related contribution is governed by $\mathbf F_n+2\mathbf F_{sv}$. Furthermore, the regularized inverse $\bar{\mathbf D}^{\mathrm{sb}}$ incorporates the modified channel correlation matrix $\mathbf H^H\mathbf M_s\mathbf H$, accounting for the effect of semi-blind subspace information on interference suppression. Unlike the MRC case, the residual training-error contribution also depends on the MMSE filtering structure through $\|\mathbf e_k^T-\mathbf f_k^{\mathrm{sb}}\bar{\mathbf D}^{\mathrm{sb}}\|_2^2$. Together, these results characterize how the statistical dependence and signal-subspace information introduced by semi-blind estimation affect the asymptotic SINR of both receivers.

\section{Simulation results}
\label{sec_sim}
This section presents numerical simulations to validate the asymptotic SINR expressions and evaluate receiver performance. We consider an uplink low Earth orbit (LEO) satellite communication system in which a satellite equipped with $M$ antennas serves $K$ single-antenna users. Channel realizations are generated based on the multipath model in~\cite{ying2024deep}, which incorporates both LoS and non-line-of-sight NLoS propagation components. The satellite employs a uniform circular array (UCA), while each user is equipped with a single antenna. Unless otherwise specified, the simulation parameters follow those in~\cite{ying2024deep}, including a satellite altitude of $1000$ km, a carrier frequency of $5$ GHz, a bandwidth of $25$ MHz, satellite and user antenna gains of $17$ and $3$ dBi, respectively, and a Rician factor of $5$. To validate the asymptotic SINR expressions, the channel remains constant within each observation block, consistent with the block-fading assumption adopted in the theoretical analysis. The impact of temporal channel variations is separately investigated in Section~IV-D, where Doppler-induced phase variations are considered throughout both pilot and data transmissions. The pilot sequences $\mathbf S_p$ are chosen to be orthogonal with a power scaling factor $a=1$, while the data symbols $\mathbf S_d$ with $P_s=1$ are drawn from a QPSK constellation. The SNR is calculated as $\mathrm{SNR}=\frac{a\beta+P_s(1-\beta)\|\mathbf{G}\|_F^2}{\sigma_v^2}$. To ensure statistical reliability, each data point is averaged over 1000 independent Monte Carlo trials.

We assess the accuracy of the asymptotic approximation to the SINR using the normalized SINR approximation error, defined as
\begin{align}
    \mathrm{NMSE} = \frac{\mathbb{E}\big[|{\rm SINR}_{\rm true}-\overline{\rm SINR}_{\rm asy}|^2\big]}{\mathbb{E}\big[{\rm SINR}_{\rm true}\big]^2},
\end{align}
where ${\rm SINR}_{\mathrm{true}}$ denotes the SINR computed using MRC or MMSE receiver with a training-based or semi-blind channel estimation method, respectively, and ${\overline{\rm SINR}}_{\mathrm{asy}}$ represents the corresponding asymptotic expression.

\subsection{Experiment for accuracy}
\begin{figure}[t]
\centering
\includegraphics[width=0.92\linewidth]{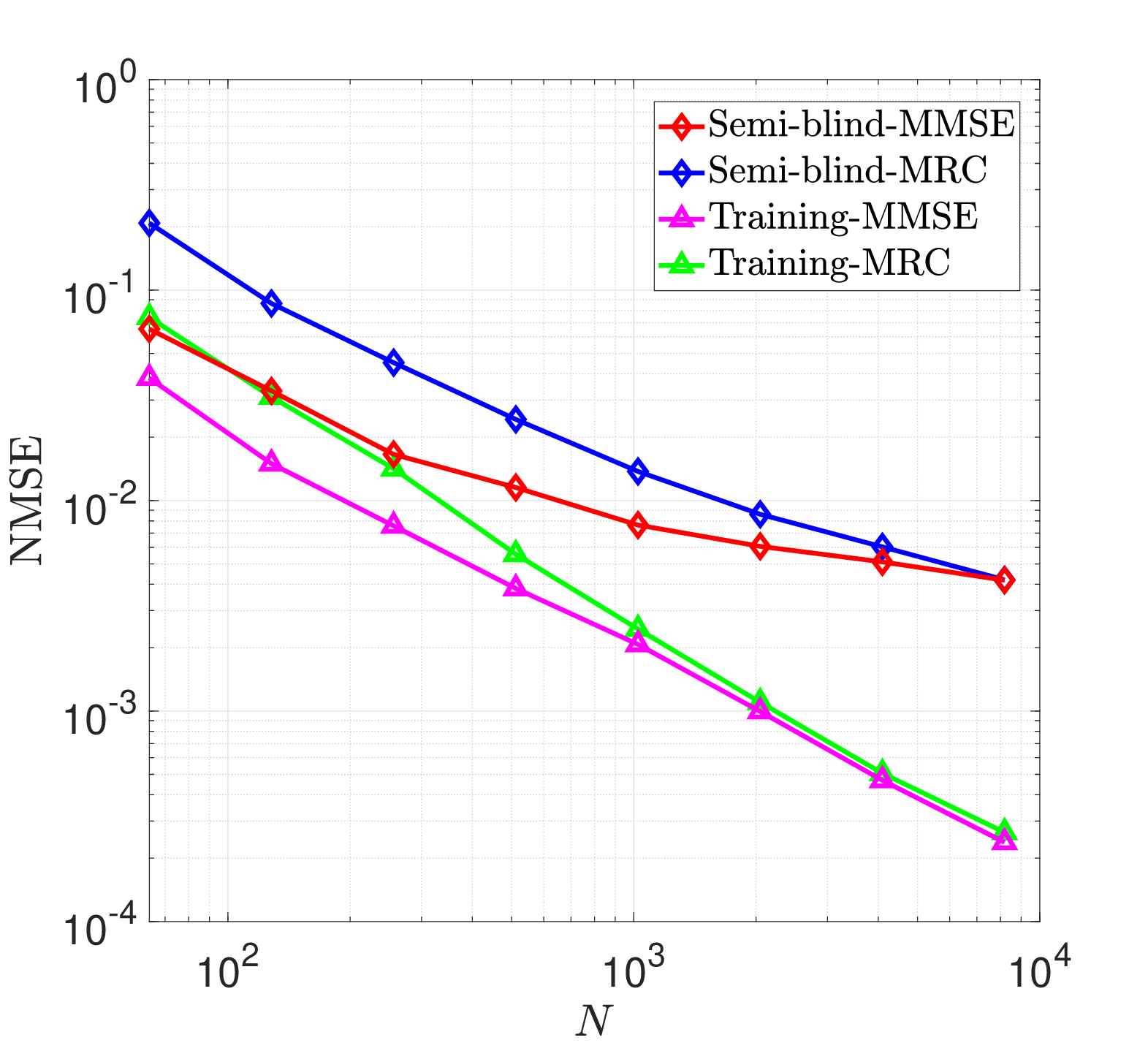}
\caption{$\mathrm{NMSE}$ versus the length $N$ of the whole transmission block with $\mathrm{SNR}=10$\,dB, $\alpha=1/4$, $\beta=1/4$, and $K=3$.}
\label{fig_NMSE_N}
\end{figure}

\begin{figure}[t]
\centering
\includegraphics[width=0.92\linewidth]{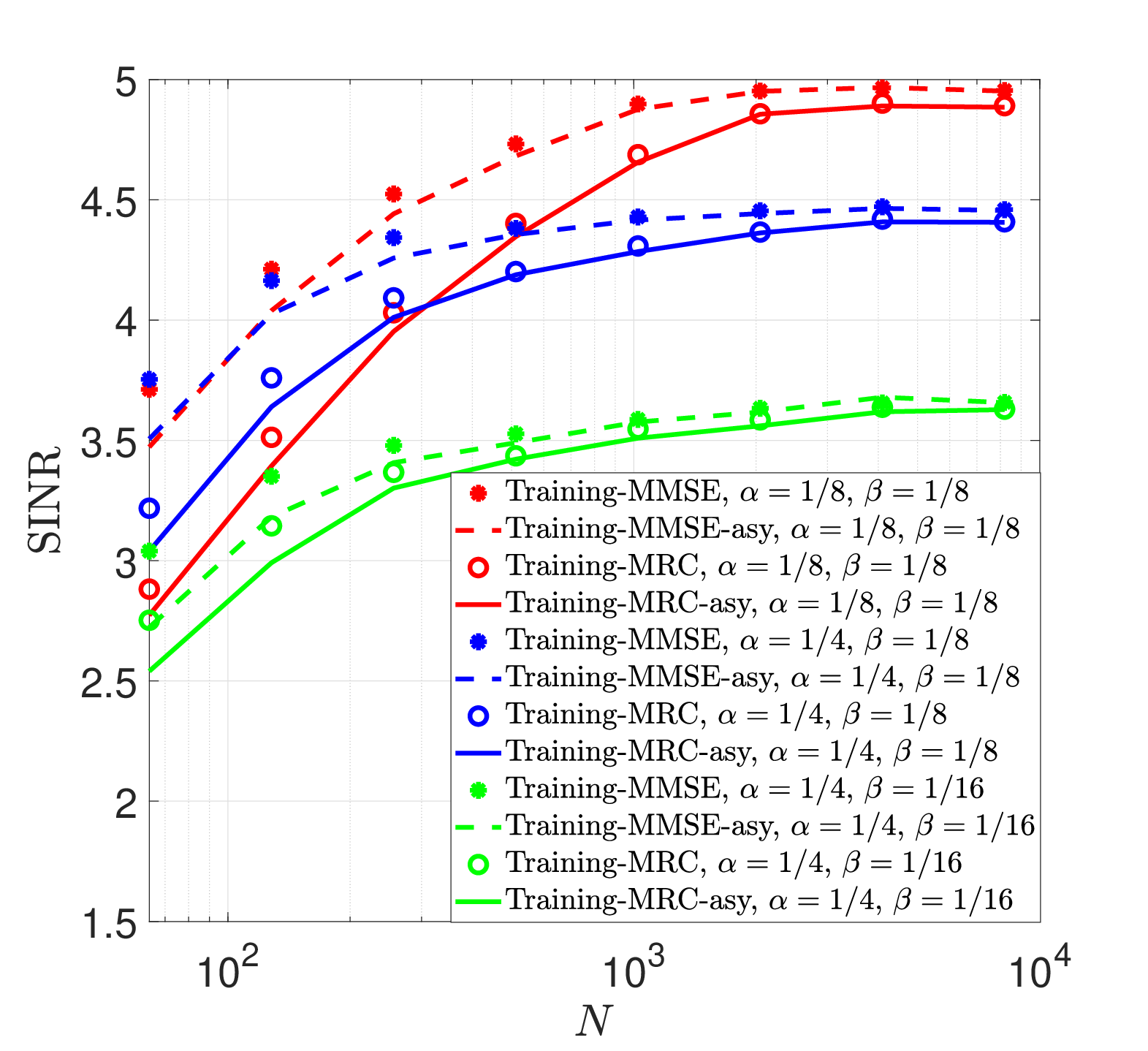}
\caption{$\mathrm{SINR}$ versus the length $N$ of the whole transmission block based on training-based method with $\mathrm{SNR}=10$\,dB and $K=3$.}
\label{fig_NMSE_alpha_bata_tra}
\end{figure}

\begin{figure}[t]
\centering
\includegraphics[width=0.92\linewidth]{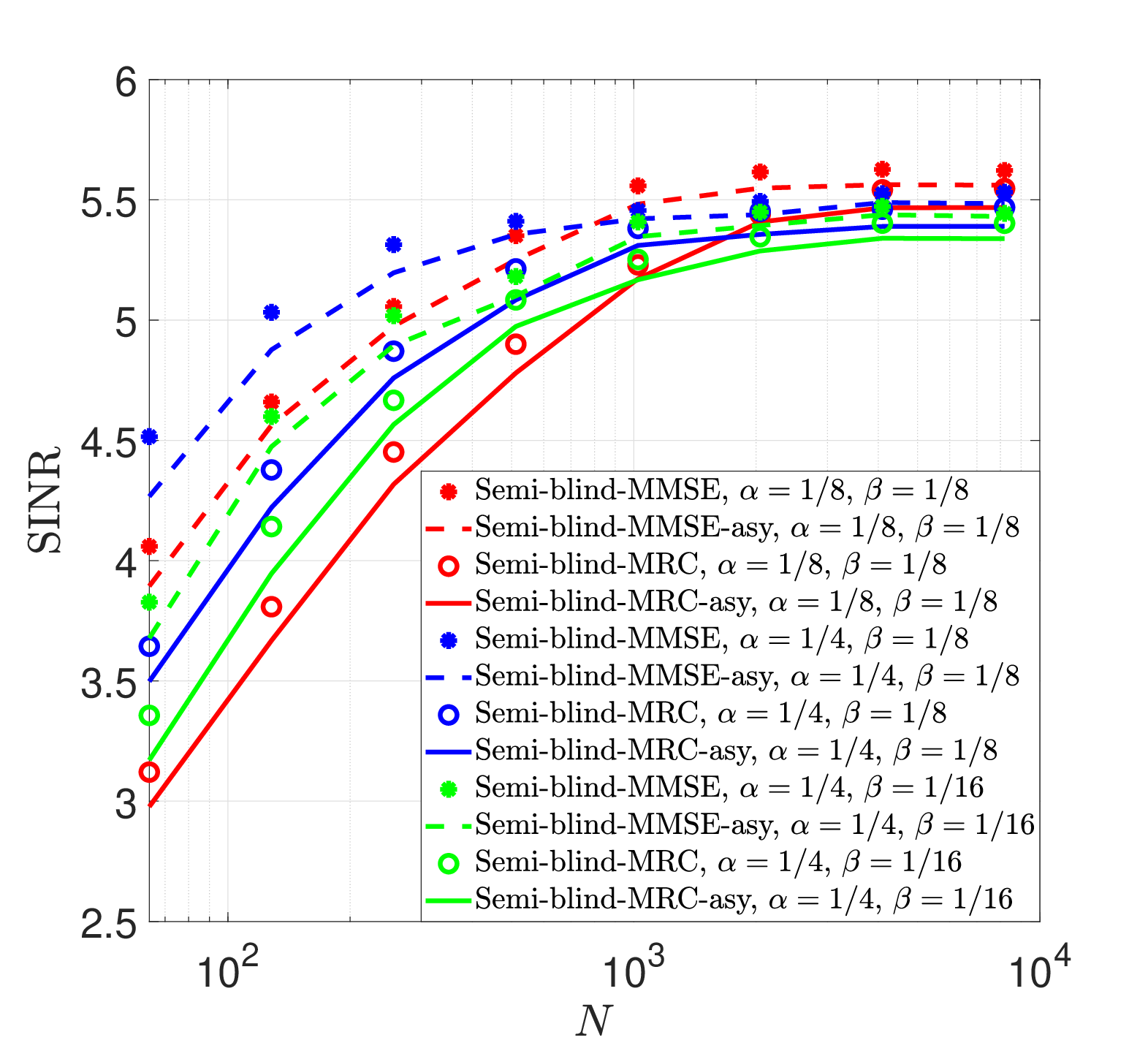}
\caption{$\mathrm{SINR}$ versus the length $N$ of the whole transmission block based on semi-blind method with $\mathrm{SNR}=10$\,dB and $K=3$.}
\label{fig_NMSE_alpha_beta_sb}
\end{figure}

In the first experiment, we evaluate the accuracy of the proposed asymptotic SINR expressions using both MRC and MMSE detectors, considering training-based and semi-blind channel estimation methods. Specifically, we compare the NMSE predicted by the asymptotic expressions with their corresponding exact values. The number of users is fixed at $K = 3$, the value of SNR is set to 10\,dB, and the parameters $\alpha$ and $\beta$ are both set to $1/4$. The transmission block length $N$ is varied from $2^6$ to $2^{13}$. As shown in Fig.~\ref{fig_NMSE_N}, the NMSE decreases with increasing $N$ for all SINR values, indicating the improved accuracy of the asymptotic expressions as the block length grows.
We also change the considered values of $\alpha$ and $\beta$, as specified in Assumption~\ref{ass2_parameter}. As the system dimension increases, the estimation error decreases accordingly. This behavior is illustrated in Fig.~\ref{fig_NMSE_alpha_bata_tra} for the training-based method and in Fig.~\ref{fig_NMSE_alpha_beta_sb} for the semi-blind method, respectively.

\subsection{Experiment for SINR}
\begin{figure}[t]
\centering
\includegraphics[width=0.92\linewidth]{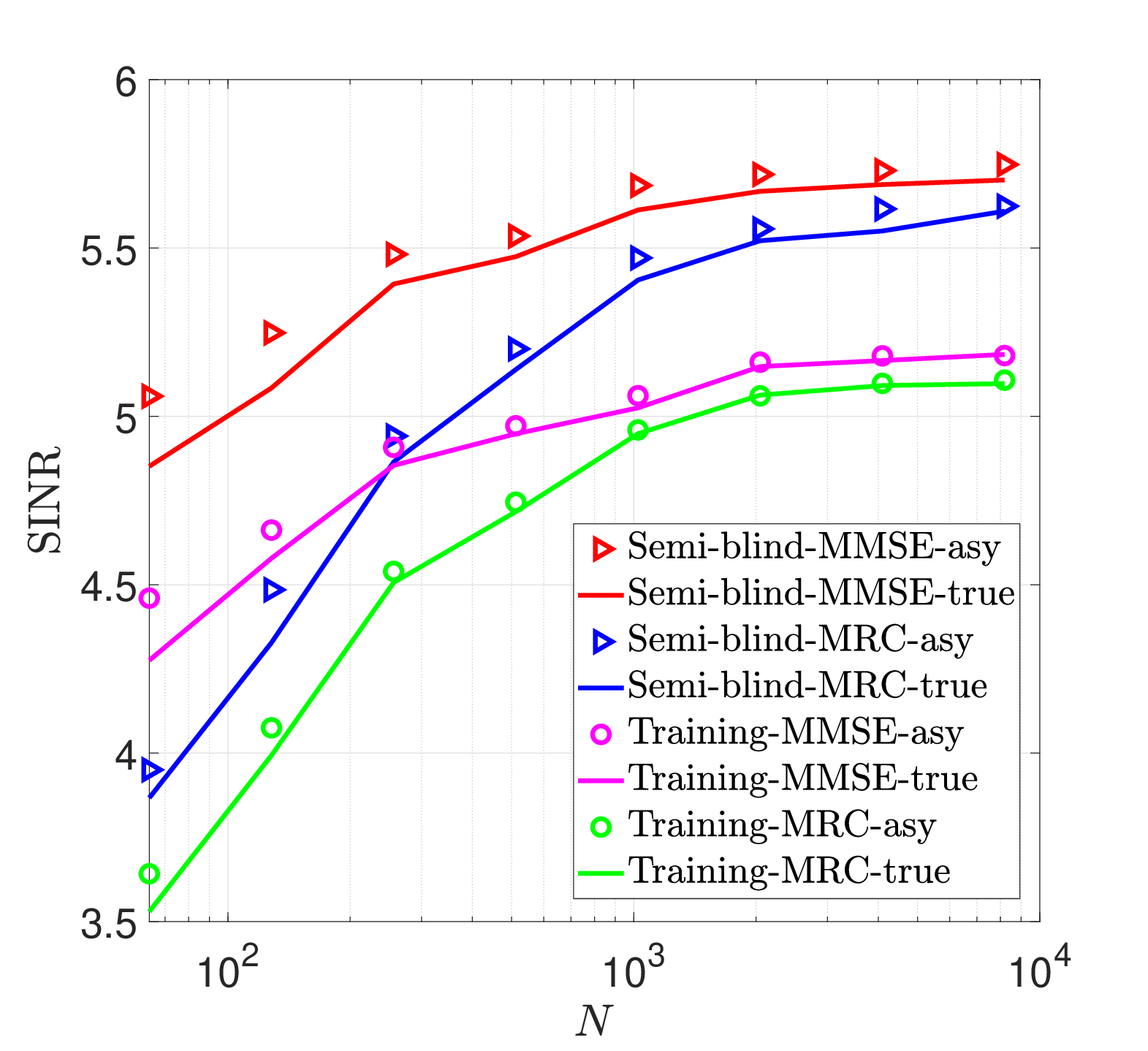}
\caption{SINR versus the length $N$ of the whole transmission block under with $\mathrm{SNR}=10$\,dB, $\alpha=1/4$, $\beta=1/4$, and $K=3$.}
\label{fig_SINR_N}
\end{figure}

In Fig.~\ref{fig_SINR_N}, we examine the behavior of the proposed asymptotic SINR expressions as a function of the total transmission block length $N$, under the regime defined by Assumption~\ref{ass2_parameter}. The system parameters are set to $\mathrm{SNR} = 10$\,dB, $\alpha = 1/4$, $\beta = 1/4$, and $K = 3$. As $N$ increases, the SINR values converge to their asymptotic limits predicted by Theorems~1-4. Furthermore, the MMSE receiver consistently achieves higher SINR than the MRC receiver for both training-based and semi-blind methods. For both receivers, the semi-blind channel estimation method yields the higher SINR value compared to the training-based method, demonstrating its improved performance.

\begin{figure}[t]
\centering
\includegraphics[width=0.92\linewidth]{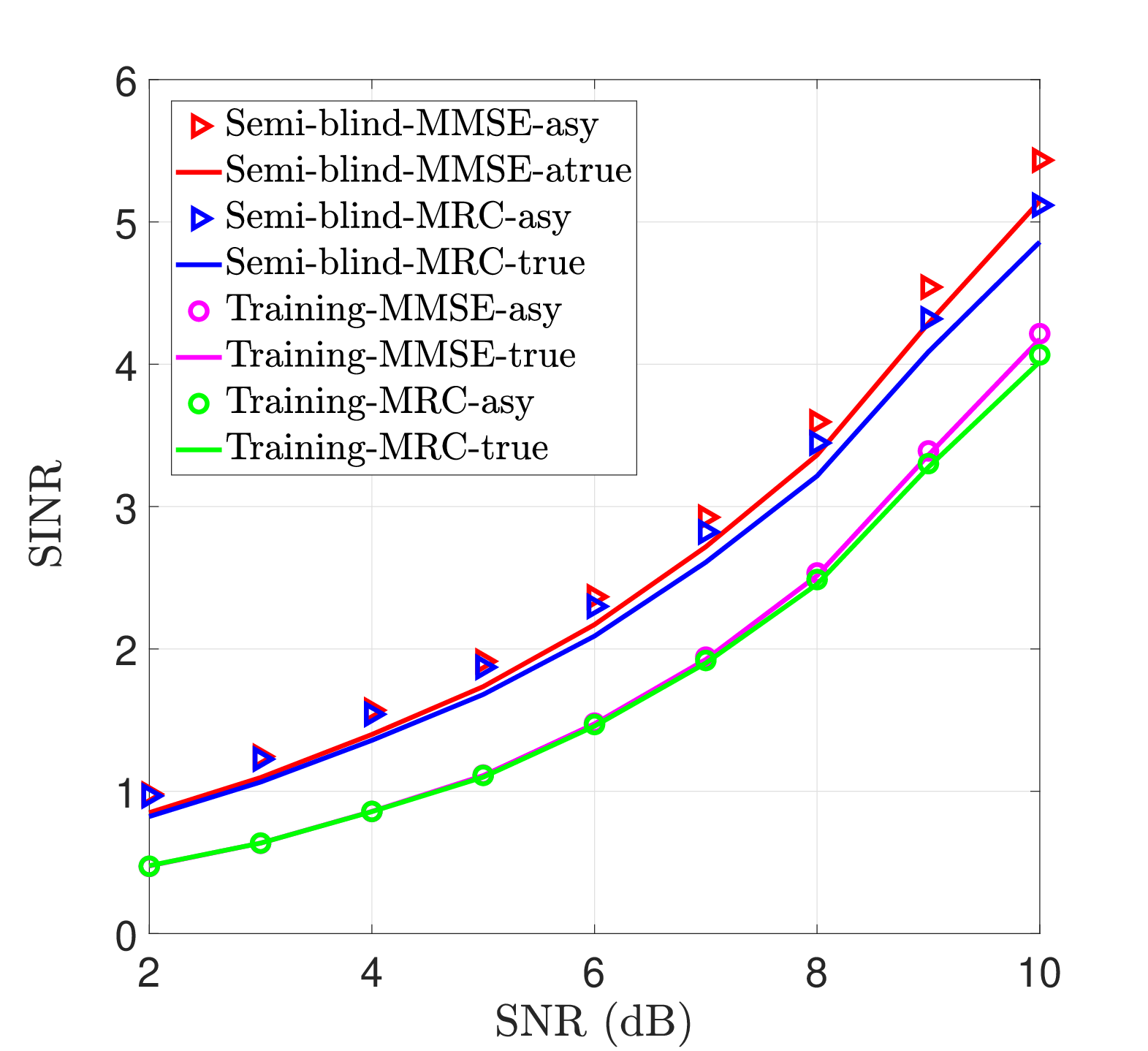}
\caption{SINR versus $\mathrm{SNR}$ under $K=3$, $\alpha=1/4$, $\beta=1/8$, and $N=512$.}
\label{fig_SINR_SNR}
\end{figure}

\begin{figure}[t]
\centering
\includegraphics[width=0.92\linewidth]{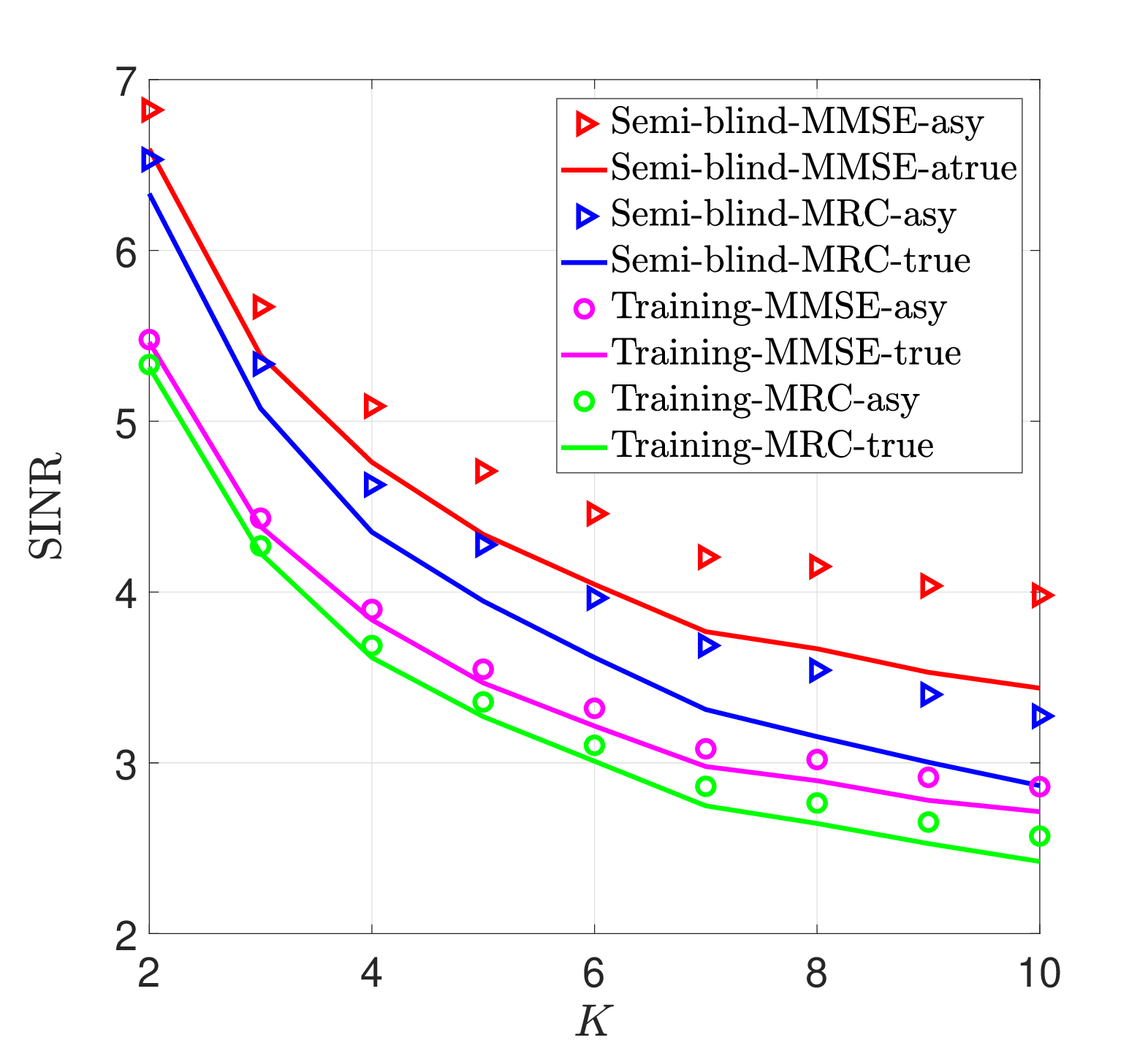}
\caption{SINR versus the number $K$ of users under $\mathrm{SNR}=10$\,dB, $\alpha=1/4$, $\beta=1/8$, and $N=1024$.}
\label{fig_SINR_K}
\end{figure}

Furthermore, we investigate the asymptotic behavior of the SINR as the SNR and the number of users increase. Fig.~\ref{fig_SINR_SNR} shows the SINR performance for SNR values ranging from 2\,dB to 10\,dB with an increment of $1$\,dB, with the number of users fixed at 3, the length of the whole transmission block set to 512, and parameters $\alpha=1/4$ and $\beta=1/16$. It can be seen that the SINR of all receivers increases with SNR. Notably, the MMSE receiver combined with the semi-blind channel estimation method consistently outperforms the other receivers across the entire range of SNR values. Fig.~\ref{fig_SINR_K} depicts the variation of the SINR with respect to the number of users $K$, ranging from $2$ to $10$, under $\mathrm{SNR}=10$\,dB, $\alpha = 1/4$, $\beta = 1/8$, and $N=1024$. It is observed that the SINR of all schemes decreases as $K$ increases, owing to the increased inter-user interference. Among all receivers, the semi-blind MMSE receiver achieves the highest SINR across all user counts, followed by the training-based MMSE receiver.

\subsection{Experiment for required pilots}
\begin{figure}[t]
\centering
\includegraphics[width=0.92\linewidth]{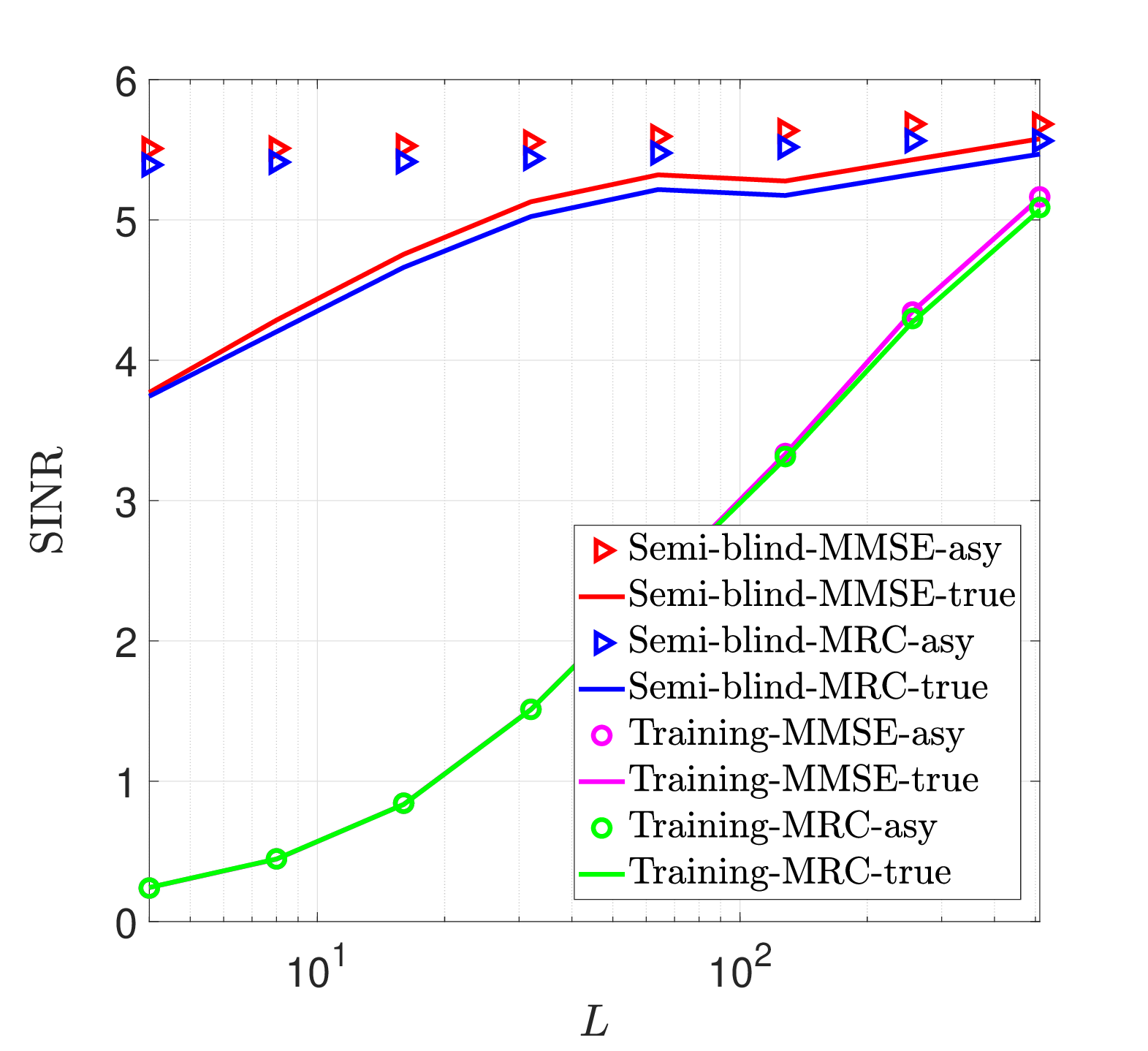}
\caption{SINR versus the length $L$ of the pilot symbols with  $N=1024$, and $K=3$, $\alpha=3/4$, and $\mathrm{SNR}=10$\,dB.}
\label{fig_SINR_L}
\end{figure}

Next, we demonstrate the practical application of the derived SINR expressions. For a given target SINR value, these expressions can be used to determine the number of pilot symbols required to achieve the desired SINR for MRC or MMSE receivers under both training-based and semi-blind estimation, while keeping all other parameters $N$, $M$, and $K$ constant. Fig.~\ref{fig_SINR_L} illustrates the relationship between SINR and the pilot length $L$, with system parameters set to $N = 1024$, $\alpha = 3/4$, $\mathrm{SNR}=10$ dB, and $K = 3$. As $L$ increases, the SINR improves for all cases, and the asymptotic results closely match the simulated ones, validating the accuracy of the derived expressions. Notably, the semi-blind approach achieves the same target SINR with less than half the number of pilot symbols required by the training-based scheme, highlighting its superior pilot efficiency and practical advantage in reducing training overhead.

\begin{figure}[t]
\centering
\includegraphics[width=0.92\linewidth]{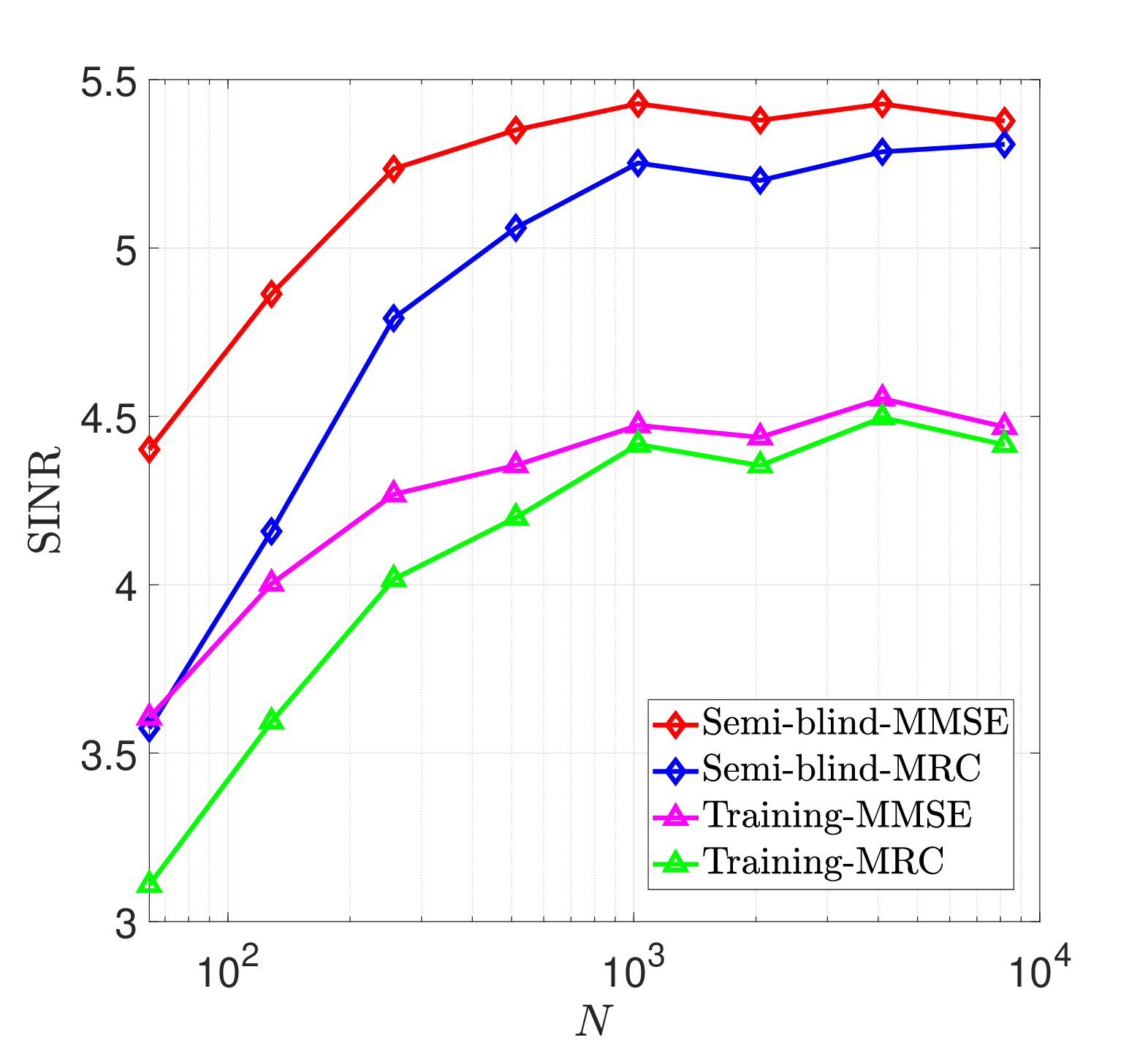}
\caption{SINR versus the length $N$ of the whole transmission block with $K=3$, $\alpha=1/4$, $\beta=1/8$, $\mathrm{SNR}=10$\,dB.}
\label{fig_SINR_time_varing_channel}
\end{figure}

\subsection{Experiment for time-varying channel}
Finally, we evaluate the SINR performance under a Doppler-induced time-varying channel, as shown in Fig.~8. Following the channel model in~\cite{ying2024deep}, the Doppler shift of each propagation path consists of a satellite-induced component and a terminal-induced component. Assuming that the dominant satellite Doppler shift is compensated, we consider residual satellite Doppler shifts of $5$, $-5$, and $10$ Hz for the three users, respectively, and a maximum terminal Doppler shift of $20$ Hz. The channel evolves continuously throughout both pilot and data transmissions due to the Doppler-induced phase variations of the LoS and NLoS components, with a symbol duration of $1\,\mu$s. Unlike the block-fading assumption adopted in the theoretical analysis, this experiment evaluates the robustness of the considered receivers to temporal channel variations. The parameters are set to $\alpha=1/4$, $\beta=1/8$, $K=3$, and $\mathrm{SNR}=10$\,dB. As illustrated in Fig.~\ref{fig_SINR_time_varing_channel}, the semi-blind receivers maintain higher SINR than their training-based counterparts under the considered Doppler-induced channel variations, demonstrating their robustness to mild temporal channel fluctuations.

\par

\section{Conclusion}
\label{sec_conclusion}
We have presented a unified analysis of the uplink performance of linear detectors in TDD massive MIMO systems. Under the large-system limit, we derived asymptotically tight approximations of the SINR for a general channel model with imperfect channel estimation. These approximations were shown to remain accurate for practical system dimensions and provide a powerful tool for studying realistic effects, such as antenna correlation, spacing, and aperture, without relying solely on simulations. Our results in realistic NTN environments demonstrate that receivers based on semi-blind channel estimation achieve superior performance compared to those relying on training-based methods. Moreover, we determined the number of pilot symbols required to achieve a target SINR in the asymptotic regime, offering valuable guidance for pilot design in massive MIMO systems.

\appendices

\par

\section{Proof of Theorem~\ref{theorem_tra_MRC}}\label{app_tra_MRC}
1) Deterministic equivalent for signal power: Using Lemma in \cite{hutchinson1989stochastic}, the term $\mathbb{E}[|(\hat{\mathbf{g}}_{\mathrm{tr},k}^{\mathrm{MRC}})^H\mathbf{h}_ks_k(n)|^2]$ yields
\begin{align}
    &\mathbb{E}[|(\hat{\mathbf{g}}_{\mathrm{tr},k}^{\mathrm{MRC}})^H\mathbf{h}_ks_k(n)|^2] \notag\\
    &\quad= P_s(\mathbf{h}_k^H\mathbf{h}_k)^2 + \frac{P_s}{a^2L^2}\mathbb{E}[\mathbf{z}_k^H\mathbf{V}_p^H\mathbf{h}_k\mathbf{h}_k^H\mathbf{V}_p\mathbf{z}_k] \notag\\  
    &\quad= P_s(\mathbf{h}_k^H\mathbf{h}_k)^2 + \frac{P_s\sigma_v^2}{aL}\mathbf{h}_k^H\mathbf{h}_k \asymp P_s\|\mathbf{h}_k\|^4.
\end{align}
where the last equivalence follows from the inequality ${\rm tr}({\bf H}_k{\bf H}_k^{H})\leq K\|{\bf H}{\bf H}^{H}\|_F$ together with the fact that $K/L\to 0$, under the considered asymptotic regime stated in Assumption \ref{ass2_parameter}. 

2) Deterministic equivalent for interference power:  The related term is given by 
\begin{align}
    &\sum_{i=1,i\neq k}^K\mathbb{E}[|(\hat{\mathbf{g}}_{\mathrm{tr},k}^{\mathrm{MRC}})^H\mathbf{h}_is_i(n)|^2]= P_s\mathbb{E}[(\hat{\mathbf{h}}_k^{\mathrm{tr}})^H\mathbf{H}_k\mathbf{H}_k^H\hat{\mathbf{h}}_k^{\mathrm{tr}}] \notag\\
    &= P_s\mathbf{h}_k^H\mathbf{H}_k\mathbf{H}_k^H\mathbf{h}_k + \frac{P_s}{a^2L^2}\mathbb{E}[\mathbf{z}_k^H\mathbf{V}_p^H\mathbf{H}_k\mathbf{H}_k^H\mathbf{V}_p\mathbf{z}_k]\notag\\
    &= P_s\|\mathbf{h}_k^H\mathbf{H}_k\|_2^2 + \frac{P_s\sigma_v^2}{aL}\mathrm{tr}(\mathbf{H}_k\mathbf{H}_k^H) \asymp P_s\|\mathbf{h}_k^H\mathbf{H}_k\|_2^2.
\end{align}

3) Deterministic equivalent for noise power: Taking the expectation with respect to the noise distribution, the noise  term  $\mathbb{E}[|(\hat{\mathbf{g}}_{\mathrm{tr},k}^{\mathrm{MRC}})^H\mathbf{v}(n)|^2]$ reduces to:
\begin{align}
    &\mathbb{E}[|(\hat{\mathbf{g}}_{\mathrm{tr},k}^{\mathrm{MRC}})^H\mathbf{v}(n)|^2] = \sigma_v^2\mathbf{h}_k^H\mathbf{h}_k + \frac{\sigma_v^2}{a^2L^2}\mathbb{E}[\mathbf{z}_k^H\mathbf{V}_p^H\mathbf{V}_p\mathbf{z}_k] \notag\\
    &\qquad= \sigma_v^2\mathbf{h}_k^H\mathbf{h}_k + \frac{\sigma_v^4M}{aL} \asymp \sigma_v^2\|\mathbf{h}_k\|_2^2 + \frac{\sigma_v^4\alpha}{a\beta}. 
\end{align}
By substituting the asymptotic expressions for the desired signal power, interference power, and noise power into the SINR expression $\mathrm{SINR}_{\mathrm{tr},k}^{\mathrm{MRC}}$, we obtain its corresponding deterministic equivalent (\ref{training_MRC_asy}).

\section{Proof of Theorem~\ref{theorem_tra_MMSE}}\label{app_tra_MMSE}
Using Woodbury matrix identity~\cite{higham2002accuracy}, $\hat{\mathbf{C}}^{\mathrm{tr}}$ can be rewritten as 
\begin{align}
    \hat{\mathbf{C}}^{\mathrm{tr}} = \frac{P_s}{\sigma_v^2}\mathbf{I}_M - \frac{P_s}{\sigma_v^2}\hat{\mathbf{H}}^{\mathrm{tr}}\hat{\mathbf{D}}^{\mathrm{tr}}(\hat{\mathbf{H}}^{\mathrm{tr}})^H,  
\end{align}
where $\hat{\mathbf{D}}^{\mathrm{tr}} = \Big((\hat{\mathbf{H}}^{\mathrm{tr}})^H\hat{\mathbf{H}}^{\mathrm{tr}}+\frac{\sigma_v^2}{P_s}\mathbf{I}_K\Big)^{-1}$. 
Then we have:  

1) Deterministic equivalent for signal power: With fixed $K$ and $M\rightarrow\infty$, we can obtain $(\hat{\mathbf{h}}_k^{\mathrm{tr}})^H\hat{\mathbf{h}}_k^{\mathrm{tr}} \asymp \mathbf{h}_k^H\mathbf{h}_k + \frac{\sigma_v^2\alpha}{a\beta}$, $(\hat{\mathbf{h}}_k^{\mathrm{tr}})^H\hat{\mathbf{h}}_i^{\mathrm{tr}}\asymp \mathbf{h}_k^{H}\mathbf{h}_i$, $(\hat{\mathbf{H}}^{\mathrm{tr}})^H\hat{\mathbf{H}}^{\mathrm{tr}}\asymp \mathbf{H}^H\mathbf{H}+\frac{\sigma_v^2\alpha}{a\beta}\mathbf{I}_K$, $(\hat{\mathbf{h}}_k^{\mathrm{tr}})^H\hat{\mathbf{H}}^{\mathrm{tr}}\rightarrow\mathbf{h}_k^H\mathbf{H}+\frac{\sigma_v^2\alpha}{a\beta}\mathbf{e}_k^T\triangleq\mathbf{f}_k^{\mathrm{tr}}$, $\hat{\mathbf{D}}^{\mathrm{tr}}\rightarrow \left(\mathbf{H}^H\mathbf{H}+(\frac{\sigma_v^2\alpha}{a\beta}+\frac{\sigma_v^2}{P_s})\mathbf{I}_K\right)^{-1}\triangleq\bar{\mathbf{D}}^{\mathrm{tr}}$. These results yields
\begin{align}
    &\mathbb{E}[(\hat{\mathbf{h}}_k^{\mathrm{tr}})^H\hat{\mathbf{C}}^{\mathrm{tr}}\mathbf{h}_k] \notag\\
    &\qquad= \frac{P_s}{\sigma_v^2}\mathbb{E}[(\hat{\mathbf{h}}_k^{\mathrm{tr}})^H\mathbf{h}_k] - \frac{P_s}{\sigma_v^2}\mathbb{E}[(\hat{\mathbf{h}}_k^{\mathrm{tr}})^H\hat{\mathbf{H}}^{\mathrm{tr}}\hat{\mathbf{D}}^{\mathrm{tr}}(\hat{\mathbf{H}}^{\mathrm{tr}})^H\mathbf{h}_k] \notag\\
    &\qquad\asymp \frac{P_s}{\sigma_v^2}\mathbf{h}_k^H\mathbf{h}_k - \frac{P_s}{\sigma_v^2}\mathbb{E}[(\hat{\mathbf{h}}_k^{\mathrm{tr}})^H\hat{\mathbf{H}}]\mathbb{E}[\hat{\mathbf{D}}^{\mathrm{tr}}]\mathbb{E}[(\hat{\mathbf{H}}^{\mathrm{tr}})^H\mathbf{h}_k] \notag \\
    &\qquad= \frac{P_s}{\sigma_v^2}\mathbf{h}_k^H\mathbf{h}_k - \frac{P_s}{\sigma_v^2} \mathbf{f}_k^{\mathrm{tr}}\bar{\mathbf{D}}^{\mathrm{tr}}\mathbf{H}^H\mathbf{h}_k.
\end{align}
According to uniform integrability~\cite{rudin1987real}, the deterministic equivalent of the signal power is given by  
\begin{align}
    &\mathbb{E}[|(\hat{\mathbf{g}}_{\mathrm{tr},k}^{\mathrm{MMSE}})^H\mathbf{h}_ks_k(n)|^2] = P_s\mathbb{E}[|(\hat{\mathbf{h}}_k^{\mathrm{tr}})^H\hat{\mathbf{C}}^{\mathrm{tr}}\mathbf{h}_k|^2] \notag\\
    &\quad \asymp P_s|\mathbb{E}[(\hat{\mathbf{h}}_k^{\mathrm{tr}})^H\hat{\mathbf{C}}^{\mathrm{tr}}\mathbf{h}_k]|^2 \notag\\
    &\quad= \frac{P_s^3}{\sigma_v^4}|\mathbf{h}_k^H\mathbf{h}_k-\mathbf{f}_k^{\mathrm{tr}}\bar{\mathbf{D}}^{\mathrm{tr}}\mathbf{H}^H\mathbf{h}_k|^2  = \frac{P_s^3}{\sigma_v^4}|(\mathbf{t}_k^{\mathrm{tr}})^H\mathbf{h}_k|^2.
\end{align}

2) Deterministic equivalent for interference power: Following a similar approach to the signal power analysis, the interference term is approximated as 
\begin{align}
    &\sum_{i=1\atop i\neq k}^K\mathbb{E}[|(\hat{\mathbf{g}}_{\mathrm{tr},k}^{\mathrm{MMSE}})^H\mathbf{h}_is_i(n)|^2] =P_s\mathbb{E}[(\hat{\mathbf{h}}_k^{\mathrm{tr}})^H\hat{\mathbf{C}}^{\mathrm{tr}}\mathbf{H}_k\mathbf{H}_k^H\hat{\mathbf{C}}^{\mathrm{tr}}\hat{\mathbf{h}}_k^{\mathrm{tr}}] \notag\\
    %&\qquad = \frac{P_s^3}{\sigma_v^4}\mathbb{E}[(\hat{\mathbf{h}}_k^{\mathrm{tr}})^H\mathbf{H}_k\mathbf{H}_k^H\hat{\mathbf{h}}_k^{\mathrm{tr}}] \notag\\
    %&\qquad\quad- \frac{P_s^3}{\sigma_v^4}\mathbb{E}[(\hat{\mathbf{h}}_k^{\mathrm{tr}})^H\hat{\mathbf{H}}^{\mathrm{tr}}\hat{\mathbf{D}}^{\mathrm{tr}}(\hat{\mathbf{H}}^{\mathrm{tr}})^H\mathbf{H}_k\mathbf{H}_k^H\hat{\mathbf{h}}_k^{\mathrm{tr}}] \notag\\
    %&\qquad\quad - \frac{P_s^3}{\sigma_v^4}\mathbb{E}[(\hat{\mathbf{h}}_k^{\mathrm{tr}})^H\mathbf{H}_k\mathbf{H}_k^H\hat{\mathbf{H}}^{\mathrm{tr}}\hat{\mathbf{D}}^{\mathrm{tr}}(\hat{\mathbf{H}}^{\mathrm{tr}})^H\hat{\mathbf{h}}_k^{\mathrm{tr}}] \notag\\
    %&\qquad\quad + \frac{P_s^3}{\sigma_v^4}\mathbb{E}[\|(\hat{\mathbf{h}}_k^{\mathrm{tr}})^H\hat{\mathbf{H}}^{\mathrm{tr}}\hat{\mathbf{D}}^{\mathrm{tr}}(\hat{\mathbf{H}}^{\mathrm{tr}})^H\mathbf{H}_k\|_2^2]  \notag\\
    &\qquad\asymp \frac{P_s^3}{\sigma_v^4}\mathbf{h}_k^H\mathbf{H}_k\mathbf{H}_k^H\mathbf{h}_k - \frac{P_s^3}{\sigma_v^4}\mathbf{f}_k^{\mathrm{tr}}\bar{\mathbf{D}}^{\mathrm{tr}}\mathbf{H}^H\mathbf{H}_k\mathbf{H}_k^H\mathbf{h}_k \notag\\
    &\qquad\quad-\frac{P_s^3}{\sigma_v^4}\mathbf{h}_k^H\mathbf{H}_k\mathbf{H}_k^H\mathbf{H}\bar{\mathbf{D}}^{\mathrm{tr}}(\mathbf{f}_k^{\mathrm{tr}})^H +\frac{P_s^3}{\sigma_v^4}\|\mathbf{f}_k^{\mathrm{tr}}\bar{\mathbf{D}}^{\mathrm{tr}}\mathbf{H}^H\mathbf{H}_k\|^2 \notag\\
    &\qquad= \frac{P_s^3}{\sigma_v^4}\|(\mathbf{t}_k^{\mathrm{tr}})^H\mathbf{H}_k\|_2^2.
\end{align}

3) Deterministic equivalent for noise power: The related term converges almost surely to its deterministic equivalent as 
\begin{align}
    &\mathbb{E}[|(\hat{\mathbf{g}}_{\mathrm{tr},k}^{\mathrm{MMSE}})^H\mathbf{v}(n)|^2] = \sigma_v^2\mathbb{E}[(\hat{\mathbf{h}}_k^{\mathrm{tr}})^H(\hat{\mathbf{C}}^{\mathrm{tr}})^2\hat{\mathbf{h}}_k^{\mathrm{tr}}] \notag\\
    %&\quad = \frac{P_s^2}{\sigma_v^2}\mathbb{E}[\hat{\mathbf{h}}_k^{\mathrm{tr}}\hat{\mathbf{h}}_k^{\mathrm{tr}}] - \frac{2P_s^2}{\sigma_v^2}\mathbb{E}[(\hat{\mathbf{h}}_k^{\mathrm{tr}})^H\hat{\mathbf{H}}^{\mathrm{tr}}\hat{\mathbf{D}}^{\mathrm{tr}}(\hat{\mathbf{H}}^{\mathrm{tr}})^H\hat{\mathbf{h}}_k^{\mathrm{tr}}] \notag\\
    %&\qquad + \frac{P_s^2}{\sigma_v^2}\mathbb{E}[(\hat{\mathbf{h}}_k^{\mathrm{tr}})^H\hat{\mathbf{H}}^{\mathrm{tr}}\hat{\mathbf{D}}^{\mathrm{tr}}(\hat{\mathbf{H}}^{\mathrm{tr}})^H\hat{\mathbf{H}}^{\mathrm{tr}}\hat{\mathbf{D}}^{\mathrm{tr}}(\hat{\mathbf{H}}^{\mathrm{tr}})^H\hat{\mathbf{h}}_k^{\mathrm{tr}}] \notag\\
    &\quad\asymp \frac{P_s^2}{\sigma_v^2}\Big(\mathbf{h}_k^H\mathbf{h}_k+\frac{\sigma_v^2\alpha}{a\beta}\Big) - \frac{P_s^2}{\sigma_v^2}\mathbf{f}_k^{\mathrm{tr}}\bar{\mathbf{D}}^{\mathrm{tr}}\Big(\mathbf{H}^H\mathbf{h}_k+\frac{\sigma_v^2\alpha}{a\beta}\mathbf{e}_k\Big)\notag\\
    &\qquad - \frac{P_s^2}{\sigma_v^2}\Big(\mathbf{h}_k^H\mathbf{H}+\frac{\sigma_v^2\alpha}{a\beta}\mathbf{e}_k^T\Big)\bar{\mathbf{D}}^{\mathrm{tr}}(\mathbf{f}_k^{\mathrm{tr}})^H\notag\\
    &\qquad+ \frac{P_s^2}{\sigma_v^2}\mathbf{f}_k^{\mathrm{tr}}\bar{\mathbf{D}}^{\mathrm{tr}}\Big(\mathbf{H}^H\mathbf{H}+\frac{\sigma_v^2\alpha}{a\beta}\mathbf{I}_K\Big)\bar{\mathbf{D}}^{\mathrm{tr}}(\mathbf{f}_k^{\mathrm{tr}})^H \notag\\
    &\quad= \frac{P_s^2}{\sigma_v^2}\|\mathbf{t}_k^{\mathrm{tr}}\|_2^2 + \frac{P_s^2\alpha}{a\beta}\|\mathbf{e}_k^T-\mathbf{f}_k^{\mathrm{tr}}\bar{\mathbf{D}}^{\mathrm{tr}}\|_2^2.  
\end{align} 
As a result, by combining the asymptotic expressions of the signal, interference, and noise power terms, the deterministic equivalent of $\mathrm{SINR}_{\mathrm{tr},k}^{\mathrm{MMSE}}$ is expressed as (\ref{training_MMSE_asy}).

\par

%\section{Proof of Theorem~\ref{theorem_sb_MRC}}\label{app_sb_MRC}
\section{Preliminary Technical Results for Deriving SINR Deterministic Equivalents}\label{app_technical_results_basis}
The derivation of the asymptotic SINR for the semi-blind estimator is significantly more involved than in the training-based case. This increased difficulty stems from the fact that, unlike training-based estimators, the semi-blind approach relies on the computation of the signal subspace, which inherently depends on both the transmitted data symbols and the associated noise during data transmission. To address this intrinsic dependency, we first establish key intermediate results, stated as Lemma \ref{first-order-signal-noise_bis} and Lemma \ref{second-order-signal-noise_bis}, which serves as the cornerstone of our analysis. Prior to presenting this lemma, we briefly review several fundamental results on the spectral properties of spiked random matrix models, which will play a crucial role in the subsequent derivations.
\subsection{Preliminary}

Under the regime specified in Assumption~\ref{ass2_parameter}, the sample covariance matrix $\hat{\bf R}$ follows a spiked random matrix model. More precisely, when $K$ is fixed, the population covariance matrix ${\bf R}$ has $(M-K)$ eigenvalues equal to $\sigma_v^2$, while the remaining $K$ eigenvalues are given by $\sigma_v^2(1+t_k)$, for $k=1,\ldots,K$. These $K$ eigenvalues correspond to the signal components and induce a finite-rank perturbation of the noise covariance.

Let the empirical spectral distribution (ESD) of $\hat{\mathbf{R}}$ be defined as~\cite{zhang2026crb}
\begin{align}
\mu_{\hat{\mathbf{R}}}(\lambda) = \frac{1}{M}\sum_{i=1}^M\delta_{\lambda_i(\hat{\mathbf{R}})}.
\end{align}
Under the asymptotic regime of Assumption~\ref{ass2_parameter}, this empirical measure converges almost surely to a deterministic limit described by the Mar\v{c}enko--Pastur (MP) law. Specifically, the limiting distribution admits the density
\begin{align}
\mu(x) = \frac{\sqrt{(x-x^{-})(x^{+}-x)}}{2\pi c x}\,\mathbf{1}_{[x^{-},x^{+}]}(x),
\end{align}
where $x^{-}=\sigma_v^2(1-\sqrt{c})^2$ and $x^{+}=\sigma_v^2(1+\sqrt{c})^2$, and $\mathbf{1}_{[x^{-},x^{+}]}(x)$ denotes the indicator function over the support of the distribution.

Equivalently, the associated Stieltjes transform $m(z)$ is the unique solution in $\mathbb{C}^+$ of the MP equation
\begin{align}
z c\sigma_v^2 m^2(z) - (\sigma_v^2(1-c)-z)m(z) + 1 = 0.
\end{align}
In addition, the extreme eigenvalues of $\hat{\mathbf{R}}$ exhibit a phase transition behavior. More precisely, the eigenvalues of $\hat{\mathbf{R}}$, ordered in decreasing order, satisfy
\begin{align}
\lambda_i(\hat{\mathbf{R}}) \xrightarrow[]{\text{a.s.}}
\bar{\eta}_i =
\begin{cases}
\sigma_v^2\Big(1 + t_i + c\,\dfrac{1 + t_i}{t_i}\Big), & i \leq K, \\[6pt]
x^{+}, & i > K,
\end{cases}
\end{align}
provided that the spikes $t_i$ are above the detection threshold.
Moreover, for any deterministic vectors $\mathbf{a}$ and $\mathbf{b}$ with bounded norms, the eigenprojection associated with the $i$-th sample eigenvector satisfies
\begin{align}
\label{eq:convergence_quadratic_form}
\mathbf{a}^{\mathrm H}\hat{\mathbf{u}}_{si}\hat{\mathbf{u}}_{si}^{\mathrm H}\mathbf{b}
\xrightarrow[]{\mathrm{a.s.}}
\begin{cases}
\displaystyle 
z_{si}\mathbf{a}^{\mathrm H}{\mathbf{u}}_{si}{\mathbf{u}}_{si}^{\mathrm H}\mathbf{b},
& i\leq K,\; t_i>\sqrt{c}, \\[10pt]
0, 
& i>K.
\end{cases}
\end{align}
Hence, letting $\hat{\mathbf{U}}_s = [\hat{\mathbf{u}}_{s1}, \ldots, \hat{\mathbf{u}}_{sK}]$ denote the matrix of sample eigenvectors associated with the signal subspace, we obtain, for any deterministic vectors $\mathbf{a}$ and $\mathbf{b}$ with bounded norms,
\begin{align}
\mathbf{a}^{\mathrm H}\hat{\mathbf{U}}_{s}\hat{\mathbf{U}}_{s}^{\mathrm H}\mathbf{b}
-
\mathbf{a}^{\mathrm H}{\mathbf{U}}_{s}\boldsymbol{\Lambda}_s{\mathbf{U}}_{s}^{\mathrm H}\mathbf{b}
\xrightarrow[]{\mathrm{a.s.}} 0, \label{eq:projection}
\end{align}
where $\boldsymbol{\Lambda}_s = \mathrm{diag}([z_{s1}, \ldots, z_{sK}])$ collects the asymptotic projection coefficients associated with the signal eigenvectors.
\subsection{Initial results} 
\subsubsection{Terms involving single projections}
As will be shown next, evaluating the SINR of the semi-blind receiver requires deriving deterministic equivalents for the following terms involving single signal-subspace projections: 
\begin{align}
&\mathbb{E}[\mathbf a^H\hat{\mathbf U}_s\hat{\mathbf U}_s^H\mathbf b\,s_i(n)s_k^{\ast}(n)],
\,\,\,
\mathbb{E}[\mathbf a^H\hat{\mathbf U}_s\hat{\mathbf U}_s^H\mathbf v(n)\mathbf v^H(n)\mathbf b], \notag\\
&\mathbb{E}[\mathbf a^H\hat{\mathbf U}_s\hat{\mathbf U}_s^H\mathbf v(n)s_k^{\ast}(n)],
\,\,\,
\mathbb{E}[\mathbf a^H\hat{\mathbf U}_s\hat{\mathbf U}_s^H\mathbf H\mathbf s(n)\mathbf v^H(n)\mathbf b], \notag\\
&\mathbb{E}[\mathbf a^H\mathbf H\mathbf s(n)\mathbf v^H(n)
\hat{\mathbf U}_s\hat{\mathbf U}_s^H\mathbf b].
\end{align}

The first and fourth quantities can be evaluated directly using the asymptotic convergence of the estimated signal-subspace projector established in \eqref{eq:projection}. Indeed, under the stated boundedness and integrability assumptions, the dominated convergence theorem yields
\begin{align}
\mathbb{E}[\mathbf a^H\hat{\mathbf U}_s\hat{\mathbf U}_s^H
\mathbf b\,s_i(n)s_k^{\ast}(n)] &\asymp
\mathbf a^H\mathbf U_s\boldsymbol{\Lambda}_s\mathbf U_s^H
\mathbf b\,\mathbb{E}[s_i(n)s_k^{\ast}(n)] \notag\\
&=
P_s\delta_{ik}\,
\mathbf a^H\mathbf U_s\boldsymbol{\Lambda}_s\mathbf U_s^H\mathbf b. \label{eq:44}
\end{align}
Similarly, for fixed $K$, the projector convergence, combined with the bounded second moments of $s_j(n)\mathbf v^H(n)\mathbf b$ and the Cauchy--Schwarz inequality, gives
\begin{align}
&\mathbb{E}[\mathbf a^H\hat{\mathbf U}_s\hat{\mathbf U}_s^H
\mathbf H\mathbf s(n)\mathbf v^H(n)\mathbf b] \notag\\
&\quad\asymp
\mathbf a^H\mathbf U_s\boldsymbol{\Lambda}_s\mathbf U_s^H
\mathbf H\,
\mathbb{E}[\mathbf s(n)\mathbf v^H(n)\mathbf b]
=0,\label{eq:zero}
\end{align}
where the last equality follows from the independence of the transmitted symbols and the zero-mean noise.

In contrast, the second and third quantities cannot be obtained directly from \eqref{eq:projection}, because the noise vector $\mathbf v(n)$ is statistically coupled with the estimated signal-subspace projector. These terms therefore require a separate analysis that explicitly accounts for this dependence. As shown in the following, the deterministic equivalent derived for
$\mathbb{E}[\mathbf a^H\hat{\mathbf U}_s\hat{\mathbf U}_s^H
\mathbf v(n)s_k^{\ast}(n)]$
can subsequently be used to obtain that of
$\mathbb{E}[\mathbf a^H\mathbf H\mathbf s(n)\mathbf v^H(n)
\hat{\mathbf U}_s\hat{\mathbf U}_s^H\mathbf b]$.
\subsubsection{Terms involving two projections} 
In addition to the terms involving a single projection, evaluating the SINR requires deriving deterministic equivalents for the following quantities involving two signal-subspace projections:
\begin{align}
&\mathbb{E}[{\bf a}^{H}\hat{\bf U}_s\hat{\bf U}_s^{H}
{\bf h}_i s_i(n)s_k^\ast(n){\bf h}_k^{H}
\hat{\bf U}_s\hat{\bf U}_s^{H}{\bf b}], \notag\\
&\mathbb{E}[{\bf a}^{H}\hat{\bf U}_s\hat{\bf U}_s^{H}
{\bf v}(n){\bf v}^{H}(n)
\hat{\bf U}_s\hat{\bf U}_s^{H}{\bf b}], \notag\\
&\mathbb{E}[{\bf a}^{H}\hat{\bf U}_s\hat{\bf U}_s^{H}
{\bf H}{\bf s}(n){\bf v}^{H}(n)
\hat{\bf U}_s\hat{\bf U}_s^{H}{\bf b}].
\label{eq:last_term}
\end{align}

The first quantity can be evaluated directly using the asymptotic convergence of the estimated signal-subspace projector established in \eqref{eq:projection}, together with the dominated convergence theorem. Under the corresponding boundedness and integrability assumptions, we obtain
\begin{align}
&\mathbb{E}[{\bf a}^{H}\hat{\bf U}_s\hat{\bf U}_s^{H}
{\bf h}_i s_i(n)s_k^\ast(n){\bf h}_k^{H}
\hat{\bf U}_s\hat{\bf U}_s^{H}{\bf b}]
\notag\\
&\quad\asymp
{\bf a}^{H}{\bf U}_s\boldsymbol{\Lambda}_s{\bf U}_s^{H}
{\bf h}_i{\bf h}_k^{H}
{\bf U}_s\boldsymbol{\Lambda}_s{\bf U}_s^{H}{\bf b}
\,\mathbb{E}[s_i(n)s_k^\ast(n)]
\notag\\
&\quad=
P_s\delta_{ik}\,
{\bf a}^{H}{\bf U}_s\boldsymbol{\Lambda}_s{\bf U}_s^{H}
{\bf h}_i{\bf h}_k^{H}
{\bf U}_s\boldsymbol{\Lambda}_s{\bf U}_s^{H}{\bf b}.\label{eq:47}
\end{align}

The deterministic equivalent of the third quantity, given in \eqref{eq:last_term}, can be obtained from that of
$\mathbb{E}[{\bf a}^{H}\hat{\bf U}_s\hat{\bf U}_s^{H}
{\bf v}(n)s_k^\ast(n)]$,
which will be derived in Corollary \ref{cor:by_product}.
In contrast, the noise covariance term
$\mathbb{E}[{\bf a}^{H}\hat{\bf U}_s\hat{\bf U}_s^{H}
{\bf v}(n){\bf v}^{H}(n)
\hat{\bf U}_s\hat{\bf U}_s^{H}{\bf b}]$
requires a separate analysis. Indeed, the noise vector ${\bf v}(n)$ is statistically coupled with the estimated signal-subspace projector, preventing a direct application of the deterministic equivalent in \eqref{eq:projection}. Its derivation must therefore explicitly account for this dependence.

%This result highlights the fundamental structure of the spectrum: most eigenvalues (a fraction $1 - \frac{K}{N}$) form a continuous bulk that follows the Mar\v{c}enko--Pastur distribution, reflecting the noise subspace, while a finite number $K$ of eigenvalues--referred to as spikes--separate from the bulk and converge to deterministic limits outside the MP support. These isolated eigenvalues carry the information about the underlying signal subspace and play a central role in subspace-based estimation methods.

\subsection{Deterministic equivalents for $\mathbb{E}[{\bf a}^{H}\hat{\bf U}_s\hat{\bf U}_s^{H}{\bf v}(n){\bf v}^{H}(n){\bf b}]$ and $\mathbb{E}[{\bf a}^{H}\hat{\bf U}_s\hat{\bf U}_s^{H}{\bf v}(n)s_k^\ast(n)]$}
%\textcolor{blue}{Here we need to adapt this introduction}
The following lemma establishes the asymptotic behavior of $\mathbb{E}[\mathbf{a}^H\hat{\mathbf{U}}_s\hat{\mathbf{U}}_s^H{\bf v}(n)s_k^{\ast}(n)\mathbf{b}]$ 
and $\mathbb{E}[\mathbf{a}^H\hat{\mathbf{U}}_s\hat{\mathbf{U}}_s^H\mathbf{v}(n)\mathbf{v}(n)^H\mathbf{b}]$. Importantly, due to the intrinsic dependence between the estimated subspace $\hat{\mathbf{U}}_s$ and the signal and noise vectors, the result in \eqref{eq:projection} cannot be directly applied, and a refined analysis is required.
%In this subsection, we introduce the deterministic equivalent as a useful analytical tool for characterizing the asymptotic behavior of scalar (eigen-spectral) observations of large random matrices.
\begin{lemma}\label{first-order-signal-noise_bis}
For any deterministic sequences $\mathbf a$ and $\mathbf b$ with uniformly bounded norms, namely $\limsup \|\mathbf a\|<\infty$ and $\limsup \|\mathbf b\|<\infty$, where $\mathbf s(n)$ and $\mathbf v(n)$ are defined in~\eqref{received signal} and $\hat{\mathbf U}_s$ denotes the matrix collecting the $K$ sample eigenvectors of the sample covariance matrix in \eqref{eq:empirical_covariance_matrix}, the following deterministic approximations hold:
\begin{align}
&\mathbb{E}[\mathbf a^H\hat{\mathbf U}_s\hat{\mathbf U}_s^H\mathbf v(n)\mathbf v^H(n)\mathbf b]\notag\\
&\qquad\asymp\sigma_v^2\sum_{k=1}^{K}(1-c/t_k^2){\bf a}^{H}{\bf u}_{sk}{\bf u}_{sk}^{H}{\bf b}, \label{aUUvvb_bis}\\
&\mathbb{E}[\mathbf a^H\hat{\mathbf U}_s\hat{\mathbf U}_s^H\mathbf v(n)s_k^{\ast}(n)]\notag\\
&\qquad\asymp \sum_{k=1}^{K}P_sc\frac{(1-c/t_k^2)}{t_k+c}{\bf a}^{H}{\bf u}_{sj}{\bf u}_{sj}^{H}{\bf h}_k,\label{aUUvs_bis}
\end{align}
\end{lemma}
\begin{proof}
The proof is organized into three main steps. In the first step, we express the quantities of interest in terms of spectral projectors and rewrite them as contour integrals involving the resolvent matrix \({\bf Q}_n(z)\), to be defined later. 
In the second step, we derive asymptotic approximations for the quadratic forms and normalizing traces involving \({\bf Q}_n(z)\). 
In the third step, we substitute these deterministic equivalents into the contour representation and evaluate the resulting integrals by residue calculus. This yields the desired deterministic approximation of the quantities of interest.

\noindent{\bf Step 1. Expressing the quantities as contour integrals.} In this step, we approximate both quantities in \eqref{aUUvvb_bis} and \eqref{aUUvs_bis} as contour integrals involving the resolvent matrix ${\bf Q}_n(z)$ defined later.  
Let $\Gamma$ be a positively oriented contour enclosing the limiting spike eigenvalues
$\bar{\eta}_1,\ldots,\bar{\eta}_K$ and separated from the limiting bulk support.
By the separability condition and the almost sure convergence of the sample eigenvalues, we can prove using the same arguments as in \cite{couillet2011random} that
there exists $\epsilon>0$ such that, almost surely for all sufficiently large dimensions,
\begin{align}
\min_{z\in\Gamma}\min_{1\leq i\leq M}|z-\hat{\eta}_i| \geq \epsilon \ \ \text{and } \min_{z\in\Gamma}\min_{1\leq i\leq M}|z-\hat{\mu}_i|,
\end{align}
where $\mu_1,\cdots,\mu_M$ are the solutions to $1+\frac{1}{N-L}{\rm tr}({\bf Q}_N(z))=0.$
Define the event
\begin{align}
\mathcal{O}_N = \Big\{ \forall z\in\Gamma,\ \min(\min_{1\leq i\leq M}|z-\hat{\eta}_i|, \min_{1\leq i\leq M}|z-\hat{\mu}_i|)\geq \epsilon \Big\}. \notag
\end{align}
Since $\mathbb{P}(\mathcal{O}_N)\to 1$, we can restrict the analysis to this event and write
\begin{align}
&\mathbb{E}\big[{\bf a}^{H}\hat{\bf U}_s\hat{\bf U}_s^{H}\,{\bf v}(n)s_k^\ast(n)\big]
\notag\\
&\qquad\asymp
\mathbb{E}\big[{\bf 1}_{\mathcal{O}_N}\,{\bf a}^{H}\hat{\bf U}_s\hat{\bf U}_s^{H}{\bf v}(n)  s_k^\ast(n)\big], \label{eq:aus_ss}\\
&\mathbb{E}\big[{\bf a}^{H}\hat{\bf U}_s\hat{\bf U}_s^{H}{\bf v}(n){\bf v}(n)^{H}{\bf b}\big] \notag\\
&\qquad\asymp
\mathbb{E}\big[{\bf 1}_{\mathcal{O}_N}{\bf a}^{H}\hat{\bf U}_s\hat{\bf U}_s^{H}{\bf v}(n){\bf v}(n)^{H}{\bf b}\big],\label{eq:aus_vv}
\end{align}
where ${\bf 1}_{\mathcal{O}_N}$ denotes the indicator function of the event $\mathcal{O}_N$. 
 Using Cauchy's integral formula, the following equality holds:
\begin{align}
&\mathbb{E}[{\bf 1}_{\mathcal{O}_N}{\bf a}^{H}\hat{\bf U}_s\hat{\bf U}_s^{H}{\bf v}(n)s_k^\ast(n)]\nonumber \\
&=-\frac{1}{2\pi\imath}\oint_{\Gamma} \mathbb{E}[{\bf 1}_{\mathcal{O}_N}{\bf a}^{H}{\bf Q}(z){\bf v}(n)s_k^\ast(n)] .\label{eq:Cauchy}
\end{align}
Since $\mathbf{Q}(z)$ depends on both $\mathbf{s}(n)$ and $\mathbf{v}(n)$, it is necessary to separate their contributions. To this end, let $\mathbf{Y}_d = \mathbf{B}[\mathbf{S}_d^H,\mathbf{V}_d^H]^H$ with $\mathbf{B}=[\mathbf{H},\mathbf{I}_M]$, and denote by ${\bf x}_d(i), i=1,\cdots,N-L$ the columns of $[{\bf S}_d^{H},{\bf V}_d^{H}]^{H}$. Define the matrix $\hat{\bf R}_n$ as:
\begin{align}
    \hat{\mathbf{R}}_{n} = \frac{1}{N-L}\sum_{i=1\atop i\neq n}^{N-L}\mathbf{B}\mathbf{x}_d(i)\mathbf{x}_d^H(i)\mathbf{B}^H.
\end{align}
and associate with it  the resolvent  $\mathbf{Q}_n(z)= (\hat{\mathbf{R}}_n-z\mathbf{I}_M)^{-1}$.  Using matrix inversion lemma~\cite{silverstein1995empirical}, the resolvent of $\mathbf{Q}(z)=(\hat{\bf R}-z{\bf I}_M)^{-1}$ can be rewritten as 
\begin{align}
    &\mathbf{Q}(z)= (\hat{\mathbf{R}}-z\mathbf{I}_M)^{-1} \notag\\
    &\quad= \left(\hat{\mathbf{R}}_{n}-z\mathbf{I}_M + \frac{1}{N-L}\mathbf{B}\mathbf{x}_d(n)\mathbf{x}_d^H(n)\mathbf{B}^H\right)^{-1} \notag\\
    &\quad= \mathbf{Q}_n(z) - \frac{\frac{1}{N-L}\mathbf{Q}_n(z)\mathbf{B}\mathbf{x}_d(n)\mathbf{x}_d^H(n)\mathbf{B}^H\mathbf{Q}_n(z)}{1+\frac{1}{N-L}\mathbf{x}_d^H(n)\mathbf{B}^H\mathbf{Q}_n(z)\mathbf{B}\mathbf{x}_d(n)}. \label{eq:QNN}
\end{align}
%We begin by noting that the top-$K$ empirical eigenvalues $\hat{\eta}_k$ of $\hat{\mathbf{R}}$ converge almost surely to distinct limits as $M$ and $N-L$ grow large. Consequently, for a positively oriented (i.e., counterclockwise) contour $\Gamma$ that encloses only the $K$ largest eigenvalues of $\hat{\mathbf{R}}$, we can obtain (\ref{eq:s_a}) and (\ref{eq:s_b}) by using Cauchy's integral formula. 
Substituting (\ref{eq:QNN}) into \eqref{eq:Cauchy}, we obtain (\ref{eq:s_b_bis}).
\begin{figure*}
\begin{subequations}
\begin{align}
    &\mathbb{E}[{\bf 1}_{\mathcal{O}_N}\mathbf{a}^H\hat{\mathbf{U}}_s\hat{\mathbf{U}}_s^H{\bf v}(n)s_k^{\ast}(n)] = -\frac{1}{2\imath\pi}\oint_{\Gamma}\mathbb{E}[{\bf 1}_{\mathcal{O}_N}\mathbf{a}^H\mathbf{Q}(z){\bf v}(n)s_k^\ast(n)] dz  \label{eq:s_a_bis}\\
    &= -\frac{1}{2\imath\pi}\oint_{\Gamma}\mathbb{E}[\mathbf{a}^H\mathbf{Q}_n(z){\bf v}(n)s_k^{\ast}(n){\bf 1}_{\mathcal{O}_N}] dz - \frac{1}{2\imath\pi}\oint_{\Gamma}\mathbb{E}\Big[ z\tilde{q}_{nn}(z)\frac{{\bf 1}_{\mathcal{O}_N}}{N-L}\mathbf{a}^H\mathbf{Q}_n(z)\mathbf{B}\mathbf{x}_d(n)\mathbf{x}_d^H(n)\mathbf{B}^H\mathbf{Q}_n(z){\bf v}(n)s_k^\ast(n) \Big]dz,\label{eq:s_b_bis}\\
    &\asymp -\frac{1}{2\imath \pi}\oint_{\Gamma}\mathbb{E}[\mathbf{a}^H\mathbf{Q}_n(z){\bf v}(n)s_k^{\ast}(n){\bf 1}_{\mathcal{O}_N}] dz + \frac{1}{2\imath\pi}\oint_{\Gamma}\mathbb{E}\Big[ \frac{\frac{{\bf 1}_{\mathcal{O}_N}}{N-L}\mathbf{a}^H\mathbf{Q}_n(z)\mathbf{B}\mathbf{x}_d(n)\mathbf{x}_d^H(n)\mathbf{B}^H\mathbf{Q}_n(z){\bf v}(n)s_k^\ast(n)}{1+\frac{\sigma_v^2}{N-L}{\rm tr}({\bf Q}_n(z))} \Big]dz\\
    &\asymp -\frac{1}{2\imath \pi}\oint_{\Gamma}\mathbb{E}[\mathbf{a}^H\mathbf{Q}_n(z){\bf v}(n)s_k^{\ast}(n){\bf 1}_{\mathcal{O}_N}] dz + \frac{1}{2\imath\pi}\oint_{\Gamma}\mathbb{E}\Big[ \frac{{\bf 1}_{\mathcal{O}_N}\mathbf{a}^H\mathbf{Q}_n(z)\mathbf{B}\mathbf{x}_d(n)s_k^\ast(n)\frac{\sigma_v^2}{N-L}{\rm tr}({\bf Q}_n(z))}{1+\frac{\sigma_v^2}{N-L}{\rm tr}({\bf Q}_n(z))} \Big]dz\label{eq:Bis}\\
    &\asymp \frac{1}{2\imath\pi} \oint_{\Gamma}\mathbb{E}\Big[\frac{P_s{\bf a}^{H}{\bf Q}_n(z){\bf h}_k\frac{\sigma_v^2}{N-L}{\rm tr}({\bf Q}_n(z))}{1+\frac{\sigma_v^2}{N-L}{\rm tr}({\bf Q}_n(z))}\Big] \label{eq:expectation_s}
    %&\asymp -\frac{1}{2\pi i}\oint_{\Gamma}P_s\mathbf{a}^H\mathbf{Q}_n(z)\mathbf{b} dz \label{eq:s_4}\\ %+ \frac{1}{2\pi i}\oint_{\Gamma}\frac{\frac{P_s}{N-L}\mathbf{a}^H\mathbf{Q}_n(z)(P_s\mathbf{H}\mathbf{FH}^H+\sigma_v^2\mathbf{I}_M)\mathbf{Q}_n(z)\mathbf{b}}{1+\frac{\sigma_v^2}{N-L}\mathrm{tr}(\mathbf{Q}_n(z))} dz = -\frac{1}{2\pi i}\oint_{\Gamma}P_s\mathbf{a}^H\mathbf{Q}_n(z)\mathbf{b} dz \label{eq:s_c}\\
    %&= \frac{1}{2\pi i}\oint_{\Gamma}P_s   \sum_{k=1}^K\frac{m_v(z)(1+zm_v(z))t_k}{(1+zm_v(z))t_k+1}\mathbf{a}^H\mathbf{u}_{sk}\mathbf{u}_{sk}^H\mathbf{b}dz \label{eq:s_d}\\
   % &= P_s\sum_{k=1}^K\lim_{z\rightarrow\bar{\eta}_k}\frac{(z-\bar{\eta}_k)m_v(z)(1+zm_v(z))t_k}{(1+zm_v(z))t_k+1}\mathbf{a}^H\mathbf{u}_{sk}\mathbf{u}_{sk}^H\mathbf{b}dz = P_s\sum_{k=1}^K\frac{1-c/t_k^2}{1+c/t_k}\mathbf{a}^H\mathbf{u}_{sk}\mathbf{u}_{sk}^H\mathbf{b}. \label{eq:s_e}
\end{align}
\end{subequations}
\hrulefill
\end{figure*}
To handle the second term in \eqref{eq:s_b_bis}, we note that
$$
\frac{1}{1+\frac{1}{N-L}{\bf x}_d^{H}(n){\bf B}^{H}{\bf Q}_n(z){\bf B}{\bf x}_d(n)}=-z\tilde{q}_{nn}(z),
$$
where $\tilde{q}_{nn}(z)$ is the $n$-th diagonal elements of $\tilde{\bf Q}(z)=({\bf Y}_d^{H}{\bf Y}_d-z{\bf I}_{N-L})^{-1}$. Hence:
$$
\sup_{z\in \Gamma}\Big|\frac{1}{1+\frac{1}{N-L}{\bf x}_d^{H}(n){\bf B}^{H}{\bf Q}_n(z){\bf B}{\bf x}_d(n)}\Big|\leq \sup_{z\in \Gamma}\frac{|z|}{\epsilon}.
$$
Using the fact that:
$$
\sup_{z\in {\Gamma}}\Big|\frac{1}{N-L}{\bf x}_d^{H}{\bf B}^{H}{\bf Q}_n(z){\bf B}{\bf x}_d-\frac{\sigma_v^2}{N-L}{\rm tr}({\bf Q}_n(z))\Big|\xrightarrow[]{a.s.}0,
$$
we thus obtain:
$$
\sup_{z\in \Gamma}|-z\tilde{q}_{nn}(z)-\frac{1}{1+\frac{\sigma_v^2}{N-L}{\rm tr}({\bf Q}_n(z))}|\xrightarrow[]{a.s.} 0
$$
Similarly, we also have:
$$
\sup_{z\in\Gamma}|\frac{1}{N-L}{\bf x}_d^{H}(n){\bf B}^{H}{\bf Q}_n(z){\bf v}(n)-\frac{\sigma_v^2}{N-L}{\rm tr}({\bf Q}_n(z)) |\xrightarrow[]{a.s.}0
$$
%Using the fact that $\mathbb{E}[|{\bf a}^{H}{\bf Q}_n(z){\bf B}{\bf x}_d(n)|^2]$, we can thus approximate $\mathbb{E}[{\bf 1}_{\mathcal{O}_N}{\bf a}^{H}\hat{\bf U}_s\hat{\bf U}_s^{H}{\bf v}(n)s_k^\star(n)]$ as in \eqref{eq:Bis}.% 
Taking the expectation with respect to ${\bf s}_n$ we thus obtain \eqref{eq:expectation_s}.
%We can thus find a deterministic equivalent for the second term in \eqref{eq:s_b_bis}.
%We can thus upper-bound the second term as %(\ref{eq:expectation_control}).
%\begin{figure*}
%\color{red}
%\begin{align}\label{eq:expectation_control_bis}
%&\left|\frac{1}{2\pi i}\oint_{\Gamma}\mathbb{E}\left[\frac{{\bf 1}_{\mathcal{O}_N}\{\mathbf{a}^H\mathbf{Q}_n(z)\mathbf{B}\frac{\mathbf{x}_d(n)\mathbf{x}_d^H(n)}{N-L}\mathbf{B}^H\mathbf{Q}_n(z){\bf v}(n)s_k^\star(n)\}}{1+\frac{1}{N-L}\{\mathbf{x}_d^H(n)\mathbf{B}^H\mathbf{Q}_n(z)\mathbf{B}\mathbf{x}_d(n)\}} \right]\right|dz \notag\\
%&\qquad\leq \sup_{z\in\Gamma}\frac{|z|}{\epsilon}\sqrt{\mathbb{E}\Big[{\bf a}^{H}{\bf Q}_n(z){\bf B}\frac{{\bf x}_d(n){\bf x}_d(n)^{H}}{N-L}{\bf B}^{H}{\bf Q}_n^{H}(z){\bf a}\Big]}\cdot\sqrt{\mathbb{E}\Big[{\bf v}(n)^{H}{\bf Q}_n(z){\bf B}\frac{{\bf x}_d(n){\bf x}_d(n)^{H}}{N-L}{\bf Q}_n^{H}(z){\bf v}(n)|s_k(n)|^2\Big]}.
%\end{align}
%\end{figure*}
%Taking the expectation of the upper-bound with respect to ${\bf x}_d(n)$, we can easily see that both expectations vanish uniformly on the contour $\Gamma$. This shows that the second term vanishes.
%We thus obtain:
%\begin{align}
%&\mathbb{E}\big[{\bf a}^{H}\hat{\bf U}_s\hat{\bf U}_s^{H}s_i(n)s_k^\ast(n){\bf b}\big]\asymp -\frac{1}{2\pi\imath}\oint_{\Gamma} P_s {\bf a}^{H}{\bf Q}_n(z){\bf b} dz. \notag
%\end{align}
Similarly, we employ the same approach to the quantity $\mathbb{E}[{\bf 1}_{\mathcal{O}_N}{\bf a}^{H}\hat{\bf U}_s\hat{\bf U}_s^{H}{\bf v}(n){\bf v}(n)^{H}{\bf b}]$ by using Cauchy's integral formula and using the matrix inversion Lemma to find \eqref{eq:v_b_bis} and \eqref{eq:v_c_bis}.
\begin{figure*}
\begin{subequations}
\begin{align}
    &\mathbb{E}[{\bf 1}_{\mathcal{O}_N}\mathbf{a}^H\hat{\mathbf{U}}_s\hat{\mathbf{U}}_s^H\mathbf{v}(n)\mathbf{v}(n)^H\mathbf{b}] = -\frac{1}{2\imath\pi}\oint_{\Gamma}\mathbb{E}[{\bf 1}_{\mathcal{O}_N}\mathbf{a}^H\mathbf{Q}(z)\mathbf{v}(n)\mathbf{v}(n)^H\mathbf{b}] dz  \label{eq:v_a}\\
    &= -\frac{1}{2\imath\pi}\oint_{\Gamma}\mathbb{E}[{\bf 1}_{\mathcal{O}_N}\mathbf{a}^H\mathbf{Q}_n(z)\mathbf{v}(n)\mathbf{v}(n)^H\mathbf{b}] dz + \frac{1}{2\imath\pi}\oint_{\Gamma}\mathbb{E}\Big[\frac{\frac{{\bf 1}_{\mathcal{O}_N}}{N-L}\mathbf{a}^H\mathbf{Q}_n(z)\mathbf{B}\mathbf{x}_d(n)\mathbf{x}_d^H(n)\mathbf{B}^H\mathbf{Q}_n(z)\mathbf{v}(n)\mathbf{v}(n)^H\mathbf{b}}{1+\frac{1}{N-L}\mathbf{x}_d^H(n)\mathbf{B}^H\mathbf{Q}_n(z)\mathbf{B}\mathbf{x}_d(n)}\Big] dz \label{eq:v_b_bis}\\
    &\asymp -\frac{\sigma_v^2}{2\imath\pi}\oint_{\Gamma}\mathbb{E}[\mathbf{a}^H\mathbf{Q}_n(z)\mathbf{b}{\bf 1}_{\mathcal{O}_N}]
    dz + \frac{1}{2\imath\pi}\oint_{\Gamma}\mathbb{E}\Big[\frac{\frac{\sigma_v^4{\bf 1}_{\mathcal{O}_N}}{N-L}\mathbf{a}^H\mathbf{Q}_n(z)\mathbf{b}\mathrm{tr}(\mathbf{Q}_n(z))}{1+\frac{\sigma_v^2}{N-L}\mathrm{tr}(\mathbf{Q}_n(z))}\Big] dz = -\frac{1}{2\imath\pi}\oint_{\Gamma}\mathbb{E}\Big[\frac{{\bf 1}_{\mathcal{O}_N}\sigma_v^2\mathbf{a}^H\mathbf{Q}_n(z)\mathbf{b}}{1+\frac{\sigma_v^2}{N-L}\mathrm{tr}(\mathbf{Q}_n(z))}\Big] dz.\label{eq:v_c_bis}
    %&\asymp \frac{1}{2\pi i}\oint_{\Gamma}\sigma_v^2   \sum_{k=1}^K\frac{m_v(z)(1+zm_v(z))t_k}{((1+zm_v(z))t_k+1)(1+\sigma_v^2cm_v(z))}\mathbf{a}^H\mathbf{u}_{sk}\mathbf{u}_{sk}^H\mathbf{b}dz \label{eq:v_d}\\    %&\asymp\sigma_v^2\sum_{k=1}^K\lim_{z\rightarrow\bar{\eta}_k}\frac{(z-\bar{\eta}_k)m_v(z)(1+zm_v(z))t_k}{((1+zm_v(z))t_k+1)(1+\sigma_v^2cm_v(z))}\mathbf{a}^H\mathbf{u}_{si}\mathbf{u}_{si}^H\mathbf{b}dz = \textcolor{blue}{\sigma_v^2\sum_{k=1}^K\Big(1-c/t_k^2\Big)\mathbf{a}^H\mathbf{u}_{sk}\mathbf{u}_{sk}^H\mathbf{b}}.  \label{eq:v_e} 
\end{align}    
\end{subequations}
\hrulefill
\end{figure*}
%However, here both terms in  \eqref{eq:v_b_bis} contribute and lead to the expression in \eqref{eq:v_c_bis}. %To handle the final term, we use the following inequality:
%\begin{align}
%&{\bf 1}_{\mathcal{O}_N}\Big|-z\tilde{q}_{nn}(z)-\frac{1}{1+\frac{1}{N-L}{\rm tr}({\bf Q}_n(z))}\Big|\nonumber \\
%&\leq \frac{|z|^2}{\epsilon^2}\Big|\frac{1}{N-L}{\bf x}_d^H{\bf B}^{H}{\bf Q}_n(z){\bf B}{\bf x}_d-\frac{\sigma_v^2}{N-L}{\rm tr}({\bf Q}_n(z))\Big|.\notag
%&\times \sup_{z\in \Gamma}|\frac{1}{1+\frac{1}{N-L}{\rm tr}({\bf Q}_N(z))}|
%\end{align}
%Using the fact that:
%$$
%\sup_{z\in {\Gamma}}\Big|\frac{1}{N-L}{\bf x}_d^{H}{\bf B}^{H}{\bf Q}_n(z){\bf B}{\bf x}_d-\frac{\sigma_v^2}{N-L}{\rm tr}({\bf Q}_n(z))\Big|\xrightarrow[]{a.s.}0,
%$$
%we thus obtain the approximation in \eqref{eq:v_c}. 

\noindent{\bf Step 2. Deriving asymptotic approximations for quadratic forms and normalized traces involving ${\bf Q}_n(z)$.}
Applying Woodbury matrix identity~\cite{higham2002accuracy} yields:
\begin{align}\label{eq:woodbury_Qn(z)}
    \mathbf{Q}_n(z) &= \left(\mathbf{W}_n\pmb{\Theta}_n\mathbf{W}_n^H+\frac{1}{N-L}\mathbf{V}_n\mathbf{V}_n^H-z\mathbf{I}_M\right)^{-1} \notag\\
    &= \mathbf{Q}_v(z) - \mathbf{Q}_v(z)\mathbf{W}_n\pmb{\Xi}(z)\mathbf{W}_n^H\mathbf{Q}_v(z),
\end{align}
where $\pmb{\Xi}(z)=\Big(\pmb{\Theta}_n^{-1}+\mathbf{W}_n^H\mathbf{Q}_v(z)\mathbf{W}_n\Big)^{-1}$, $\mathbf{Q}_v(z)=\Big(\frac{1}{N-L}\mathbf{V}_n\mathbf{V}_n^H-z\mathbf{I}_M\Big)^{-1}$, $\mathbf{W}_n = [\begin{array}{cc}
        \mathbf{H} & \frac{\mathbf{V}_n\mathbf{S}_n^H}{N-L} \\
    \end{array}]$, and 
\begin{align}
    \pmb{\Theta}_n = \left[\begin{array}{cc}
        \frac{{\bf S}_n{\bf S}_n^{H}}{N-L} & \mathbf{I}_K \\
        \mathbf{I}_k & \mathbf{0}_K
    \end{array}\right].
\end{align}
 We then use this identity to derive deterministic equivalents for  $\mathbf{a}^H\mathbf{Q}_n(z)\mathbf{b}$. Using \eqref{eq:woodbury_Qn(z)}, we obtain:
\begin{align}
&{\bf a}^{H}{\bf Q}_n(z){\bf b}={\bf a}^{H}{\bf Q}_v(z){\bf b}-{\bf a}^{H}{\bf Q}_v(z){\bf W}_n\pmb{\Xi}(z){\bf W}_n^{H}{\bf Q}_v(z){\bf b} \notag\\
&={\bf a}^{H}{\bf Q}_v(z){\bf b}-{\rm tr}(\pmb{\Xi}(z){\bf W}_n^{H}{\bf Q}_v(z){\bf b}{\bf a}^{H}{\bf Q}_v(z){\bf W}_n).\label{eq:77}
\end{align}
Since the dimension of $\pmb{\Xi}$ is $2K$, its asymptotic behavior can be characterized by analyzing the limiting behavior of its individual entries. Proceeding in the same way, let \(m_v(z)\) denote the limiting Stieltjes transform
of the empirical eigenvalue distribution of
$
\frac{1}{N-L}{\bf V}_n{\bf V}_n^{H}.
$
Then,
$
{\bf Q}_v(z)\asymp m_v(z){\bf I}_M .
$
Moreover, using the resolvent identity
$
\frac{1}{N-L}{\bf V}_n^{H}{\bf Q}_v(z){\bf V}_n
=
{\bf I}_{N-L}
+
z\widetilde{\bf Q}_v(z),
$
we obtain
$
{\bf I}_{N-L}
-
\frac{1}{N-L}{\bf V}_n^{H}{\bf Q}_v(z){\bf V}_n
=
-z\tilde{\bf Q}_v(z).
$
where $\tilde{\bf Q}_v(z):=(\frac{1}{N-L}{\bf V}_n^{H}{\bf V}_n-z{\bf I}_{N-L})^{-1}$.
Therefore,
\begin{align}
&\frac{1}{N-L}{\bf S}_n
\Big(
{\bf I}_{N-L}
-
\frac{1}{N-L}{\bf V}_n^{H}{\bf Q}_v(z){\bf V}_n
\Big)
{\bf S}_n^{H}
\notag\\
&\qquad\asymp
-P_sz\tilde{m}_v(z){\bf I}_K,
\end{align}
where \(\tilde m_v(z)\) is the limiting Stieltjes transform associated with
\(\tilde{\bf Q}_v(z)\).
Since the nonzero eigenvalues of
\(\frac{1}{N-L}{\bf V}_n{\bf V}_n^{H}\) and
\(\frac{1}{N-L}{\bf V}_n^{H}{\bf V}_n\) coincide, their limiting Stieltjes transforms satisfy
\[
\tilde{m}_v(z)=c\,m_v(z)-\frac{1-c}{z}.
\]
Using the MP equation satisfied by \(m_v(z)\),
{
\begin{align}
    z c\sigma_v^2 m_v^2(z) - (\sigma_v^2(1-c)-z)m(z) + 1 = 0.
\end{align}}
we obtain
\[
z\tilde m_v(z)
=
c z m_v(z)-(1-c)
=
-\frac{z m_v(z)+1}{\sigma_v^2}m_v(z).
\]
Consequently,
\begin{align}
&\frac{1}{N-L}{\bf S}_n
\Big(
{\bf I}_{N-L}
-
\frac{1}{N-L}{\bf V}_n^{H}{\bf Q}_v(z){\bf V}_n
\Big)
{\bf S}_n^{H}
\notag\\
&\qquad\asymp
\frac{z m_v(z)+1}{\sigma_v^2m_v(z)}P_s{\bf I}_K .
\end{align}
Plugging these approximations into $\pmb{\Xi}(z)$, we get:
\begin{align}
    \pmb{\Xi}(z) &\asymp \left[\begin{array}{cc}
        m_v(z)\mathbf{H}^H\mathbf{H} & \mathbf{I}_K \\
        \mathbf{I}_K & -\frac{zm_v(z)+1}{\sigma_v^2m_v(z)}P_s\mathbf{I}_K  
    \end{array}\right]^{-1} \notag\\
    &= \left[\begin{array}{cc}
        \mathbf{D}_1(z) & \mathbf{D}_2(z) \\
        \mathbf{D}_2(z) & \mathbf{D}_3(z) 
    \end{array}\right],
\end{align}    
where 
\begin{align}
    \mathbf{D}_1(z) =& \left(m_v(z)\mathbf{H}^H\mathbf{H}+\frac{\sigma_v^2m_v(z)}{(1+zm_v(z))P_s}\mathbf{I}_K\right)^{-1}, \notag\\
    \mathbf{D}_2(z) =& \frac{\sigma_v^2m_v(z)}{(1+zm_v(z))P_s}\mathbf{D}_1, \mathbf{D}_3(z) = -m_v(z)\mathbf{H}^H\mathbf{H}\mathbf{D}_2(z). \notag
\end{align}
Similarly, using the fact that 
\begin{align}
&{\bf W}_n^{H}{\bf Q}_v(z){\bf b} \asymp m_v(z)[{\bf b}^{H}{\bf H} \ \ \  {\bf 0}_K]^{H}, \label{eq:app_1}\\  
&{\bf a}^{H}{\bf Q}_v(z){\bf W}_n\asymp m_v(z)[{\bf a}^{H }{\bf H} \ \ {\bf 0}_K], \label{eq:app_2}
\end{align}
we can approximate $\mathbf{W}_n^H\mathbf{Q}_v(z)\mathbf{b}\mathbf{a}^H\mathbf{Q}_v(z)\mathbf{W}_n $ as:
\begin{align}
\mathbf{W}_n^H\mathbf{Q}_v(z)\mathbf{b}\mathbf{a}^H\mathbf{Q}_v(z)\mathbf{W}_n \asymp \left[\begin{array}{cc}
m_v^2(z)\mathbf{H}^H\mathbf{b}\mathbf{a}^H\mathbf{H} & \mathbf{0}_K \\
        \mathbf{0}_K & \mathbf{0}_K
    \end{array}\right]. \notag
\end{align}
Pluggin the above   approximations along with $\mathbf{Q}_v(z)\asymp m_v(z)\mathbf{I}_M$~\cite{wang2026large} into \eqref{eq:77}, we obtain:
\begin{align}
&{\bf a}^{H}{\bf Q}_n(z){\bf b}\asymp m_v(z){\bf a}^{H}{\bf b}-m_v^2(z){\bf a}^{H}{\bf H}{\bf D}_1(z){\bf H}^{H}{\bf b} \notag\\
&\quad=m_v(z){\bf a}^{H}\Big({\bf I}_M+\frac{(1+zm_v(z))P_s}{\sigma_v^2}{\bf H}{\bf H}^{H}\Big)^{-1}{\bf b}, \label{eq:88}
\end{align}
where the last equality follows from Woodbury matrix identity. Using the fact that $P_s{\bf H}{\bf H}^{H}=\sigma_v^2\sum_{k=1}^{K}t_k{\bf u}_{sk}{\bf u}_{sk}^{H}$, we thus obtain:
\begin{align}
&{\bf a}^{H}{\bf Q}_n(z){\bf b}\asymp m_v(z){\bf a}^{H}{\bf b}\nonumber\\
&\qquad\qquad+\sum_{k=1}^{K}\frac{-m_v(z)(1+zm_v(z))t_k}{(1+zm_v(z))t_k+1}{\bf a}^{H}{\bf u}_{sk}{\bf u}_{sk}^{H}{\bf b}. \label{eq:quadratic_form_approx_bis}
\end{align}
Finally, using finite-rank perturbation lemma, we obtain for $\epsilon$ sufficiently small:
$$
\Big|\frac{1}{N-L}{\rm tr}({\bf Q}_n(z))-\frac{1}{N-L}\mathrm{tr}({\bf Q}_v(z))\Big| \leq \frac{2K}{\epsilon(N-L)}.
$$
Hence,
\begin{equation}
\frac{1}{N-L}{\rm tr}({\bf Q}_n(z))\asymp cm_v(z). \label{eq:normalized_trace_bis}
\end{equation}
\noindent{\bf Step 3. Simplifying derivations using residue calculations.} Using \eqref{eq:quadratic_form_approx_bis} and \eqref{eq:normalized_trace_bis} into \eqref{eq:s_b_bis}, we obtain: %and \eqref{eq:v_c_bis}, we obtain
\begin{align}
&\mathbb{E}[{\bf a}^{H}\hat{\bf U}_s\hat{\bf U}_s^{H}{\bf v}(n)s_k^\ast(n)]\nonumber \\
&\asymp\frac{1}{2\imath \pi} \oint_{\Gamma} \sum_{j=1}^K \frac{P_s\sigma_v^2c(m_v(z))^2{\bf a}^{H}{\bf u}_{sj}{\bf u}_{sj}^{H}{\bf h}_k}{((1+zm_v(z))t_j+1)(1+\sigma_v^2cm_v(z))}dz. \notag
%&=-\frac{\delta_{ik}}{2\imath\pi}P_s\oint_{\Gamma}m_v(z){\bf a}^{H}{\bf b}dz\nonumber\\
%&\quad+\frac{\delta_{ik}}{2\imath\pi}P_s\oint_{\Gamma} \sum_{k=1}^{K}\frac{m_v(z)(1+zm_v(z))t_k}{(1+zm_v(z))t_k+\sigma_v^2}{\bf a}^{H}{\bf u}_{sk}{\bf u}_{sk}^{H}{\bf b}dz\notag\\
%&=\frac{\delta_{ik}}{2\imath\pi}P_s\oint_{\Gamma} \sum_{k=1}^{K}\frac{m_v(z)(1+zm_v(z))t_k}{(1+zm_v(z))t_k+1}{\bf a}^{H}{\bf u}_{sk}{\bf u}_{sk}^{H}{\bf b}dz\notag\\
%&=P_s\delta_{ik}\sum_{k=1}^{K}\text{Residue}\Big(\frac{m_v(z)(1+zm_v(z))t_k}{(1+zm_v(z))t_k+1},z=\overline{\eta}_k\Big)\notag\\
%&\quad\cdot{\bf a}^{H}{\bf u}_{sk}{\bf u}_{sk}^{H}{\bf b}.
%&=P_s\sum_{k=1}^{K}\frac{1-c/t_k^2}{1+c/t_k}{\bf a}^{H}{\bf u}_{sk}{\bf u}_{sk}^{H}{\bf b}.
\end{align}
To evaluate the residues, we use the pole condition
\[
1+t_k\phi(\overline{\eta}_k)=0,
\qquad
\phi(z):=1+zm_v(z),
\]
which gives $\phi(\overline{\eta}_k)
=
1+\overline{\eta}_k m_v(\overline{\eta}_k)
=
-\frac{1}{t_k}$. Moreover, from the MP equation,
$
z=\sigma_v^2\Big(1-\frac{1}{\phi(z)}\Big)\left(1-c\phi(z)\right),
$
and therefore
$
\phi'(z)
=
\frac{1}{\sigma_v^2\left(\phi(z)^{-2}-c\right)}.
$
Since 
$    m_v(z) = \frac{\phi(z)}{\sigma_v^2(1-c\phi(z))},
$
we can obtain 
\begin{align}
&\text{Residue}\Big(\frac{(m_v(z))^2}{((1+zm_v(z))t_k+1)(1+\sigma_v^2cm_v(z))},z=\overline{\eta}_k\Big) \notag\\
&=\text{Residue}\Big(\frac{(m_v(z))^2}{(\phi(z)t_k+1)(1+\sigma_v^2cm_v(z))},z=\overline{\eta}_k\Big)\notag\\
&=\frac{(m_v(\overline{\eta}_k))^2}{t_k\phi'(\overline{\eta}_k)(1+\sigma_v^2cm_v(\overline{\eta}_k))}= \frac{1-c/t_k^2}{\sigma_v^2(t_k+c)}.      
\end{align}
Hence,
$$
\mathbb{E}\Big[{\bf a}^{H}\hat{\bf U}_s\hat{\bf U}_s^{H}{\bf v}(n)s_k^\ast(n)\Big]\asymp\sum_{j=1}^{K}P_sc\frac{(1-c/t_j^2)}{t_j+c}{\bf a}^{H}{\bf u}_{sj}{\bf u}_{sj}^{H}{\bf h}_k.
$$
%\begin{align}
 %   &\text{Residue}\Big(\frac{m_v(z)(1+zm_v(z))t_k}{(1+zm_v(z))t_k+1},z=\overline{\eta}_k\Big) \notag\\
  %  &= \text{Residue}\Big(\frac{m_v(z)\phi(z)t_k}{\phi(z)t_k+1},z=\overline{\eta}_k\Big) \notag\\
   % &= \frac{m_v(\bar{\eta}_k)\phi(\bar{\eta}_k)}{\phi^{\prime}(\bar{\eta}_k)} = \frac{1/(\sigma_v^2t_k(t_k+c))}{1/(\sigma_v^2(t_k^2-c))} = \frac{1-c/t_k^2}{1+c/t_k}. 
%\end{align}

%These identities are then substituted into the residue expression to find:
%\begin{align}
%&\mathbb{E}[{\bf a}^{H}\hat{\bf U}_s\hat{\bf U}_s^{H}s_i(n)s_k^{\ast}(n){\bf b}]\nonumber \\
%&\qquad\asymp P_s\delta_{ik}\sum_{k=1}^{K}\frac{1-c/t_k^2}{1+c/t_k}{\bf a}^{H}{\bf u}_{sk}{\bf u}_{sk}^{H}{\bf b}.
%\end{align}
The term $\mathbb{E}[{\bf a}^H\hat{\bf U}_s\hat{\bf U}_s^{H}{\bf v}(n){\bf v}(n)^{H}{\bf b}]$ is treated similarly. Continuing from \eqref{eq:v_c_bis} and using \eqref{eq:quadratic_form_approx_bis}, we obtain:
\begin{align}
&\mathbb{E}[{\bf a}^{H}\hat{\bf U}_s\hat{\bf U}_s^{H}{\bf v}(n){\bf v}(n)^{H}{\bf b}]\nonumber\\
&\asymp \frac{1}{2\pi\imath}\oint_{\Gamma}\sigma_v^2\sum_{k=1}^{K}\frac{m_v(z)(1+zm_v(z))t_k{\bf a}^{H}{\bf u}_{sk}{\bf u}_{sk}^{H}{\bf b}}{((1+zm_v(z))t_k+1)(1+\sigma_v^2cm_v(z))}dz\nonumber\\
&=\sigma_v^2\sum_{k=1}^{K}(1-c/t_k^2){\bf a}^{H}{\bf u}_{sk}{\bf u}_{sk}^{H}{\bf b}.
\end{align}
%Similarly, \eqref{aUUvs_bis}--\eqref{aHsvUUb_bis} can be derived following the same procedure.
\end{proof}
\begin{corollary}\label{corollary1}
Under the setting of Lemma \ref{first-order-signal-noise_bis}, the following approximations holds true:
\begin{align}
&\mathbb{E}[{\bf a}^{H}{\bf H}{\bf s}(n){\bf v}^{H}(n)\hat{\bf U}_s\hat{\bf U}_s^{H}{\bf b}] \notag\\
&\qquad\asymp c\sigma_v^2 \sum_{j=1}^{K}\frac{1-c/t_j^2}{1+c/t_j}{\bf a}^{H}{\bf u}_{sj}\hat{\bf u}_{sj}^{H}{\bf b},\label{eq:first_1}\\
&\mathbb{E}[{\bf a}^{H}\hat{\bf U}_s\hat{\bf U}_s^{H}{\bf H}{\bf s}(n){\bf v}(n)^{H}\hat{\bf U}_s\hat{\bf U}_s^{H}{\bf b}] \notag\\
&\qquad\asymp c\sigma_v^2 {\bf a}^{H}{\bf U}_s\boldsymbol{\Lambda}_s{\bf U}_s^{H}{\bf b}.\label{eq:first_2}
\end{align}
\label{eq:by_product}
\label{cor:by_product}
\end{corollary}
\begin{proof}
The approximation in follows by decomposing ${\bf H}{\bf s}(n)=\sum_{k=1}^{K}{\bf h}_k s_k(n)$ and then computing the expectation of $\mathbb{E}[s_k(n){\bf v}^{H}(n)\hat{\bf U}_s\hat{\bf U}_s^{H}{\bf b}]$ using Lemma \ref{first-order-signal-noise_bis}.
\end{proof}

\begin{lemma}\label{second-order-signal-noise_bis}
For all deterministic sequences $\mathbf{a}$ and $\mathbf{b}$  with bounded norms, i.e., $\limsup\|\mathbf{a}\|<\infty$, $\limsup\|\mathbf{b}\|<\infty$, the vectors $\mathbf{s}(n)$ and $\mathbf{v}(n)$ are defined in (\ref{received signal}), and $\hat{\mathbf U}_s$ denotes the matrix collecting the $K$ sample
eigenvectors of the sample covariance matrix in
\eqref{eq:empirical_covariance_matrix}, the following deterministic approximations hold:\begin{align}
  %\label{eq:first_equation}  &\mathbb{E}[\mathbf{a}^H\hat{\mathbf{U}}_s\hat{\mathbf{U}}_s^H\mathbf{h}_is_i(n)s_k^{\ast}(n)\mathbf{h}_k^H\hat{\mathbf{U}}_s\hat{\mathbf{U}}_s^H\mathbf{a}] \notag\\
  %&\qquad\asymp P_s\delta_{ik}|\mathbf{a}^H\mathbf{U}_s\pmb{\Lambda}_s\mathbf{U}_s^H\mathbf{h}_k|^2, \\
&\mathbb{E}[\mathbf
a^H\hat{\mathbf{U}}_s\hat{\mathbf{U}}_s^H\mathbf{v}(n)\mathbf{v}(n)^H\hat{\mathbf{U}}_s\hat{\mathbf{U}}_s^H\mathbf
{b}]\notag\\
&\asymp \sum_{k=1}^{K}\frac{(t_k^2-c)(t_k(t_k+c)^2+c^2(t_k^2-c))}{t_k^3(t_k+c)^2} {\bf a}^{H}{\bf u}_{sk}{\bf u}_{sk}^{H}{\bf b}\label{eq:desired}.
%&\qquad\asymp \sigma_v^2\mathbf{a}^H\mathbf{U}_s\bar{\pmb{\Lambda}}_n\mathbf{U}_s^H\mathbf{b},  \label{eq:second_equation}\\
%&\textcolor{red}{\mathbb{E}[\mathbf{a}^H\hat{\mathbf{U}}_s\hat{\mathbf{U}}_s^H\mathbf{Hs}(n)\mathbf{v}(n)^H\hat{\mathbf{U}}_s\hat{\mathbf{U}}_s^H\mathbf{b}]} \notag\\
%&\qquad\textcolor{red}{\asymp c\sigma_v^2
%\mathbf{a}^{H}\mathbf{U}_{s}\pmb{\Lambda}_s^2\mathbf{U}_{s}^{H}\mathbf{b}}. \label{eq:third_equation}
\end{align}
%where $\bar{\pmb{\Lambda}}_n=\mathrm{diag}([\bar{z}_{n1},\ldots,\bar{z}_{nK}])$ with $\bar{z}_{nk}= \frac{(t_k^2-c)(t_k(t_k+c)^2+c^2(t_k^2-c))}{t_k^3(t_k+c)^2}$.
\end{lemma}
%\begin{remark}
%The same remark as in Lemma 5 applies here. In particular, the first term can be directly obtained from \eqref{eq:projection} if the dependence of $s(n)$ on 
%$\hat{\bf U}_s$
 %is neglected. This is, however, no longer true for ${\bf v(n)}$, which seems to introduce a stronger coupling, thereby potentially altering the asymptotic behavior.
%\end{remark}
\begin{proof}
The proof follows along the same lines as that of Lemma \ref{first-order-signal-noise_bis}, starting from a writing of the desired contours involving the resolvent matrix ${\bf Q}_n(z)$ and deriving asymptotic approximations of quadratic forms and normalizing traces involving ${\bf Q}_n(z)$ to find deterministic equivalents of the contour representation. For brevity, we outline the key steps associated with each term.

%\noindent{\bf Proof of \eqref{eq:first_equation}.} 
Let $\Gamma_1$ and $\Gamma_2$ two positively oriented contours enclosing the limiting spikes $\overline{\eta}_1,\cdots,\overline{\eta}_K$. 
Using Cauchy's integral formula, the following equality holds asymptotically true:
%\begin{align}
%&\mathbb{E}[{\bf a}^{H}\hat{\bf U}_s\hat{\bf U}_s^{H}{\bf h}_is_i(n)s_k^\ast(n){\bf h}^{H}\hat{\bf U}_s\hat{\bf U}_s^H{\bf a}]=\frac{1}{(2\pi\imath)^2}\nonumber\\
%&\cdot\oint_{\Gamma_1}\oint_{\Gamma_2}\mathbb{E}\big[{\bf a}^{H}\!{\bf Q}(z_1){\bf h}_is_i(n)s_k^\ast(n){\bf h}_k^{H}\!{\bf Q}(z_2){\bf a}{\bf 1}_{\mathcal{O}_{N}}\big]dz_1dz_2. \notag
%&=\frac{1}{(2\pi\imath)^2}\oint_{\Gamma_1}\oint_{\Gamma_2} \mathbb{E}\big[{\bf a}^{H}{\bf Q}(z_1){\bf b}{\bf v}(n){\bf v}^{H}(n){\bf b}^{H}{\bf Q}(z_2){\bf a}{\bf 1}_{\mathcal{O}_{N}}\big]dz_1dz_2
%\end{align}
%Similarly, using the matrix inversion Lemma as in \eqref{eq:QNN} and controlling the terms in the cross-product, we ultimately obtain:
%\begin{align}
%&\mathbb{E}[{\bf a}^{H}\hat{\bf U}_s\hat{\bf U}_s^{H}{\bf h}_is_i(n)s_k^\ast(n){\bf h}_k^{H}\hat{\bf U}_s\hat{\bf U}_s^{H}{\bf a}] \notag\\
%&\asymp -\frac{P_s\delta_{ik}}{4\pi}\oint_{\Gamma_1}\oint_{\Gamma_2} {\bf a}^{H}{\bf Q}_n(z_1){\bf h}_k{\bf h}_k^{H}{\bf Q}_n(z_2){\bf a}dz_1dz_2.
%\end{align}
%After residue calculations, we thus obtain:
%\begin{align}
%&\mathbb{E}[{\bf a}^{H}\hat{\bf U}_s\hat{\bf U}_s^{H}{\bf h}_is_i(n)s_k^\ast(n){\bf h}_k^{H}\hat{\bf U}_s\hat{\bf U}_s^{H}{\bf a}] \notag\\
%&\qquad\qquad\asymp P_s\delta_{ik}|{\bf a}^{H}{\bf U}_s\boldsymbol{\Lambda}_s{\bf U}_s^{H}{\bf h}_k|^2.
%\end{align}
%\noindent{\bf Proof of \eqref{eq:second_equation}}. Considering the same contours as in the proof of \eqref{eq:first_equation}, we use Cauchy's integral formula to obtain:
\begin{align}
&\mathbb{E}[{\bf a}^{H}\hat{\bf U}_s\hat{\bf U}_s^{H}{\bf v}(n){\bf v}^{H}(n)\hat{\bf U}_s\hat{\bf U}_s^{H}{\bf b}] \notag\\
&=\frac{1}{(2\pi\imath)^2}\oint_{\Gamma_1}\oint_{\Gamma_2}\mathbb{E}[{\bf a}^{H}{\bf Q}(z_1){\bf v}(n){\bf v}^{H}(n){\bf Q}(z_2){\bf b}]dz_1dz_2. \notag
\end{align}
To remove the dependence of ${\bf v}(n)$ on ${\bf Q}(z_1)$ and ${\bf Q}(z_2)$, we use the matrix inversion Lemma and control the vanishing terms  to obtain after simplifications:
\begin{align}
\mathbb{E}[{\bf a}^{H}\hat{\bf U}_s\hat{\bf U}_s^{H}{\bf v}(n){\bf v}^{H}(n)\hat{\bf U}_s\hat{\bf U}_s^{H}{\bf b}]\asymp {I}_1+I_2,
\end{align}
where $I_1$ and $I_2$ are given by:
\begin{align}
 &I_1=-\frac{1}{4\pi}\oint_{\Gamma_1}\oint_{\Gamma_2}\frac{\sigma_v^2\mathbb{E}[{\bf a}^{H}{\bf Q}_n(z_1){\bf Q}_n(z_2){\bf b}]}{(1+c\sigma_v^2m_v(z_1))(1+c\sigma_v^2m_v(z_2))}dz_1dz_2,\nonumber\\
 &I_2=-\frac{1}{4\pi}\oint_{\Gamma_1}\oint_{\Gamma_2}\frac{\sigma_v^4P_sc^2m_v(z_1)m_v(z_2)}{(1+c\sigma_v^2m_v(z_1))(1+c\sigma_v^2m_v(z_2))}\nonumber\\
 &\qquad\cdot \mathbb{E}[{\bf a}^{H}{\bf Q}_n(z_1){\bf H}{\bf H}^{H}{\bf Q}_n(z_2){\bf b}]dz_1 dz_2.\notag
\end{align}
Since $K$ is fixed, we may use \eqref{eq:88} to obtain the following approximation:
\begin{align}
&P_s\mathbb{E}[{\bf a}^{H}{\bf Q}_n(z_1){\bf H}{\bf H}^{H}{\bf Q}_n(z_2){\bf b}]\nonumber\\
&\asymp\sum_{k=1}^{K}\frac{\sigma_v^2m_v(z_1)m_v(z_2)t_k {\bf a}^{H}{\bf u}_{sk}{\bf u}_{sk}^{H}{\bf b}}{(1+(1+z_1m_v(z_1))t_k)(1+(1+z_2m_v(z_2))t_k)}.\notag
\end{align}
Using this approximation, $I_2$ simplifies as:
$$
I_2\asymp\sigma_v^2\sum_{k=1}^{K}\frac{c^2(t_k^2-c)^2}{t_k^3(t_k+c)^2}{\bf a}^{H}{\bf u}_{sk}{\bf u}_{sk}^{H}{\bf b}.
$$
Computing the term $I_1$ requires to find deterministic equivalents for $\mathbb{E}[{\bf a}^{H}{\bf Q}_n(z_1){\bf Q}_n(z_2){\bf b}]$. Using \eqref{eq:woodbury_Qn(z)}, we obtain:
\begin{align}
&{\bf a}^{H}{\bf Q}_n(z_1){\bf Q}_n(z_2){\bf b}={\bf a}^{H}{\bf Q}_v(z_1){\bf Q}_v(z_2){\bf b}\nonumber\\
&\qquad-{\bf a}^{H}{\bf Q}_v(z_1){\bf W}_n\boldsymbol{\Xi}(z_1){\bf W}_n^{H}{\bf Q}_v(z_1){\bf Q}_v(z_2){\bf b}\nonumber\\
&\qquad-{\bf a}^{H}{\bf Q}_v(z_1){\bf Q}_v(z_2){\bf W}_n\pmb{\Xi}(z_2){\bf W}_n^{H}{\bf Q}_v(z_2){\bf b}\nonumber\\
&\qquad+{\bf a}^{H}{\bf Q}_v(z_1){\bf W}_n\pmb{\Xi}(z_1){\bf W}_n^{H}{\bf Q}_v(z_1)\notag\\
&\qquad\quad\cdot{\bf Q}_v(z_2){\bf W}_n\pmb{\Xi}(z_2){\bf W}_n^{H}{\bf Q}_v(z_2){\bf b}
\notag\\
&\quad\asymp {\bf a}^{H}{\bf Q}_v(z_1){\bf Q}_v(z_2){\bf b}\notag\\
&\qquad-m(z_1){\bf a}^{H}{\bf H}{\bf D}_1(z_1){\bf H}^{H}{\bf Q}_v(z_1){\bf Q}_v(z_2){\bf b}\nonumber\\
&\qquad-m(z_2){\bf a}^{H}{\bf Q}_v(z_1){\bf Q}_v(z_2){\bf H}{\bf D}_1(z_2){\bf H}^{H}{\bf b}\nonumber\\
&\qquad+m(z_1)m(z_2){\bf a}^{H}{\bf H}{\bf D}_1(z_1){\bf H}^{H}{\bf Q}_v(z_1)\notag\\
&\qquad\quad\cdot{\bf Q}_v(z_2){\bf H}{\bf D}_1(z_2){\bf H}^{H}{\bf b}. \label{eq:quad_double}
\end{align}
Using Lemma 6 in \cite{wang2026large}, we know that:
\begin{equation}
\label{eq:approx_double}
{\bf Q}_v(z_1){\bf Q}_v(z_2)\asymp \theta(z_1,z_2){\bf I}_M, % m_v(z_1)m_v(z_2){\bf I}_M+c\eta(z_1,z_2){\bf I}_M \label{eq:approx_double}
\end{equation}
where $\theta(z_1,z_2)=m_v(z_1)m_v(z_2)+c\eta(z_1,z_2)$ with
\begin{align}
&\eta(z_1,z_2)\notag\\
&=\frac{\sigma_v^4m_v^2(z_1)m_v^2(z_2)}{(1+c\sigma_v^2m_v(z_1))(1+c\sigma_v^2m_v(z_2))-c\sigma_v^4m_v(z_1)m_v(z_2)}. \notag
\end{align}
Plugging \eqref{eq:approx_double}  into \eqref{eq:quad_double}, we obtain:
\begin{align}
&{\bf a}^{H}{\bf Q}_n(z_1){\bf Q}_n(z_2){\bf b}\nonumber\\
&\asymp \theta(z_1,z_2){\bf a}^{H}({\bf I}_M-{\bf H}{\bf D}_1(z_1){\bf H}^{H})({\bf I}_M-{\bf H}{\bf D}_1(z_2){\bf H}^{H}){\bf b} \notag\\
&=\theta(z_1,z_2) {\bf a}^{H}({\bf I}_M-\sum_{k=1}^{K}\frac{(1+z_1m_v(z_1))t_k}{(1+z_1m_v(z_1))t_k+1}){\bf u}_{sk}{\bf u}_{sk}^{H}\nonumber\\
&\quad\cdot({\bf I}_M-\sum_{\ell=1}^{K}\frac{(1+z_2m_v(z_2))t_\ell}{(1+z_2m_v(z_2))t_\ell+1}){\bf u}_{s\ell}{\bf u}_{s\ell}^{H}){\bf b}.
\end{align}
When expanding this product in the integral terms in $I_1$, the only contributing term is given by
\begin{equation}
\begin{aligned}
&B(z_1,z_2) \notag\\
&=\sum_{k=1}^{K}\frac{\theta(z_1,z_2)(1+z_1m_v(z_1))(1+z_2m_v(z_2))t_k^2}{((1+z_1m_v(z_1))t_k+1)((1+z_2m_v(z_2))t_k+1)}\nonumber\\
&\quad\cdot {\bf a}^H{\bf u}_{sk}{\bf u}_{sk}^{H}{\bf b}.
\end{aligned}
\end{equation}
Hence:
\begin{align}
&I_1\asymp-\frac{1}{4\pi}\oint_{\Gamma_1}\oint_{\Gamma_2} \frac{\sigma_v^2B(z_1,z_2)}{(1+c\sigma_v^2m_v(z_1))(1+c\sigma_v^2m_v(z_2))} dz_1 dz_2. \notag
\end{align}
From residue calculations, we finally obtain:
$$
I_1\asymp \sigma_v^2\sum_{k=1}^{K}\frac{(t_k^2-c)}{t_k^2} {\bf a}^{H}{\bf u}_{sk}{\bf u}_{sk}^{H}{\bf b}.
$$
Taking the sum of $I_1$ and $I_2$, we thus obtain \eqref{eq:desired}.
%$$
%\mathbb{E}[{\bf a}^{H}\hat{\bf U}_s\hat{\bf U}_s^{H}{\bf v}(n){\bf v}(n)^{H}\hat{\bf U}_s\hat{\bf U}_s^{H}{\bf b}]\asymp {\bf a}^{H}{\bf U}_s\bar{\boldsymbol{\Lambda}}_n{\bf U}_s^{H}{\bf b},
%$$
%where $\bar{\boldsymbol{\Lambda}}_n$ is defined in the Lemma. 

%Similarly, \eqref{eq:third_equation} can be derived following the same procedure.
\end{proof}
\section{Proof of Theorem \ref{theorem_sb_MRC}: Deterministic Equivalents of the SINR for Semi-Blind and MRC Receivers }\label{app_sb_MRC_res}
Based on Lemmas~\ref{first-order-signal-noise_bis} and~\ref{second-order-signal-noise_bis}, Corollary~\ref{corollary1}, and the derivations in Appendix~\ref{app_sb_MRC_res}, we obtain the deterministic equivalent of the SINR achieved by the MMSE detector with semi-blind channel estimation.

1) Deterministic equivalent for signal power: The desired signal power term
$|\mathbb E[(\mathbf g_{\mathrm{sb},k}^{\mathrm{MRC}})^H\mathbf y(n)s_k^\ast(n)]|^2$
admits the following deterministic equivalent:
\begin{align}
&\left|\mathbb{E}\!\left[(\hat{\mathbf{g}}_{\mathrm{sb},k}^{\mathrm{MRC}})^H\mathbf{y}(n)s_k^{\ast}(n)\right]\right|^2 \notag\\
%&=\Big|\mathbb{E}[(\hat{\mathbf{h}}_k^{\mathrm{sb}})^H\mathbf{H}\mathbf{s}(n)s_k^{\ast}(n)]+\mathbb{E}[(\hat{\mathbf{h}}_k^{\mathrm{sb}})^H\mathbf{v}(n)s_k^{\ast}(n)]\Big|^2 \notag\\
&=\Big|\sum_{i=1}^{K}\mathbb{E}[(\hat{\mathbf{h}}_k^{\mathrm{sb}})^H\mathbf{h}_i s_i(n)s_k^{\ast}(n)]+\mathbb{E}[(\hat{\mathbf{h}}_k^{\mathrm{sb}})^H\mathbf{v}(n)s_k^{\ast}(n)]\Big|^2 \notag\\
&\overset{(a)}{\asymp}\Big|\bar{\lambda}P_s\mathbf{h}_k^H\mathbf{h}_k+(1-\bar{\lambda})\mathbb{E}[\mathbf{h}_k^H\hat{\mathbf{U}}_s\hat{\mathbf{U}}_s^H s_k(n)s_k^{\ast}(n)\mathbf{h}_k] \notag\\
&\quad+(1-\bar{\lambda})\mathbb{E}[\mathbf{h}_k^H\hat{\mathbf{U}}_s\hat{\mathbf{U}}_s^H\mathbf{v}(n)s_k^{\ast}(n)]\Big|^2 \notag\\
&\stackrel{(b)}{\asymp}|P_s(\bar{\lambda}\mathbf{h}_k^H\mathbf{h}_k+(1-\bar{\lambda})\mathbf{h}_k^H\mathbf{U}_s\mathbf{\Lambda}_s\mathbf{U}_s^H\mathbf{h}_k \notag\\
&\quad+(1-\bar{\lambda})\mathbf{h}_k^H\mathbf{U}_s\mathbf{\Lambda}_{vs}\mathbf{U}_s^H\mathbf{h}_k)|^2 = P_s^2|\mathbf{h}_k^H\mathbf{F}\mathbf{h}_k|^2,
\end{align}
where $(a)$ follows from the fact that $\hat{\lambda}-\overline{\lambda}\xrightarrow[]{a.s.} 0.$ while (b) from applying Lemma \ref{first-order-signal-noise_bis} and \eqref{eq:44}.  

2) Deterministic equivalent for the interference-plus-noise power:
Following a similar procedure to that used for the desired signal power, the deterministic equivalent of the interference-plus-noise power is obtained as
\begin{align}
&\mathbb{E}[|(\hat{\mathbf{g}}_{\mathrm{sb},k}^{\mathrm{MRC}})^H\mathbf{y}(n)|^2] =\mathbb{E}[(\hat{\mathbf{h}}_k^{\mathrm{sb}})^H\mathbf{H}\mathbf{s}(n)\mathbf{s}^H(n)\mathbf{H}^H\hat{\mathbf{h}}_k^{\mathrm{sb}}] \notag\\
&\qquad\qquad+\mathbb{E}[(\hat{\mathbf{h}}_k^{\mathrm{sb}})^H\mathbf{v}(n)\mathbf{v}^H(n)\hat{\mathbf{h}}_k^{\mathrm{sb}}] \notag\\
&\qquad\qquad+2\mathrm{Re}\left\{\mathbb{E}[(\hat{\mathbf{h}}_k^{\mathrm{sb}})^H\mathbf{H}\mathbf{s}(n)\mathbf{v}^H(n)\hat{\mathbf{h}}_k^{\mathrm{sb}}]\right\} \notag\\
&= P_s\|\mathbf{h}_k^H\mathbf{F}_s\mathbf{H}\|_2^2 + \sigma_v^2\mathbf{h}_k^H(\mathbf{F}_n+2\mathbf{F}_{sv})\mathbf{h}_k + \bar{\lambda}^2\frac{\sigma_v^4\alpha}{a\beta},
\end{align}
Indeed, applying  \eqref{eq:47}, we obtain
\begin{align}
    &\mathbb{E}[(\hat{\mathbf{h}}_k^{\mathrm{sb}})^H\mathbf{H}\mathbf{s}(n)\mathbf{s}^H(n)\mathbf{H}^H\hat{\mathbf{h}}_k^{\mathrm{sb}}] \asymp \bar{\lambda}^2P_s\mathbf{h}_k^H\mathbf{H}\mathbf{H}^H\mathbf{h}_k \notag\\
    &\qquad+ \bar{\lambda}(1-\bar{\lambda})P_s\mathbf{h}_k^H\mathbf{U}_s\pmb{\Lambda}_s\mathbf{U}_s^H\mathbf{H}\mathbf{H}^H\mathbf{h}_k\notag\\
    &\qquad+ \bar{\lambda}(1-\bar{\lambda})P_s\mathbf{h}_k^H\mathbf{H}\mathbf{H}^H\mathbf{U}_s\pmb{\Lambda}_s\mathbf{U}_s^H\mathbf{h}_k \notag\\
    &\qquad+ (1-\bar{\lambda})^2P_s\mathbf{h}_k^H\mathbf{U}_s\pmb{\Lambda}_s\mathbf{U}_s^H\mathbf{H}\mathbf{H}^H\mathbf{U}_s\pmb{\Lambda}_s\mathbf{U}_s^H\mathbf{h}_k \notag\\
    &\quad= P_s\|\bar{\lambda}\mathbf{h}_k^H\mathbf{H} + (1-\bar{\lambda})\mathbf{h}_k^H\mathbf{U}_s\pmb{\Lambda}_s\mathbf{U}_s^H\mathbf{H}\|_2^2 \notag\\
    &\quad= P_s\|\mathbf{h}_k^H\mathbf{F}_s\mathbf{H}\|_2^2, 
    \end{align}
    Similarly using Lemma \ref{first-order-signal-noise_bis} and \ref{second-order-signal-noise_bis}, we obtain:
    \begin{align}
    &\mathbb{E}[(\hat{\mathbf{h}}_k^{\mathrm{sb}})^H\mathbf{v}(n)\mathbf{v}(n)^H\hat{\mathbf{h}}_k^{\mathrm{sb}}]\asymp \bar{\lambda}^2\sigma_v^2\mathbf{h}_k^H\mathbf{h}_k \notag\\
    %&= \mathbb{E}[\hat{\lambda}^2\sigma_v^2(\hat{\mathbf{h}}_k^{\mathrm{tr}})^H\hat{\mathbf{h}}_k^{\mathrm{tr}} ] \notag\\
    %&\quad+ \mathbb{E}[\hat{\lambda}(1-\hat{\lambda})(\hat{\mathbf{h}}_k^{\mathrm{tr}})^H\hat{\mathbf{U}}_s\hat{\mathbf{U}}_s^H\mathbf{v}(n)\mathbf{v}(n)^H\hat{\mathbf{h}}_k^{\mathrm{tr}}] \notag\\  
    %&\quad+\mathbb{E}[ \hat{\lambda}(1-\hat{\lambda})(\hat{\mathbf{h}}_k^{\mathrm{tr}})^H\mathbf{v}(n)\mathbf{v}(n)^H\hat{\mathbf{U}}_s\hat{\mathbf{U}}_s^H\hat{\mathbf{h}}_k^{\mathrm{tr}}] \notag\\
    %&\quad+ \mathbb{E}[(1-\hat{\lambda})^2(\hat{\mathbf{h}}_k^{\mathrm{tr}})^H\hat{\mathbf{U}}_s\hat{\mathbf{U}}_s^H\mathbf{v}(n)\mathbf{v}(n)^H\hat{\mathbf{U}}_s\hat{\mathbf{U}}_s^H\hat{\mathbf{h}}_k^{\mathrm{tr}} ] \notag\\
    &\qquad+ \bar{\lambda}^2\frac{\sigma_v^4\alpha}{a\beta} + 2\bar{\lambda}(1-\bar{\lambda})\sigma_v^2\mathbf{h}_k^H\mathbf{U}_s\pmb{\Lambda}_n\mathbf{U}_s^H\mathbf{h}_k \notag\\
    &\qquad+ (1-\bar{\lambda})^2\sigma_v^2\mathbf{h}_k^H\mathbf{U}_s\bar{\pmb{\Lambda}}_n\mathbf{U}_s^H\mathbf{h}_k  \notag\\
    &\quad= \sigma_v^2\mathbf{h}_k^H\mathbf{F}_n\mathbf{h}_k + \bar{\lambda}^2\frac{\sigma_v^4\alpha}{a\beta},  
    \end{align}
    Using Corollary \ref{cor:by_product} along with \eqref{eq:zero}, we obtain:
    \begin{align}
&\mathbb{E}[(\hat{\mathbf{h}}_k^{\mathrm{sb}})^H\mathbf{H}\mathbf{s}(n)\mathbf{v}^H(n)\hat{\mathbf{h}}_k^{\mathrm{sb}}] \notag\\
%&\asymp\bar{\lambda}(1-\bar{\lambda})\mathbb{E}\!\left[\mathbf{h}_k^H\hat{\mathbf{U}}_s\hat{\mathbf{U}}_s^H\mathbf{H}\mathbf{s}(n)\mathbf{v}^H(n)\mathbf{h}_k\right] \notag\\
%&\quad+\bar{\lambda}(1-\bar{\lambda})\mathbb{E}\!\left[\mathbf{h}_k^H\mathbf{H}\mathbf{s}(n)\mathbf{v}^H(n)\hat{\mathbf{U}}_s\hat{\mathbf{U}}_s^H\mathbf{h}_k\right] \notag\\
%&\quad+(1-\bar{\lambda})^2\mathbb{E}\!\left[\mathbf{h}_k^H\hat{\mathbf{U}}_s\hat{\mathbf{U}}_s^H\mathbf{H}\mathbf{s}(n)\mathbf{v}^H(n)\hat{\mathbf{U}}_s\hat{\mathbf{U}}_s^H\mathbf{h}_k\right] \notag\\
%&\quad+\frac{\bar{\lambda}(1-\bar{\lambda})}{a^2L^2}\mathbb{E}\!\left[\mathbf{z}_k^H\mathbf{V}_p^H\hat{\mathbf{U}}_s\hat{\mathbf{U}}_s^H\mathbf{H}\mathbf{s}(n)\mathbf{v}^H(n)\mathbf{V}_p\mathbf{z}_k\right] \notag\\
%&\quad+\frac{\bar{\lambda}(1-\bar{\lambda})}{a^2L^2}\mathbb{E}\!\left[\mathbf{z}_k^H\mathbf{V}_p^H\mathbf{H}\mathbf{s}(n)\mathbf{v}^H(n)\hat{\mathbf{U}}_s\hat{\mathbf{U}}_s^H\mathbf{V}_p\mathbf{z}_k\right] \notag\\
%&\quad+\frac{(1-\bar{\lambda})^2}{a^2L^2}\mathbb{E}\!\left[\mathbf{z}_k^H\mathbf{V}_p^H\hat{\mathbf{U}}_s\hat{\mathbf{U}}_s^H\mathbf{H}\mathbf{s}(n)\mathbf{v}^H(n)\hat{\mathbf{U}}_s\hat{\mathbf{U}}_s^H\mathbf{V}_p\mathbf{z}_k\right] \notag\\
&\quad\asymp\bar{\lambda}(1-\bar{\lambda})\mathbb{E}[\mathbf{h}_k^H\hat{\mathbf{U}}_s\hat{\mathbf{U}}_s^H\mathbf{H}\mathbf{s}(n)\mathbf{v}^H(n)\mathbf{h}_k] \notag\\
&\qquad+\bar{\lambda}(1-\bar{\lambda})\mathbb{E}[\mathbf{h}_k^H\mathbf{H}\mathbf{s}(n)\mathbf{v}^H(n)\hat{\mathbf{U}}_s\hat{\mathbf{U}}_s^H\mathbf{h}_k] \notag\\
&\qquad+(1-\bar{\lambda})^2\mathbb{E}[\mathbf{h}_k^H\hat{\mathbf{U}}_s\hat{\mathbf{U}}_s^H\mathbf{H}\mathbf{s}(n)\mathbf{v}^H(n)\hat{\mathbf{U}}_s\hat{\mathbf{U}}_s^H\mathbf{h}_k] \notag\\
&\quad\asymp \bar{\lambda}(1-\bar{\lambda})c\sigma_v^2\mathbf{h}_k^H\mathbf U_s\pmb{\Lambda}_s\mathbf U_s^H\mathbf h_k \notag\\
&\qquad+ (1-\bar{\lambda})^2c\sigma_v^2
\mathbf{h}_k^{H}\mathbf{U}_{s}\pmb{\Lambda}_s^2\mathbf{U}_{s}^{H}\mathbf{h}_k = \sigma_v^2\mathbf{h}_k^H\mathbf{F}_{sv}\mathbf{h}_k. 
\end{align}

Consequently, by substituting the asymptotic expressions for the desired signal power, interference power, and noise power into the SINR expression $\mathrm{SINR}_{\mathrm{sb},k}^{\mathrm{MRC}}$, we obtain the final deterministic equivalent (\ref{semiblind_MRC_asy}).

\par

\section{Proof of Theorem~\ref{theorem_sb_MMSE}: Deterministic Equivalents of the SINR for Semi-blind and MMSE receivers} \label{app_sb_MMSE}
\subsection{Preliminaries:}
Using Woodbury matrix identify~\cite{higham2002accuracy}, $\hat{\mathbf{C}}^{\mathrm{sb}}$ can be rewritten as 
\begin{align}
    \hat{\mathbf{C}}^{\mathrm{sb}} = \frac{P_s}{\sigma_v^2}\mathbf{I}_M - \frac{P_s}{\sigma_v^2}\hat{\mathbf{H}}^{\mathrm{sb}}\hat{\mathbf{D}}^{\mathrm{sb}}(\hat{\mathbf{H}}^{\mathrm{sb}})^H,  
\end{align}
where
\begin{align}
    &\hat{\mathbf{D}}^{\mathrm{sb}} = \Big((\hat{\mathbf{H}}^{\mathrm{sb}})^H\hat{\mathbf{H}}^{\mathrm{sb}}+\frac{\sigma_v^2}{P_s}\mathbf{I}_K\Big)^{-1} \notag\\
    &= \Big(\hat{\lambda}^2(\hat{\mathbf{H}}^{\mathrm{tr}})^H\hat{\mathbf{H}}^{\mathrm{tr}}+\frac{\sigma_v^2}{P_s}\mathbf{I}_K + (1-\hat{\lambda}^2)(\hat{\mathbf{H}}^{\mathrm{tr}})^H\hat{\mathbf{U}}_s\hat{\mathbf{U}}_s^H\hat{\mathbf{H}}^{\mathrm{tr}}\Big)^{-1}. \notag
\end{align}
%Motivated by~\cite{couillet2011random,yang2018high}, we have
Using \eqref{eq:convergence_quadratic_form}, we obtain $(\hat{\mathbf{h}}_k^{\mathrm{tr}})^H\hat{\mathbf{u}}_{si}\hat{\mathbf{u}}_{si}^H\hat{\mathbf{h}}_k^{\mathrm{tr}} \asymp z_{si}{\bf h}_k^{H}\mathbf{u}_{si}\mathbf{u}_{si}^H{\bf h}_k$ for $i=1,\ldots,K$, thus the deterministic equivalent of $\hat{\mathbf{D}}^{\mathrm{sb}}$ is given by  
\begin{align}
    \bar{\mathbf{D}}^{\mathrm{sb}} = \Big(\mathbf{H}^H\mathbf{M}_s\mathbf{H} + \frac{\alpha\sigma_v^2}{a\beta}\bar{\lambda}^2\mathbf{I}_K + \frac{\sigma_v^2}{P_s}\mathbf{I}_K\Big)^{-1},
\end{align}
where $\mathbf{M}_s=\bar{\lambda}^2\mathbf{I}_M+(1-\bar{\lambda}^2)\mathbf{U}_s\pmb{\Lambda}_s\mathbf{U}_s^H$. Similarly, we have 
\begin{align}
    &(\hat{\mathbf{h}}_k^{\mathrm{sb}})^H\hat{\mathbf{H}}^{\mathrm{sb}} = \hat{\lambda}^2(\hat{\mathbf{h}}_k^{\mathrm{tr}})^H\hat{\mathbf{H}}^{\mathrm{tr}} + (1-\hat{\lambda}^2)(\hat{\mathbf{h}}_k^{\mathrm{tr}})^H\hat{\mathbf{U}}_s\hat{\mathbf{U}}_s^H\hat{\mathbf{H}}^{\mathrm{tr}} \notag\\
    &\qquad\asymp \bar{\lambda}^2\mathbf{h}_k^H\mathbf{H}+\bar{\lambda}^2\frac{\sigma_v^2\alpha}{a\beta}\mathbf{e}_k^T + (1-\bar{\lambda}^2)\mathbf{h}_k^H\mathbf{U}_s\pmb{\Lambda}_s\mathbf{U}_s^H\mathbf{H} \notag\\
    &\qquad= \mathbf{h}_k^H\mathbf{M}_s\mathbf{H}+\bar{\lambda}^2\frac{\sigma_v^2\alpha}{a\beta}\mathbf{e}_k^T \triangleq \mathbf{f}_k^{\mathrm{sb}}. 
\end{align}

\subsection{Derivation of the deterministic equivalent for the SINR}
Based on Lemmas~\ref{first-order-signal-noise_bis} and~\ref{second-order-signal-noise_bis}, Corollary~\ref{corollary1}, and the derivations in Appendix~\ref{app_sb_MRC_res}, we obtain the deterministic equivalent of the SINR achieved by the MMSE detector with semi-blind channel estimation.

1) Deterministic equivalent for signal power: The signal power converges almost surely to its deterministic equivalent as
\begin{align}
&\left|\mathbb{E}\!\left[(\hat{\mathbf{g}}_{\mathrm{sb},k}^{\mathrm{MMSE}})^H\mathbf{y}(n)s_k^{\ast}(n)\right]\right|^2 \notag\\
%&=\left|\mathbb{E}[(\hat{\mathbf{h}}_k^{\mathrm{sb}})^H\hat{\mathbf{C}}^{\mathrm{sb}}\mathbf{H}\mathbf{s}(n)s_k^{\ast}(n)]+\mathbb{E}[(\hat{\mathbf{h}}_k^{\mathrm{sb}})^H\hat{\mathbf{C}}^{\mathrm{sb}}\mathbf{v}(n)s_k^{\ast}(n)]\right|^2 \notag\\
%&=\Big|\sum_{i=1}^{K}\mathbb{E}[(\hat{\mathbf{h}}_k^{\mathrm{sb}})^H\hat{\mathbf{C}}^{\mathrm{sb}}\mathbf{h}_i s_i(n)s_k^{\ast}(n)]+\mathbb{E}[(\hat{\mathbf{h}}_k^{\mathrm{sb}})^H\hat{\mathbf{C}}^{\mathrm{sb}}\mathbf{v}(n)s_k^{\ast}(n)]\Big|^2 \notag\\
%&=\Big|\frac{P_s}{\sigma_v^2}\sum_{i=1}^{K}\mathbb{E}[(\hat{\mathbf{h}}_k^{\mathrm{sb}})^H\mathbf{h}_i s_i(n)s_k^{\ast}(n)] \notag\\
%&\quad-\frac{P_s}{\sigma_v^2}\sum_{i=1}^{K}\mathbb{E}[(\hat{\mathbf{h}}_k^{\mathrm{sb}})^H\hat{\mathbf{H}}^{\mathrm{sb}}\hat{\mathbf{D}}^{\mathrm{sb}}(\hat{\mathbf{H}}^{\mathrm{sb}})^H\mathbf{h}_i s_i(n)s_k^{\ast}(n)] \notag\\
%&\quad+\frac{P_s}{\sigma_v^2}\mathbb{E}[(\hat{\mathbf{h}}_k^{\mathrm{sb}})^H\mathbf{v}(n)s_k^{\ast}(n)] \notag\\
%&\quad-\frac{P_s}{\sigma_v^2}\mathbb{E}[(\hat{\mathbf{h}}_k^{\mathrm{sb}})^H\hat{\mathbf{H}}^{\mathrm{sb}}\hat{\mathbf{D}}^{\mathrm{sb}}(\hat{\mathbf{H}}^{\mathrm{sb}})^H\mathbf{v}(n)s_k^{\ast}(n)]\Big|^2 \notag\\
&\asymp \Big|\frac{P_s^2}{\sigma_v^2}(\bar{\lambda}\mathbf{h}_k^H\mathbf{h}_k+(1-\bar{\lambda})\mathbf{h}_k^H\mathbf{U}_s\mathbf{\Lambda}_s\mathbf{U}_s^H\mathbf{h}_k) \notag\\
&\quad -\frac{P_s^2}{\sigma_v^2}\mathbf{f}_k^{\mathrm{sb}}\bar{\mathbf{D}}^{\mathrm{sb}}(\bar{\lambda}\mathbf{H}^H\mathbf{h}_k+(1-\bar{\lambda})\mathbf{H}^H\mathbf{U}_s\mathbf{\Lambda}_s\mathbf{U}_s^H\mathbf{h}_k) \notag\\
&\quad + \frac{P_s^2}{\sigma_v^2}(1-\bar{\lambda})\mathbf{h}_k^H\mathbf{U}_s\mathbf{\Lambda}_{vs}\mathbf{U}_s^H\mathbf{h}_k \notag\\
&\quad -\frac{P_s^2}{\sigma_v^2}(1-\bar{\lambda})\mathbf{f}_k^{\mathrm{sb}}\bar{\mathbf{D}}^{\mathrm{sb}}\mathbf{H}^H\mathbf{U}_s\mathbf{\Lambda}_{vs}\mathbf{U}_s^H\mathbf{h}_k\Big|^2 \notag\\
&\asymp \frac{P_s^4}{\sigma_v^4}\left|(\mathbf{t}_k^{\mathrm{sb}})^H\mathbf{F}_s\mathbf{h}_k\right|^2.
\end{align}
%where $\mathbf{t}_k^{\mathrm{sb}} = \mathbf{h}_k - \mathbf{H}\bar{\mathbf{D}}^{\mathrm{sb}}(\mathbf{f}_k^{\mathrm{sb}})^H$. 

2) Deterministic equivalent for interference-plus-noise power: The deterministic equivalent of the interference-plus -noise power is given by 
\begin{align}
&\mathbb{E}[|(\hat{\mathbf{g}}_{\mathrm{sb},k}^{\mathrm{MMSE}})^H\mathbf{y}(n)|^2] \notag\\
&\qquad=\mathbb{E}[(\hat{\mathbf{h}}_k^{\mathrm{sb}})^H\hat{\mathbf{C}}^{\mathrm{sb}}\mathbf{H}\mathbf{s}(n)\mathbf{s}^H(n)\mathbf{H}^H\hat{\mathbf{C}}^{\mathrm{sb}}\hat{\mathbf{h}}_k^{\mathrm{sb}}] \notag\\
&\quad\qquad+\mathbb{E}[(\hat{\mathbf{h}}_k^{\mathrm{sb}})^H\hat{\mathbf{C}}^{\mathrm{sb}}\mathbf{v}(n)\mathbf{v}^H(n)\hat{\mathbf{C}}^{\mathrm{sb}}\hat{\mathbf{h}}_k^{\mathrm{sb}}] \notag\\
&\quad\qquad+2\mathrm{Re}\left\{\mathbb{E}[(\hat{\mathbf{h}}_k^{\mathrm{sb}})^H\hat{\mathbf{C}}^{\mathrm{sb}}\mathbf{H}\mathbf{s}(n)\mathbf{v}^H(n)\hat{\mathbf{C}}^{\mathrm{sb}}\hat{\mathbf{h}}_k^{\mathrm{sb}}]\right\} \notag\\
&\qquad= \frac{P_s^3}{\sigma_v^4}\left\|(\mathbf{t}_k^{\mathrm{sb}})^H\mathbf{F}_s\mathbf{H}\right\|_2^2 + \frac{P_s^2}{\sigma_v^2}(\mathbf{t}_k^{\mathrm{sb}})^H(\mathbf{F}_n+2\mathbf{F}_{sv})\mathbf{t}_k^{\mathrm{sb}} \notag\\
&\qquad\quad+ \bar{\lambda}^2\frac{P_s^2\alpha}{a\beta}\|\mathbf{e}_k^T-\mathbf{f}_k^{\mathrm{sb}}\bar{\mathbf{D}}^{\mathrm{sb}}\|_2^2,
\end{align}
where
\begin{align}
    &\mathbb{E}[(\hat{\mathbf{h}}_k^{\mathrm{sb}})^H\hat{\mathbf{C}}^{\mathrm{sb}}\mathbf{H}\mathbf{s}(n)\mathbf{s}^H(n)\mathbf{H}^H\hat{\mathbf{C}}^{\mathrm{sb}}\hat{\mathbf{h}}_k^{\mathrm{sb}}] \notag\\
    &=\frac{P_s^2}{\sigma_v^4}\mathbb{E}[(\hat{\mathbf{h}}_k^{\mathrm{sb}})^H\mathbf{H}\mathbf{s}(n)\mathbf{s}^H(n)\mathbf{H}^H\hat{\mathbf{h}}_k^{\mathrm{sb}}] \notag\\
    &\quad-\frac{P_s^2}{\sigma_v^4}\mathbb{E}[(\hat{\mathbf{h}}_k^{\mathrm{sb}})^H\hat{\mathbf{H}}^{\mathrm{sb}}\hat{\mathbf{D}}^{\mathrm{sb}}(\hat{\mathbf{H}}^{\mathrm{sb}})^H\mathbf{H}\mathbf{s}(n)\mathbf{s}^H(n)\mathbf{H}^H\hat{\mathbf{h}}_k^{\mathrm{sb}}] \notag\\
    &\quad-\frac{P_s^2}{\sigma_v^4}\mathbb{E}[(\hat{\mathbf{h}}_k^{\mathrm{sb}})^H\mathbf{H}\mathbf{s}(n)\mathbf{s}^H(n)\mathbf{H}^H\hat{\mathbf{H}}^{\mathrm{sb}}\hat{\mathbf{D}}^{\mathrm{sb}}(\hat{\mathbf{H}}^{\mathrm{sb}})^H\hat{\mathbf{h}}_k^{\mathrm{sb}}] \notag\\
    &\quad+\frac{P_s^2}{\sigma_v^4}\mathbb{E}[|(\hat{\mathbf{h}}_k^{\mathrm{sb}})^H\hat{\mathbf{H}}^{\mathrm{sb}}\hat{\mathbf{D}}^{\mathrm{sb}}(\hat{\mathbf{H}}^{\mathrm{sb}})^H\mathbf{H}\mathbf{s}(n)|^2] \notag\\
    &\asymp\frac{P_s^3}{\sigma_v^4}\left\|\mathbf{h}_k^H\mathbf{F}_s\mathbf{H}-\mathbf{f}_k^{\mathrm{sb}}\bar{\mathbf{D}}^{\mathrm{sb}}\mathbf{H}^H\mathbf{F}_s\mathbf{H}\right\|_2^2 \notag\\
    &=\frac{P_s^3}{\sigma_v^4}\left\|(\mathbf{t}_k^{\mathrm{sb}})^H\mathbf{F}_s\mathbf{H}\right\|_2^2, 
    \end{align}
    \begin{align}
    &\mathbb{E}[(\hat{\mathbf{h}}_k^{\mathrm{sb}})^H\hat{\mathbf{C}}^{\mathrm{sb}}\mathbf{v}(n)\mathbf{v}(n)^H\hat{\mathbf{C}}^{\mathrm{sb}}\hat{\mathbf{h}}_k^{\mathrm{sb}}] \notag\\
    %&= \frac{P_s^2}{\sigma_v^4}\mathbb{E}[(\hat{\mathbf{h}}_k^{\mathrm{sb}})^H\mathbf{v}(n)\mathbf{v}(n)^H\hat{\mathbf{h}}_k^{\mathrm{sb}}] \notag\\
    %&\quad- \frac{P_s^2}{\sigma_v^4}\mathbb{E}[(\hat{\mathbf{h}}_k^{\mathrm{sb}})^H\hat{\mathbf{H}}^{\mathrm{sb}}\hat{\mathbf{D}}^{\mathrm{sb}}(\hat{\mathbf{H}}^{\mathrm{sb}})^H\mathbf{v}(n)\mathbf{v}(n)^H\hat{\mathbf{h}}_k^{\mathrm{sb}}] \notag\\
    %&\quad- \frac{P_s^2}{\sigma_v^4}\mathbb{E}[(\hat{\mathbf{h}}_k^{\mathrm{sb}})^H\mathbf{v}(n)\mathbf{v}(n)^H\hat{\mathbf{H}}^{\mathrm{sb}}\hat{\mathbf{D}}^{\mathrm{sb}}(\hat{\mathbf{H}}^{\mathrm{sb}})^H\hat{\mathbf{h}}_k^{\mathrm{sb}}] \notag\\
    %&\quad+  \frac{P_s^2}{\sigma_v^4}\|\mathbb{E}[(\hat{\mathbf{h}}_k^{\mathrm{sb}})^H\hat{\mathbf{H}}^{\mathrm{sb}}\hat{\mathbf{D}}^{\mathrm{sb}}(\hat{\mathbf{H}}^{\mathrm{sb}})^H\mathbf{v}(n)\|_2^2] \notag\\
    &\asymp \frac{P_s^2}{\sigma_v^2}\mathbf{h}_k^H\mathbf{F}_n\mathbf{h}_k + \bar{\lambda}^2\frac{P_s^2\alpha}{a\beta} - \frac{P_s^2}{\sigma_v^2}\mathbf{f}_k^{\mathrm{sb}}\bar{\mathbf{D}}^{\mathrm{sb}}\mathbf{H}^H\mathbf{F}_n\mathbf{h}_k \notag\\
    &\quad- \bar{\lambda}^2\frac{P_s^2\alpha}{a\beta}\mathbf{f}_k^{\mathrm{sb}}\bar{\mathbf{D}}^{\mathrm{sb}}\mathbf{e}_k - \frac{P_s^2}{\sigma_v^2}\mathbf{h}_k^H\mathbf{F}_n\mathbf{H}\bar{\mathbf{D}}^{\mathrm{sb}}(\mathbf{f}_k^{\mathrm{sb}})^H \notag\\
    &\quad- \bar{\lambda}^2\frac{P_s^2\alpha}{a\beta}\mathbf{e}_k^T\bar{\mathbf{D}}^{\mathrm{sb}}(\mathbf{f}_k^{\mathrm{sb}})^H \notag\\
    &\quad+ \frac{P_s^2}{\sigma_v^2}\mathbf{f}_k^{\mathrm{sb}}\bar{\mathbf{D}}^{\mathrm{sb}}\mathbf{H}^H\mathbf{F}_n\mathbf{H}\bar{\mathbf{D}}^{\mathrm{sb}}(\mathbf{f}_k^{\mathrm{sb}})^H\notag\\
    &\quad + \bar{\lambda}^2\frac{P_s^2\alpha}{a\beta}\mathbf{f}_k^{\mathrm{sb}}\bar{\mathbf{D}}^{\mathrm{sb}}\bar{\mathbf{D}}^{\mathrm{sb}}(\mathbf{f}_k^{\mathrm{sb}})^H \notag\\
    %&= \frac{P_s^2}{\sigma_v^2}\|\mathbf{h}_k^H\mathbf{F}_n^{1/2}-\mathbf{f}_k^{\mathrm{sb}}\bar{\mathbf{D}}^{\mathrm{sb}}\mathbf{H}^H\mathbf{F}_n^{1/2}\|_2^2 \notag\\
    %&\quad+ \bar{\lambda}^2\frac{P_s^2\alpha}{a\beta}\|\mathbf{e}_k^T-\mathbf{f}_k^{\mathrm{sb}}\bar{\mathbf{D}}^{\mathrm{sb}}\|_2^2 \notag\\
    &=\frac{P_s^2}{\sigma_v^2}(\mathbf{t}_k^{\mathrm{sb}})^H\mathbf{F}_n\mathbf{t}_k^{\mathrm{sb}} + \bar{\lambda}^2\frac{P_s^2\alpha}{a\beta}\|\mathbf{e}_k^T-\mathbf{f}_k^{\mathrm{sb}}\bar{\mathbf{D}}^{\mathrm{sb}}\|_2^2, \\
&\mathbb{E}[(\hat{\mathbf{h}}_k^{\mathrm{sb}})^H
\hat{\mathbf{C}}^{\mathrm{sb}}\mathbf{H}\mathbf{s}(n)
\mathbf{v}^H(n)\hat{\mathbf{C}}^{\mathrm{sb}}
\hat{\mathbf{h}}_k^{\mathrm{sb}}] \notag\\
%&=\frac{P_s^2}{\sigma_v^4}\mathbb{E}[(\hat{\mathbf{h}}_k^{\mathrm{sb}})^H\mathbf{H}\mathbf{s}(n)\mathbf{v}^H(n)\hat{\mathbf{h}}_k^{\mathrm{sb}}] \notag\\
%&\quad-\frac{P_s^2}{\sigma_v^4}\mathbb{E}[(\hat{\mathbf{h}}_k^{\mathrm{sb}})^H\hat{\mathbf{H}}^{\mathrm{sb}}\hat{\mathbf{D}}^{\mathrm{sb}}(\hat{\mathbf{H}}^{\mathrm{sb}})^H\mathbf{H}\mathbf{s}(n)\mathbf{v}^H(n)\hat{\mathbf{h}}_k^{\mathrm{sb}}] \notag\\
%&\quad-\frac{P_s^2}{\sigma_v^4}\mathbb{E}[(\hat{\mathbf{h}}_k^{\mathrm{sb}})^H\mathbf{H}\mathbf{s}(n)\mathbf{v}^H(n)\hat{\mathbf{H}}^{\mathrm{sb}}\hat{\mathbf{D}}^{\mathrm{sb}}(\hat{\mathbf{H}}^{\mathrm{sb}})^H\hat{\mathbf{h}}_k^{\mathrm{sb}}] \notag\\
%&\quad+\frac{P_s^2}{\sigma_v^4}\mathbb{E}[(\hat{\mathbf{h}}_k^{\mathrm{sb}})^H\hat{\mathbf{H}}^{\mathrm{sb}}\hat{\mathbf{D}}^{\mathrm{sb}}(\hat{\mathbf{H}}^{\mathrm{sb}})^H\mathbf{H}\mathbf{s}(n)\mathbf{v}^H(n)\hat{\mathbf{H}}^{\mathrm{sb}}\hat{\mathbf{D}}^{\mathrm{sb}}(\hat{\mathbf{H}}^{\mathrm{sb}})^H\hat{\mathbf{h}}_k^{\mathrm{sb}}]\\
&\asymp \frac{P_s^2}{\sigma_v^2}\mathbf{h}_k^H\mathbf{F}_{sv}\mathbf{h}_k -\frac{P_s^2}{\sigma_v^2}\mathbf{f}_k^{\mathrm{sb}}\bar{\mathbf{D}}^{\mathrm{sb}}\mathbf{H}^H\mathbf{F}_{sv}\mathbf{h}_k \notag\\
&-\frac{P_s^2}{\sigma_v^2}\mathbf{h}_k^H\mathbf{F}_{sv}\mathbf{H}\bar{\mathbf{D}}^{\mathrm{sb}}(\mathbf{f}_k^{\mathrm{sb}})^H + \frac{P_s^2}{\sigma_v^2}\mathbf{f}_k^{\mathrm{sb}}\bar{\mathbf{D}}^{\mathrm{sb}}\mathbf{H}^H\mathbf{F}_{sv}\mathbf{H}\bar{\mathbf{D}}^{\mathrm{sb}}(\mathbf{f}_k^{\mathrm{sb}})^H \notag\\
&=\frac{P_s^2}{\sigma_v^2}(\mathbf{t}_k^{\mathrm{sb}})^H\mathbf{F}_{sv}\mathbf{t}_k^{\mathrm{sb}}.
\end{align}

As a result, by combining the asymptotic expressions of the signal, interference, and noise power terms, the deterministic equivalent of $\mathrm{SINR}_{\mathrm{sb},k}^{\mathrm{MMSE}}$ is expressed as (\ref{semiblind_MMSE_asy}). 

\bibliographystyle{IEEEtran}
\bibliography{references}

\end{document}